# Data-knowledge dual-driven intelligent framework for full-chain, experiment-efficient synthesis of 2D dendrites

Wenqiang Huang[1,2,†], Xuhang Gu[3,†], Susu Fang[2,†,*], Shen'ao Xue[1,2], Huanhuan Xing[2], Junjie Jiang[1,2], Junying Zhang[1,2], Shen Zhou[4], Zheng Luo[2], Jin Zhang[3], Fangping Ouyang[1,5,6*] & Shanshan Wang[2,3,7*]

[1]*School of Physics, Hunan Key Laboratory for Super-Microstructure and Ultrafast Process, and Hunan Key Laboratory of Nanophotonics and Devices, Central South University, Changsha 410083, China*

[2]*College of Aerospace Science and Engineering, National University of Defense Technology, Changsha 410000, China*

[3]*School of Advanced Materials, Peking University Shenzhen Graduate School, Shenzhen 518055, China*

[4]*College of Science, National University of Defense Technology, Changsha 410000, China*

[5]*School of Physics and Technology, State Key Laboratory of Chemistry and Utilization of Carbon Based Energy Resources, Xinjiang University, Urumqi 830046, China*

[6]*State Key Laboratory of Powder Metallurgy, and Powder Metallurgy Research Institute, Central South University, Changsha 410083, China*

[7]*Guangdong Provincial Key Laboratory of Nano-Micro Materials Research, Peking University Shenzhen Graduate School, Shenzhen 518055, China*

†These authors contributed equally to this work.

*Corresponding authors, Email: fangsusugfkd@nudt.edu.cn (Susu Fang); oyfp@csu.edu.cn (Fangping Ouyang); wangshanshan08@nudt.edu.cn (Shanshan Wang)

## Graphical abstract

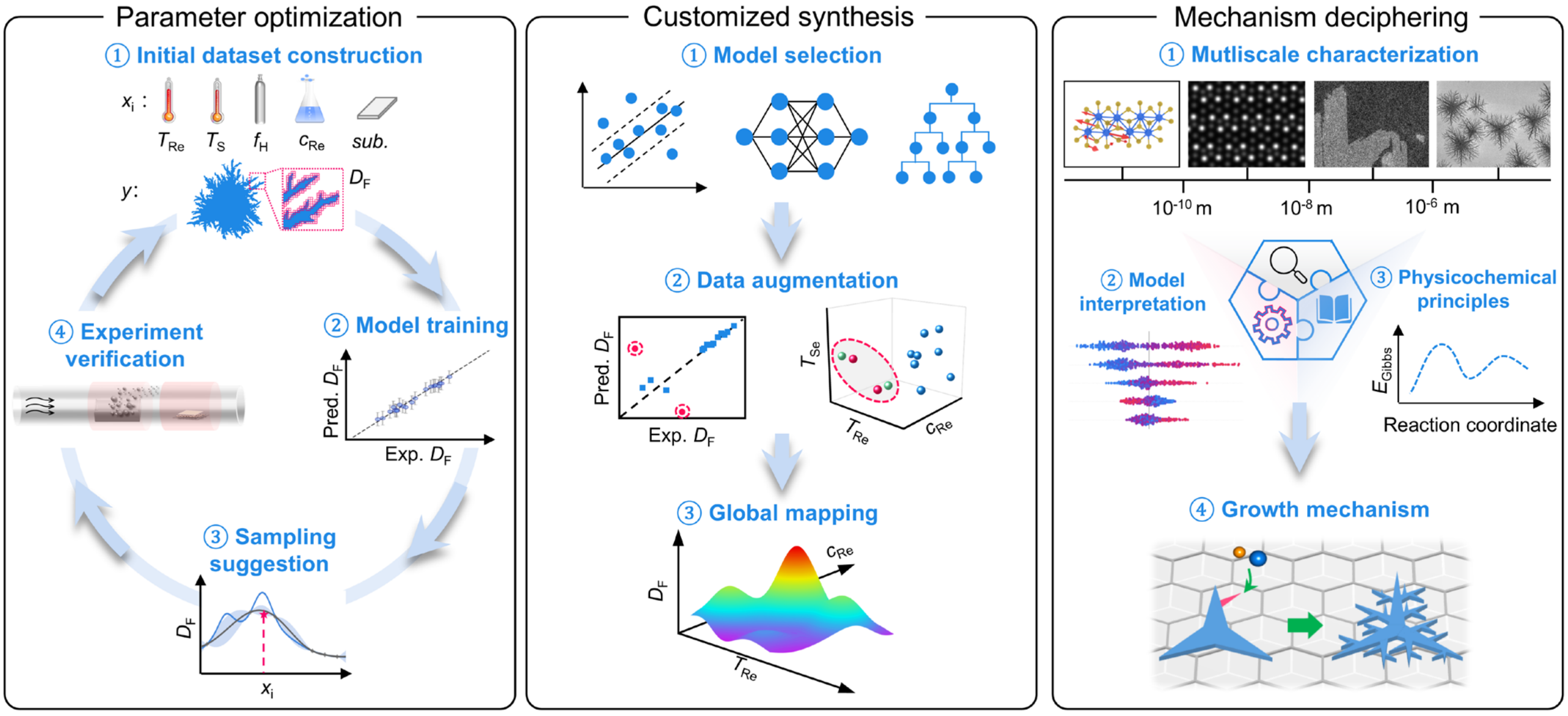

Two-dimensional (2D) dendrites feature atomic thinness, abundant unsaturated sites and exotic edge states, holding great promise in catalysis [1,2], nonlinear optics [3,4], and dielectric memristors [5]. Chemical vapor deposition (CVD) is ideal for their scalable preparation, yet its numerous process parameters and complex mechanisms render traditional trial-and-error methods inefficient [6–9]. Machine learning (ML) excels at solving high-dimensional problems and has been increasingly utilized in materials research [10–14]. However, current applications in synthesis are often fragmented, lacking a holistic framework that integrates process optimization, customized synthesis, and multifactorial mechanism deciphering, especially under data scarcity (Table S1 online).

Here we report a small-data machine intelligence framework for full-chain synthesis, using 2D rhenium diselenide ($ReSe_2$) dendrites as a model system—a representative low-symmetry material that involves diverse preparation parameters and complex growth kinetics. The framework integrates active learning, data-augmented XGBoost algorithm, and data-knowledge dual-driven ML to achieve three tightly connected objectives (see graphic abstract): (i) rapid process optimization for highly-branched dendrites within just one week, (ii) construction of a nonlinear recipe–morphology mapping for customized synthesis using fewer than 70 experiments (<1.5% of total parameter combinations), and (iii) interpretable multifactorial mechanism decoding by combining SHapley Additive exPlanations (SHAP) analysis [15], cross-scale characterizations, and phase-field simulations. Our work establishes a human-AI synergistic paradigm that intertwines data-driven induction with knowledge-driven reasoning, universally applicable to broader materials synthesis.

The ML framework comprises three modules: process optimization, customized synthesis, and mechanism deciphering. For process optimization, five key CVD parameters are selected as input variables, including Re source temperature ($T_{Re}$), Se temperature ($T_{Se}$), molar concentration of Re precursor solution ($c_{Re}$), $H_2$ flow rate ($f_{H2}$), and substrate type (*sub.*) (Tables S2-S5 online). The perimetric fractal dimension $D_F$ is used as the target output to quantify the branching degree of dendrites (Figs. S1-S3 online). Bayesian optimization (BO) with Gaussian process regression as the surrogate model and max-value entropy search as the acquisition function is adopted for active learning. After only 4 iterations (60 experiments, <1.3% of total parameter combinations), the median $D_F$ increases from 1.36 to 1.61 (Fig. 1a), yielding a 69.4% boost in the feasible range ($D_F \in (1, 2)$) with a maximum of 1.71.

Visualization of the active learning process reveals a CVD parameter-dependent sampling strategy (Fig. 1b). Dominant parameters ($T_{Re}$ and *sub.*) converge rapidly via exploitation within the first iteration to establish a high-performance baseline ($T_{Re} \approx 620$°C, *sub.* = $Al_2O_3$), while variables such as $T_{Se}$ and $f_{H2}$ remain non-convergent to map complex interactions. Notably, $c_{Re}$ follows an "exploration-first, exploitation-later" trajectory, narrowing down to ≈ 0.75 mol/L after iteration 3. This feature-dependent sampling steers the search from a dispersed distribution to a focused cluster, maximizing $D_F$ with minimized uncertainty (Fig. 1c, Figs. S4-S9 online).

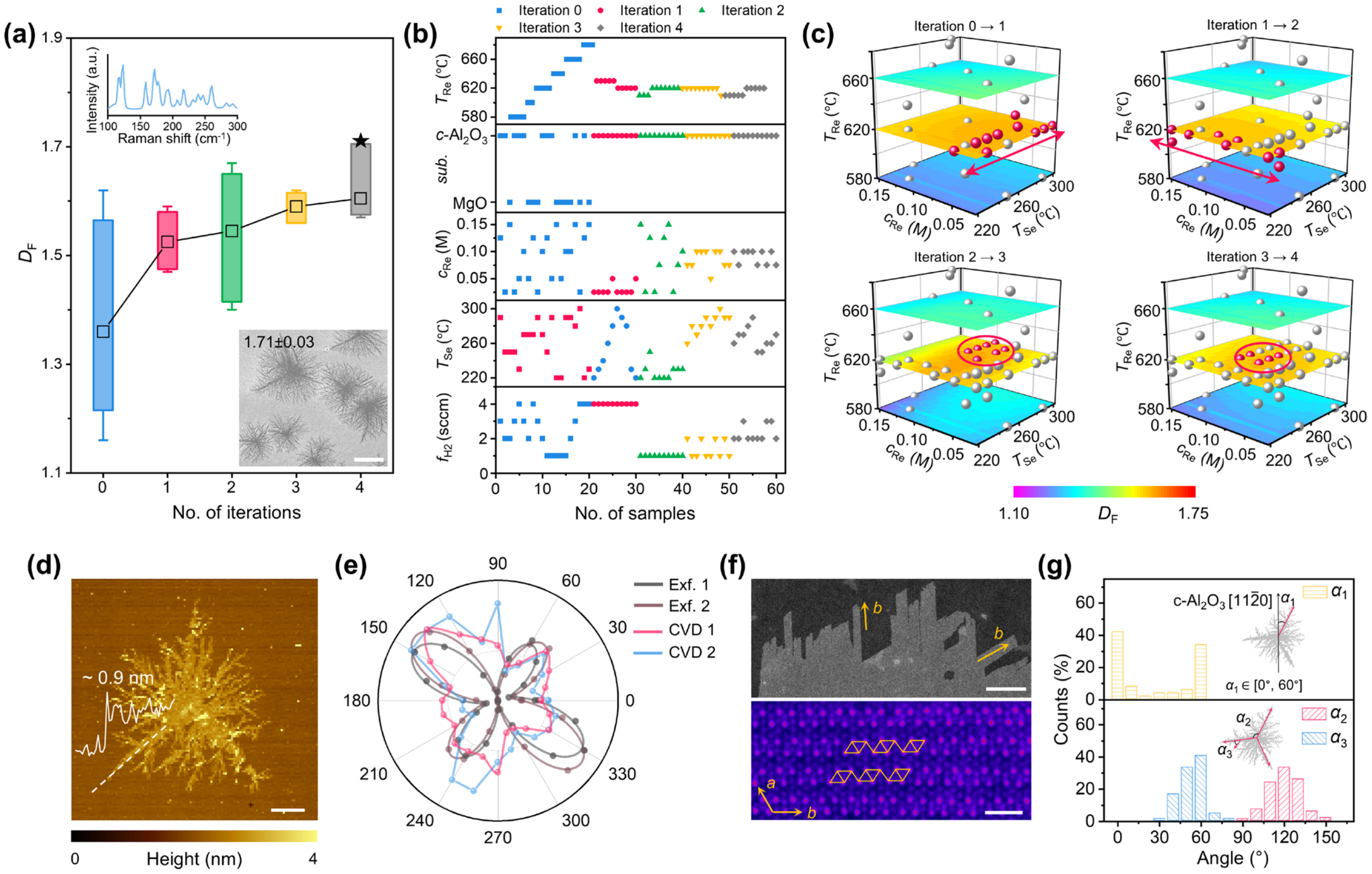

**Fig. 1.** Active learning-guided process optimization and structural characterization of 2D $ReSe_2$ dendrites. (a) Box plot showing the improvement of the $D_F$ with iterative cycles. Insets show the Raman spectra (top) and the typical scanning electron microscopy image of the $ReSe_2$ dendrites with a maximum $D_F$ = 1.71±0.03 achieved in the final iteration (bottom, black star). (b) Evolution of five process parameters' values across iterations. Iteration 0 contains 20 initial experiments, and each subsequent iteration adds 10 experiments, shown in different colors. (c) Visualization of the Bayesian optimization process by exhibiting the 3D CVD parameter-$D_F$ plot and the sampling strategies in every iteration $s \rightarrow (s+1)$ ($s$=0,1,2,3). CVD parameters: $T_{Re}$, $c_{Re}$, and $T_{Se}$. Contours denote predicted $D_F$, gray dots indicate existing samples in iteration $s$, and red dots indicate samples suggested for iteration ($s$+1). (d) Atomic force microscopy image of a monolayer $ReSe_2$ dendrite. Scale bar: 1 μm. (e) Angle-resolved Raman intensities of the 162 $cm^{-1}$ mode measured at different positions on CVD-grown dendritic $ReSe_2$ and exfoliated single-crystalline $ReSe_2$. (f) Medium- and high-magnification ADF-STEM images of $ReSe_2$ dendrites. Yellow arrows indicate the $b$-axis of $ReSe_2$ (top), while the rhombus markers represent the Re4 chains (bottom). Scale bars: 100 nm (top), 1 nm (bottom). (g) Angle distribution of $\alpha_1$, $\alpha_2$, and $\alpha_3$ for 2D $ReSe_2$ grown on $c$-$Al_2O_3$. $\alpha_1$: the minimum angle between $c$-$Al_2O_3$ [11$\bar{2}$0] and the primary branch of $ReSe_2$. $\alpha_2$: the minimum angle between the primary branches. $\alpha_3$: the minimum angle between the primary and secondary branches. Insets define $\alpha_1$, $\alpha_2$, and $\alpha_3$.

Multiple characterizations confirm that the product exhibits a monolayer, snowflake-like morphology with characteristic $ReSe_2$ vibrational modes and favorable hydrogen evolution activity (insets of Fig. 1a, Fig. 1d, Figs. S10-S12 online). Angle-resolved Raman spectroscopy and annular dark-field scanning transmission electron microscopy (ADF-STEM) reveal that the CVD-grown dendrites are mesoscale polycrystalline with well-ordered Re4 chains at the atomic scale (Fig. 1e and f). Crucially, their growth orientations are dictated by substrate symmetry, yielding preferred branches at 0°/60° on *c*-$Al_2O_3$ ($C_{3v}$) and 90° on MgO ($C_{4v}$) (Fig. 1g, Figs. S13 and S14 online).

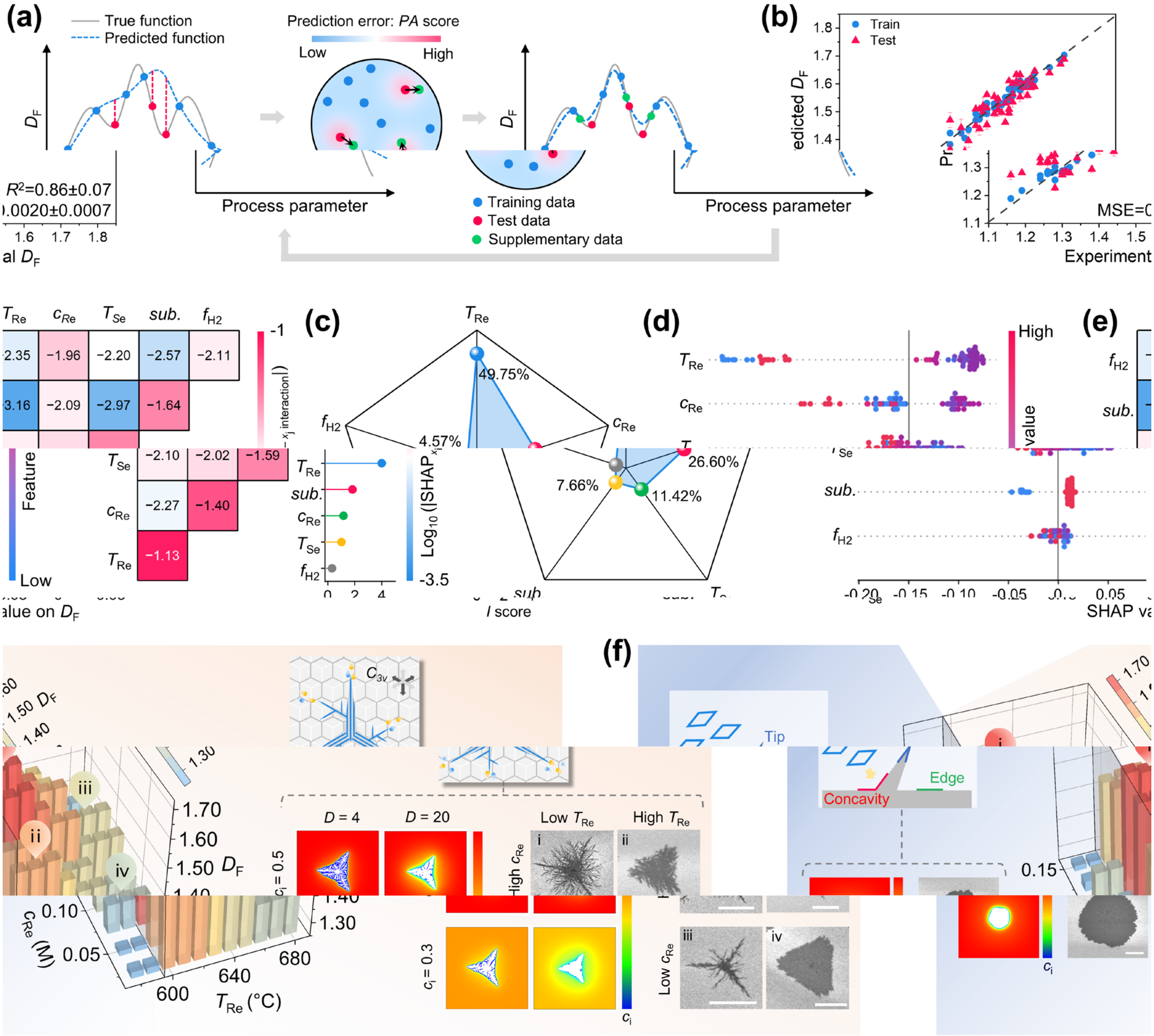

**Fig. 2.** Prediction accuracy (*PA*)-guided model refinement and data–knowledge dual-driven growth mechanism analysis of 2D $ReSe_2$ dendrites. (a) Schematic of the *PA*-guided data augmentation strategy. (b) Prediction performance of the XGBoost model after data augmentation. (c) Feature importance of five CVD parameters on $D_F$ determined by normalized SHAP values. (d) Beeswarm plot of SHAP values. (e) SHAP interaction value matrix in the logarithmic form obtained from the data-augmented XGBoost model. Diagonal elements denote $SHAP_{x_i\ main}$ of individual CVD parameters on $D_F$, whereas off-diagonal elements denote $SHAP_{x_i\text{-}x_j\ interaction}$ between different parameters. The bottom right panel shows the *I* scores of the five parameters. (f) Data-knowledge dual-driven multifactorial growth mechanism deciphering of 2D $ReS_2$ dendrites. Middle panel: Correlation between the most important CVD parameters ($T_{Re}$ and $c_{Re}$) and $D_F$ of 2D $ReS_2$ crystals grown on *c*-$Al_2O_3$ predicted by XGBoost. Left panel: Attachment-limited growth mechanism when $T_{Re}$ is below 600°C. Right panel: Diffusion-limited growth mechanism when $T_{Re}$ rises above 600°C. Phase-field simulations coupled with experimental observations are displayed in the bottom right panels. Scale bars: 2 μm.

To achieve on-demand synthesis of desired $D_F$, we screened six ML models and identified XGBoost regressor as the optimal candidate with an initial coefficient of determination ($R^2$) of 0.74 (Figs. S15 and S16 online). To overcome limited global accuracy from uneven data distribution, we proposed a prediction accuracy (*PA*)-guided data augmentation strategy to supplement experiments near the worst predicted points (*i.e.*, highest *PA* score, Fig. 2a):

$$PA\ \text{score} = \frac{1}{n}\sum_{i=1}^{n}\left(y_{\text{pred.}} - y_{\text{exp.}}\right)^2, \tag{1}$$

where $y_{exp.}$ and $y_{pred.}$ represent experimental and predicted $D_F$, respectively, and $n$ denotes the number of 5-fold cross-validation iterations. By iteratively compensating the blind spots of parameter space with merely 9 supplementary experiments over 3 rounds, the model accuracy increased to $R^2$= 0.86 with a 41.2% mean square error (MSE) reduction (Fig. 2b, Figs. S17-S19 and Tables S6-S12 online). This established a high-fidelity CVD parameter-$D_F$ mapping (Fig. 2f, middle), enabling the inverse design of $ReSe_2$ dendrites with user-defined branching degrees and marking a paradigm shift from "finding one recipe" to "mastering the synthesis space" (Figs. S20 and S21, Tables S13 and S14 online).

SHAP analysis identified $T_{Re}$ and $c_{Re}$ as the dominant determinants of $D_F$, contributing ~50% and 27%, respectively (Fig. 2c, Fig. S22 online). Notably, $T_{Re}$ exhibits a non-monotonic influence: intermediate temperatures promote $D_F$, whereas extremes suppress it, aligning with experimental validation (Fig. 2d and Fig. S13 online). We further quantified the contribution independence of different features to $D_F$ via a novel $I$ score, defined as the ratio of main to interaction effects (Eq. 2, Supplementary materials online):

$$I \text{ score} = \frac{\left|\text{SHAP}_{x_i \text{ main}}\right|}{\sum_{i \neq j}\left|\text{SHAP}_{x_i\text{-}x_j \text{ interaction}}\right|}. \tag{2}$$

High-$I$ features ($T_{Re}$ and *sub.*) maintain consistent optimality regardless of other variables, whereas low-$I$ parameters (e.g., $T_{Se}$ and $f_{H2}$) display conditional dependencies and non-convergent BO behavior, indicating strong entanglement with other parameters that jointly modulates $D_F$ (Fig. 2e, Fig. S23 online).

Combining interpretable ML results and material characterizations in Fig. 1, we establish a data-knowledge dual-driven growth model of 2D $ReSe_2$ dendrites. The XGBoost-derived CVD parameter-$D_F$ mapping (Fig. 2f, middle) reveals that $T_{Re}$, $c_{Re}$ and *sub.*, the top-ranked features in importance and contribution independence, jointly dominant the crystal morphology via distinct growth regimes. Below 600 °C, attachment-limited growth prevails, where precursor adsorption to concavities minimizes surface energy, suppressing dendrites and yielding isotropic polycrystals with small $D_F$ (Fig. 2f, left and Fig. S24 online). Above 600°C, diffusion-limited growth takes over, with adatoms diffusion anisotropy on *c*-$Al_2O_3$, leading to dendritic morphologies and dictating branch orientations aligned with its $C_{3v}$ symmetry (Fig. 2f, right and Fig. S25 online). Phase-field simulations quantitatively corroborate the proposed mechanism: the adatom diffusion coefficient ($D$) and initial adatom concentration ($c_i$), two critical simulation factors, are predominantly modulated by the CVD parameters $T_{Re}$ and $c_{Re}$, respectively. Small $D$ (low $T_{Re}$) paired with high $c_i$ (high $c_{Re}$) hinders adatom migration and produces a narrow depletion zone with steep negative precursor concentration gradients, facilitating branch elongation and elevated $D_F$. In contrast, large $D$ with low $c_i$ yields the reverse trend, matching experimental observations (Fig. 2f, bottom right panels, Figs. S26 and S27 online).

In summary, this work develops a human-AI collaborative, data-knowledge dual-driven small-data ML framework for the controllable synthesis of low-symmetry 2D dendrites. It achieves rapid process optimization within one week, constructs high-fidelity recipe-morphology mappings, and deciphers multifactorial growth mechanisms. This workflow can be extended to diverse low-dimensional materials and general CVD systems by resetting optimization targets to material thickness, surface coverage and phase/stacking-related features, as well as incorporating extra process variables like chamber pressure and heating rate. We also discuss realistic challenges in complex synthesis environments and viable solutions: (i) long-cycle performance measurements cause delayed data feedback and hinder iterative optimization, which can be mitigated by cross-modal learning that bridges time-consuming performance metrics and rapidly obtainable microscopic/spectral features; (ii) high-throughput autonomous experimental platforms are highly desirable for multi-objective and highly nonlinear tasks, as reliable modeling cannot be achieved with very limited data, and (iii) for parameters that cannot be arbitrarily recommended by algorithms due to experimental limitations (e.g., substrate surface configuration), judgement from human scientists based on mechanistic understanding and practical experience remain essential. Overall, this study holds great potential for promoting the methodological paradigm shift in materials synthesis.

## Conflict of interest

The authors declare that they have no conflict of interest.

## Acknowledgments

Shanshan Wang acknowledges support from the National Natural Science Foundation of China (22494643), Major Fundamental Research Project of Hunan Province (2025JC0005), Shenzhen Science and Technology Innovation Commission Project (KQTD20221101115627004), and Innovation Research Foundation of National University of Defense Technology (ZZCX-ZZGC-01-07). Susu Fang acknowledges support from the National Natural Science Foundation of China 22503112 and Innovation Research Foundation of National University of Defense Technology. Fangping Ouyang acknowledges support from the Key Project of the Natural Science Program of Xinjiang Uygur Autonomous Region (2023D01D03) and the computing resources at the High Performance Computing Center of Central South University.

## Author contributions

Shanshan Wang initiated the project and generated experimental protocols. Wenqiang Huang wrote the code. Shanshan Wang prepared the STEM samples and captured images. Xuhang Gu grew 2D crystals. Shen'ao Xue and Susu Fang prepared the samples and conducted measurements. Huanhuan Xing, Junjie Jiang, Junying Zhang, Shen Zhou, Zheng Luo, Jin Zhang, and Fangping Ouyang discussed the work and gave suggestions. All authors contributed to the data analysis, manuscript writing, and revision of the manuscript.

## Code availability

All data supporting this study and the code are available on GitHub (https://github.com/csuhwq0421/ML4Dendrites).

## Appendix A. Supplementary material

Supplementary data to this article can be found online.

## References

[1] Ji QQ, Zhang Y, Shi JP, et al. Morphological engineering of CVD-grown transition metal dichalcogenides for efficient electrochemical hydrogen evolution. Adv Mater 2016;28:6207–12.

[2] Shao GL, Xue XX, Zhou XL, et al. Shape-engineered synthesis of atomically thin 1T-$SnS_2$ catalyzed by potassium halides. ACS Nano 2019;13:8265–74.

[3] Chen JY, Zhao XX, Grinblat G, et al. Homoepitaxial growth of large-scale highly organized transition metal dichalcogenide patterns. Adv Mater 2018;30:1704674.

[4] Wang BL, Ma B, Wang K, et al. Fractal growth of 2D $NbSe_2$ for broadband nonlinear optical limiting. Adv Funct Mater 2024;34:2401490.

[5] Kang Y, Zhai XY, Yang Q, et al. Conductive dendrite engineering of single-crystalline two-dimensional dielectric memristors. Innovation 2025;6:100885.

[6] Wang MH, Huang M, Luo D, et al. Single-crystal, large-area, fold-free monolayer graphene. Nature 2021;596:519–24.

[7] Moon D, Lee W, Lim C, et al. Hypotaxy of wafer-scale single-crystal transition metal dichalcogenides. Nature 2025;638:957–64.

[8] Jiang JJ, Wang X, Wang DY, et al. Chirality-transferred epitaxy of circular polarization-sensitive $ReS_2$ monolayer single crystals. Nat Commun 2025;16:7119.

[9] Li JZ, Chen MG, Zhang CH, et al. Fractal-theory-based control of the shape and quality of CVD-grown 2D materials. Adv Mater 2019;31:1902431.

[10] Chen TL, Pang ZQ, He SM, et al. Machine intelligence-accelerated discovery of all-natural plastic substitutes. Nat Nanotechnol 2024;19:782-91.

[11] Moon J, Beker W, Siek M, et al. Active learning guides discovery of a champion four-metal perovskite oxide for oxygen evolution electrocatalysis. Nat Mater 2024;23:108–15.

[12] Huang WQ, Jin YC, Li ZM, et al. Auto-resolving the atomic structure at van der Waals interfaces using a generative model. Nat Commun 2025;16:2927.

[13] Luo Z, Feng M, Gao ZJ, et al. Exploring structure diversity in atomic resolution microscopy with graph. Adv Mater 2025;37:2417478.

[14] Cao B, Qin Y, Luo Y, et al. Spatial-adaptive active learning identifies ultra-durable and highly active catalysts for acidic oxygen evolution reaction. Sci Bull 2026;71:1084–92.

[15] Lundberg SM, Erion G, Chen H, et al. From local explanations to global understanding with explainable AI for trees. Nat Mach Intell 2020;2:56–67.

Supporting Information

# Data-knowledge dual-driven intelligent framework for full-chain, experiment-efficient synthesis of 2D dendrites

Wenqiang Huang[1,2,†], Xuhang Gu[3,†], Susu Fang[2,†,*], Shen'ao Xue[1,2], Huanhuan Xing[2], Junjie Jiang[1,2], Junying Zhang[1,2], Shen Zhou[4], Zheng Luo[2], Jin Zhang[3], Fangping Ouyang[1,5,6*] & Shanshan Wang[2,3,7*]

[1]*School of Physics, Hunan Key Laboratory for Super-Microstructure and Ultrafast Process, and Hunan Key Laboratory of Nanophotonics and Devices, Central South University, Changsha 410083, China*

[2]*College of Aerospace Science and Engineering, National University of Defense Technology, Changsha 410000, China*

[3]*School of Advanced Materials, Peking University Shenzhen Graduate School, Shenzhen 518055, China*

[4]*College of Science, National University of Defense Technology, Changsha 410000, China*

[5]*School of Physics and Technology, State Key Laboratory of Chemistry and Utilization of Carbon Based Energy Resources, Xinjiang University, Urumqi 830046, China*

[6]*State Key Laboratory of Powder Metallurgy, and Powder Metallurgy Research Institute, Central South University, Changsha 410083, China*

[7]*Guangdong Provincial Key Laboratory of Nano-Micro Materials Research, Peking University Shenzhen Graduate School, Shenzhen 518055, China*

[†]These authors contributed equally to this work.

*Corresponding authors, Email: fangsusugfkd@nudt.edu.cn (Susu Fang); oyfp@csu.edu.cn (Fangping Ouyang); wangshanshan08@nudt.edu.cn (Shanshan Wang)

## Supplementary methods

### Preparation and transfer of 2D $ReSe_2$

**(i) Synthesis of 2D $ReSe_2$.** The growth of 2D $ReSe_2$ was performed with a spatially confined chemical vapor deposition (CVD) system [1]. The spin-coated substrate served as the bottom chip, with another $c$-$Al_2O_3$ or MgO substrate placed face-to-face on top of it to form a space-confined reaction microchamber, which was positioned at the center of the high-temperature zone of the furnace (580-680°C). A 0.025-0.15 M solution of ammonium perrhenate ($NH_4ReO_4$, 99%, RHAWN) was applied as a rhenium source and spin-coated onto the plasma-treated substrate surface using 30 μL dosage at 5000 rpm for 30 seconds. The temperature of this substrate is the rhenium source temperature ($T_{Re}$). $C$-$Al_2O_3$ or MgO was treated with oxygen plasma to create a hydrophilic surface. For selenium (Se) source, 80 mg of Se powder (99.99%, RHAWN) was put in a ceramic boat at the center of the temperature zone of the furnace (220-300°C), and this temperature is defined as the $T_{se}$. Before 2D $ReSe_2$ synthesis, the CVD system was flushed with argon (Ar) gas flow for 10 min to remove the air from the tube. Subsequently, a mixed gas of Ar/$H_2$ was introduced, with Ar and $H_2$ gas flow rates ($f_{H2}$) of 100 sccm and 1~4 sccm, respectively. The CVD system was maintained under the above-mentioned conditions for 20 minutes to ensure the growth of $ReSe_2$ on the target substrate. After the growth period, the furnace was naturally cooled to room temperature under a continuous flow of Ar at ~500 sccm.

**(ii) Transfer of 2D $ReSe_2$.** A thin film of poly methyl methacrylate (PMMA) was spin-coated onto the $ReSe_2$/substrate surface. The coated substrates were baked at 80 °C for 3 minutes to strengthen the interaction between the PMMA film and the $ReSe_2$ and then floated on a 2 M potassium hydroxide (KOH) solution to separate the PMMA/$ReSe_2$ film from the substrate. The detached PMMA/$ReSe_2$ film was transferred to deionized water three times to remove the residual etchants. Then, the film was transferred to a TEM grid or a glass carbon electrode, dried naturally in air, and baked on a hotplate at 100°C for 15 min. The sample was immersed in acetone for 8 h to remove the PMMA.

### Characterizations

**(i) Raman spectroscopy.** Raman spectroscopy was employed to study structural properties, utilizing a Renishaw inVia confocal Raman system with an excitation wavelength of 532 nm. The laser power was kept below 100 μW, and the system was calibrated with the Raman peak of silicon at 520.7 $cm^{-1}$. The experimentally measured data of $A_{1g}$-like mode at 162 $cm^{-1}$ with parallel polarization $\bar{x}(yy)x$ or configurations are plotted as the blue and red dots, respectively. Under parallel polarization configuration, the angular dependence of the Raman intensity for this mode follows equations [2,3]:

$$\begin{aligned} I^{\parallel}_{A_g} \propto & |a|^2\cos^4\theta + |d|^2\sin^4\theta + 4|b|^2\sin^2\theta\cos^2\theta \\ & + 2|a||d|\cos\phi_{ad}\sin^2\theta\cos^2\theta \\ & + 4|a||b|\cos\phi_{ab}\sin\theta\cos^3\theta \\ & + 4|d||b|\cos\phi_{db}\sin^3\theta\cos\theta. \end{aligned} \tag{S1}$$

Here, $I^{\parallel}_{A_g}$ is the Raman intensity for the $A_g$-like (162 $cm^{-1}$) mode in parallel polarization configurations. $|a|$,$|b|$ and $|d|$are the value of Raman tensor element $a$, $b$, and $d$. $\theta$ is the angle between the incident laser polarization and the Re-chain direction (b-axis). $\phi_{ad}=\phi_a-\phi_d$, $\phi_{ab}=\phi_a-\phi_b$, and $\phi_{db}=\phi_d-\phi_b$.

**(ii) Annular dark-field scanning transmission electron microscopy (ADF-STEM) imaging.** The measurements were conducted at room temperature on aberration-corrected Titan Cubed Themis G2 300. STEM was operated under an accelerating voltage of 300 kV. The experimental setup included a condenser lens with an aperture diameter of 50 mm. The convergence semi-angle was set to 21.3 mrad, and the collection angle ranged from 39 to 200 mrad. For each frame, the dwell time was set to 2 μs per pixel. Imaging was performed at a pixel size of 0.012 nm and a beam current of 30 pA.

### Electrochemical measurements

Hydrogen evolution reaction (HER) measurements for 2D $ReSe_2$ with different fractal dimension ($D_F$) were performed at room temperature using a CHI 760E electrochemical workstation. The experimental setup consisted of a glassy carbon disk as the working electrode, a platinum plate as the counter electrode, and a saturated Ag/AgCl electrode as the reference, forming a standard three-electrode configuration. Linear sweep voltammetry (LSV) was conducted at a scan rate of 5 mV $s^{-1}$ in a 0.5 M $H_2SO_4$ electrolyte. Tafel slopes were obtained from the linear portions of the LSV curves using the Tafel equation, with current densities normalized based on the material's exposed surface area. Potentials were referenced to the reversible hydrogen electrode (RHE) using the relation $E$ (vs RHE) = $E$ (vs Ag/AgCl) + 0.197 V + 0.0592 pH.

### Machine learning (ML) models

**(i) Bayesian optimization (BO)**. BO operates by constructing a surrogate model to approximate the objective function and employs an acquisition function to guide subsequent experimental sampling, thereby identifying the global optimum with minimal experimental iterations. We set up the BO framework through the package Emukit in Python [4,5]. The Gaussian process regression (GPR) model was selected as the surrogate model [6]. GPR infers the distribution of the objective function across the parameter space using a limited number of data points and provides uncertainty estimates for predictions. Its mathematical formulation is as follows:

$$F(\mathrm{x}) \sim N\,(\mu(x), K(x, x')), \quad \text{(S2)}$$

where $\mu(x)$ is the mean function and $K(x, x')$ is the covariance function (kernel function).

The acquisition function recommends the experimental sampling points of the next iteration based on the surrogate model's predictions, balancing exploration (sampling regions with high uncertainty) and exploitation (regions predicted to yield near-optimal values). We employ the max-value entropy search (MES) acquisition function, which selects sampling points by maximizing the information gain associated with the optimal value of the objective function [7]. Its mathematical formulation is as follows:

$$\mathrm{MES}(x)=\mathrm{H}(p(f(\mathrm{x})))-\mathrm{E}_{p(f(x^*))}[\mathrm{H}\,(p(f(x)|f(x^*)))], \quad \text{(S3)}$$

while $f(x)$ and $f(x^*)$ represent the predicted value and the predicted maximum value of the GPR, respectively. It is worth noting that all data were normalized prior to being fed into the model.

**(ii) Model evaluation.** Prior to feature analysis, ML models were evaluated using five repetitions of 5-repeated 5-fold cross-validation. Two indicators are selected to evaluate the model's performance, which are $R^2$ and MSE. $R^2$ is a statistical measure that provides information about the goodness of fit of a model. In the context of regression, it is a statistical measure of how well the regression line approximates the actual data. $R^2$ is $(-\infty,1]$, where a value closer to 1 indicates better model performance. MSE measures the average squared difference between predicted and actual values, which disproportionately weights larger deviations, resulting in greater sensitivity to outlier values. $R^2$ and MSE were utilized to assess model performance, with their calculations shown as follows:

$$R^2 = 1\text{-}\sum_{i=1}^{n}(y_{\mathrm{pred.}}-y_{\mathrm{exp.}})^2 / \sum_{i=1}^{n}(y_{\mathrm{pred.}}-y_{\mathrm{mean}})^2, \quad \text{(S4)}$$

$$\mathrm{MSE} = \frac{1}{n}\sum_{i=1}^{n}(y_{\mathrm{pred.}}\text{-}y_{\mathrm{exp.}})^2, \quad \text{(S5)}$$

where $y_{\mathrm{pred.}}$, $y_{\mathrm{exp.}}$ and $y_{\mathrm{mean}}$ are the predicted $D_{\mathrm{F}}$ value, the experimental $D_{\mathrm{F}}$ value and the average of all the measured values, respectively.

**(iii) eXtreme gradient boosting (XGBoost) model.** As an advanced implementation of gradient-boosted trees, XGBoost distinguishes itself through its efficient regularization techniques and parallel computing capabilities [8]. Its core principle involves iteratively constructing multiple decision tree models and integrating them into a robust ensemble framework to enhance predictive accuracy. Its mathematical formulation can be expressed as:

$$y_{\mathrm{pred.}} = \sum_{m=1}^{M} f_m(x_{\mathrm{i}}), \quad \text{(S6)}$$

where $f_m$ denotes the base estimator, $y_{\mathrm{pred.}}$ donates the predicted $D_{\mathrm{F}}$ value, and $x_{\mathrm{i}}$ donates the growth parameters value. $m$=1,2,…, $M$.

The objective function comprises a loss function and a regularization term, mathematically expressed as:

$$\mathrm{Obj}(\theta) = \sum_{i=1}^{N} L\,(y_{\mathrm{pred.}}, y_{\mathrm{exp.}}) + \sum_{m=1}^{M}\Omega(f_{\mathrm{m}}), \quad \text{(S7)}$$

where $L(y_{\mathrm{pred.}}, y_{\mathrm{exp.}})$ represents the MSE loss function, and $\sum_{m=1}^{M}\Omega(f_{\mathrm{m}})$ is the regularization term designed to control model complexity and mitigate overfitting. $i$ = 1, 2,…, $N$.

### Model interpretability

**(i) Quantification of feature importance.** We employed SHapley Additive exPlanations (SHAP) method to quantitatively assess the contribution of each input feature to predictions in the XGBoost model, thereby enhancing ML model's interpretability [9]. Feature importance values are assigned by SHAP through calculating the marginal contributions of all possible feature subsets for each data instance. Global feature importance is derived by aggregating the mean absolute

SHAP values across the entire dataset. SHAP also enables local interpretability by explaining predictions for individual instances through SHAP values. Features with larger mean absolute SHAP values exhibit higher importance and vice versa.

**(ii) Quantification of the independence level of feature's impact on $D_F$.** SHAP values can be decomposed into the independent effects of features (main effects, $SHAP_{x_i\ main}$) and the mutual influences between features (interaction effects, $SHAP_{x_i\text{-}x_j\ interaction}$, $i{\neq}j$) [10]. For any model prediction $f(x)$, SHAP guarantees the following additive decomposition:

$$f(x) = \phi_0 + SHAP_{x_i\ main} + SHAP_{x_i\text{-}x_j\ interaction}, \quad (S8)$$

where $\phi_0$ is the baseline value, i.e., the model prediction when no features are considered. Therefore, using Eq. 2 in the main text, which is the ratio of the main effect to the interaction effect, to quantify the independence level of a feature's impact on $D_F$ is reasonable and easy to understand.

## Range of $D_F$

In this paper, we focus on the perimeter fractal dimension of the dendritic domain boundary, which characterizes the geometric complexity of the extracted edge contour. For the boundaries of two-dimensional crystals, the perimeter fractal dimension ($D_F$) lies between 1 and 2. Specifically, a value closer to 1 indicates a smoother grain boundary, whereas a value closer to 2 indicates a rougher boundary. The $D_F$ medians rose from 1.36 (iteration 0, initial dataset) to 1.61 (iteration 4). The improvement amount can be calculated using the normalization principle expressed as: (1.61-1.36)/(1.36-1)×100%= 69.4%.

## Phase-field simulation of 2D $ReSe_2$ morphological evolution

To elucidate the morphological evolution of $ReSe_2$ dendrites during fabricate process, a continuous spatial-temporal phase-field model was developed, coupling the interfacial growth dynamics with the surface adatom diffusion. The system is described by two continuous field variables: the non-conserved order parameter $\psi(\vec{r},t)$, where $\psi$=1 and $\psi$=-1 represent the solid $ReSe_2$ crystal and the bare substrate respectively, and the conserved adatom concentration field $c(\vec{r},t)$. The morphological evolution of the $ReSe_2$ domains is governed by a phenomenological Allen-Cahn formulation incorporating crystallographic anisotropy and thermodynamic driving forces:

$$\tau_\psi \frac{\partial \psi}{\partial t} = \nabla\cdot\{[w^2(\theta)I_0 + w(\theta)w'(\theta)I_1]\nabla\psi\} + \sin(\pi\psi) + \Gamma(c-c_0)(1+\cos(\pi\psi))$$
$$+\alpha\eta(\vec{r},t)\sqrt{|\nabla\psi|^2}, \quad (S9)$$

where $\tau_\psi$ is the characteristic interfacial relaxation time, $I_0$ and $I_1$ are the diagonal and antisymmetric matrices respectively. The interface width function $w(\theta)=w_0[1+g\cos(N\theta)]$ captures the intrinsic in-plane anisotropy of the distorted $ReSe_2$ lattice and substrate, with $N$=3 dictating the symmetry factor and g denoting the anisotropy strength. The driving force is determined by the local supersaturation ($c$-$c_0$), scaled by the coupling coefficient Γ. A Gaussian white noise term $\eta(\vec{r},t)$ with amplitude α is introduced to simulate thermal fluctuations, which is essential for triggering side-branching during dendritic growth. Concurrently, the diffusion and consumption of Re-containing adatoms on the substrate are captured by a modified diffusion equation:

$$\frac{\partial c}{\partial t} = D\nabla^2 c + \frac{1-\psi}{2}F - \frac{1-\psi}{2}\frac{c}{\tau_s} - \frac{1}{2}\frac{\partial \psi}{\partial t}, \quad (S10)$$

where $D$ is the effective surface diffusion coefficient of adatoms. The second and third terms on the right-hand side represent the incoming precursor flux ($F$) and the adatom desorption process characterized by the mean lifetime ($\tau_s$), respectively. The prefactor $\frac{1-\psi}{2}$ ensures these processes exclusively occur on the bare substrate. The last sink term $-\frac{1}{2}\frac{\partial \psi}{\partial t}$ strictly couples the mass conservation, denoting the depletion of adatoms as the solid phase expands.

The coupled partial differential equations were numerically solved using the finite element method (FEM). The 2D simulation domain size was set to $600\Delta x \times 600\Delta y$. To ensure numerical stability and accuracy across the diffuse interface, a highly refined free triangular mesh was applied with a grid spacing of $\Delta x=\Delta y=0.01\sim0.05w_0$ (where the normalized interface width $w_0$=1). A fixed time step of $\Delta t=0.1\tau_\psi$ (with $\tau_\psi$=1) was utilized for temporal integration. Initially, a small circular nucleus with a radius of $r=0.1w_0$ ($\psi$=1) was placed at the center of the domain to initiate growth, while the rest of the area was set as the bare substrate ($\psi$=-1).

Supplementary figures

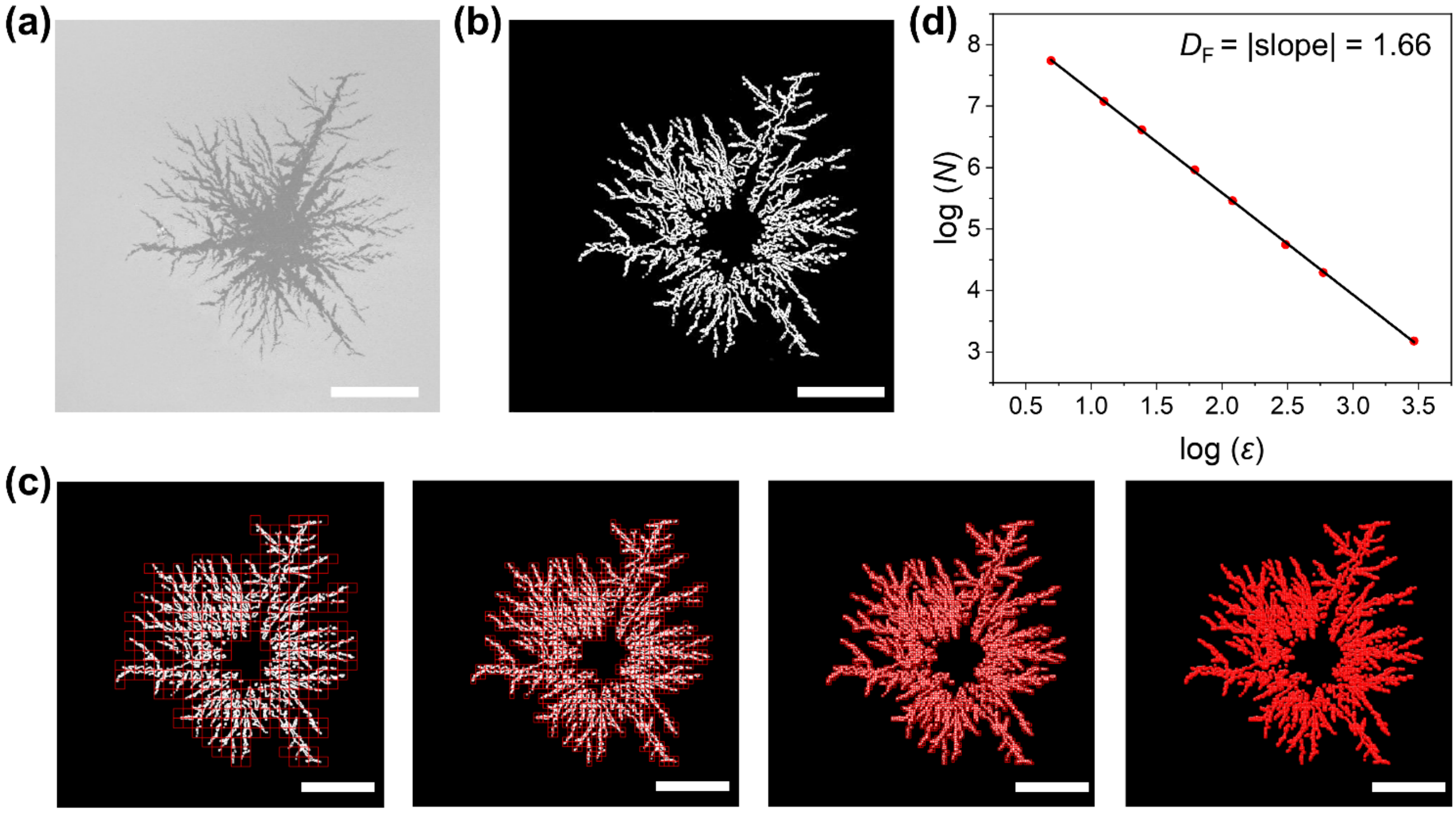

**Fig. S1.** Calculation of the perimeter fractal dimension ($D_F$) of 2D $ReSe_2$ domain boundaries using standard box-counting evaluation [11]. (a) SEM image of a 2D $ReSe_2$ domain. (b) Contour of the $ReSe_2$ domain. (c) Schematic diagram of the filling of the domain edge contour with squares of gradually decreasing size. The sizes of the boxes are 2, 4, 8, and 16, respectively. Scale bars: 2 µm. (d) $D_F$ determination of $ReSe_2$ domain edges via ln($N$)-ln($\varepsilon$) linear regression. $\varepsilon$: box size; $N$: box count covering domain edge.

For irregular shapes, the $D_F$ was calculated using the box-counting method implemented in the Fractal Box Count plugin of ImageJ software. Box sizes were systematically varied, and the number of covering boxes ($N$) was recorded. The slope ($D_F$) was determined by linear regression of ln($N$) versus ln($\varepsilon$).

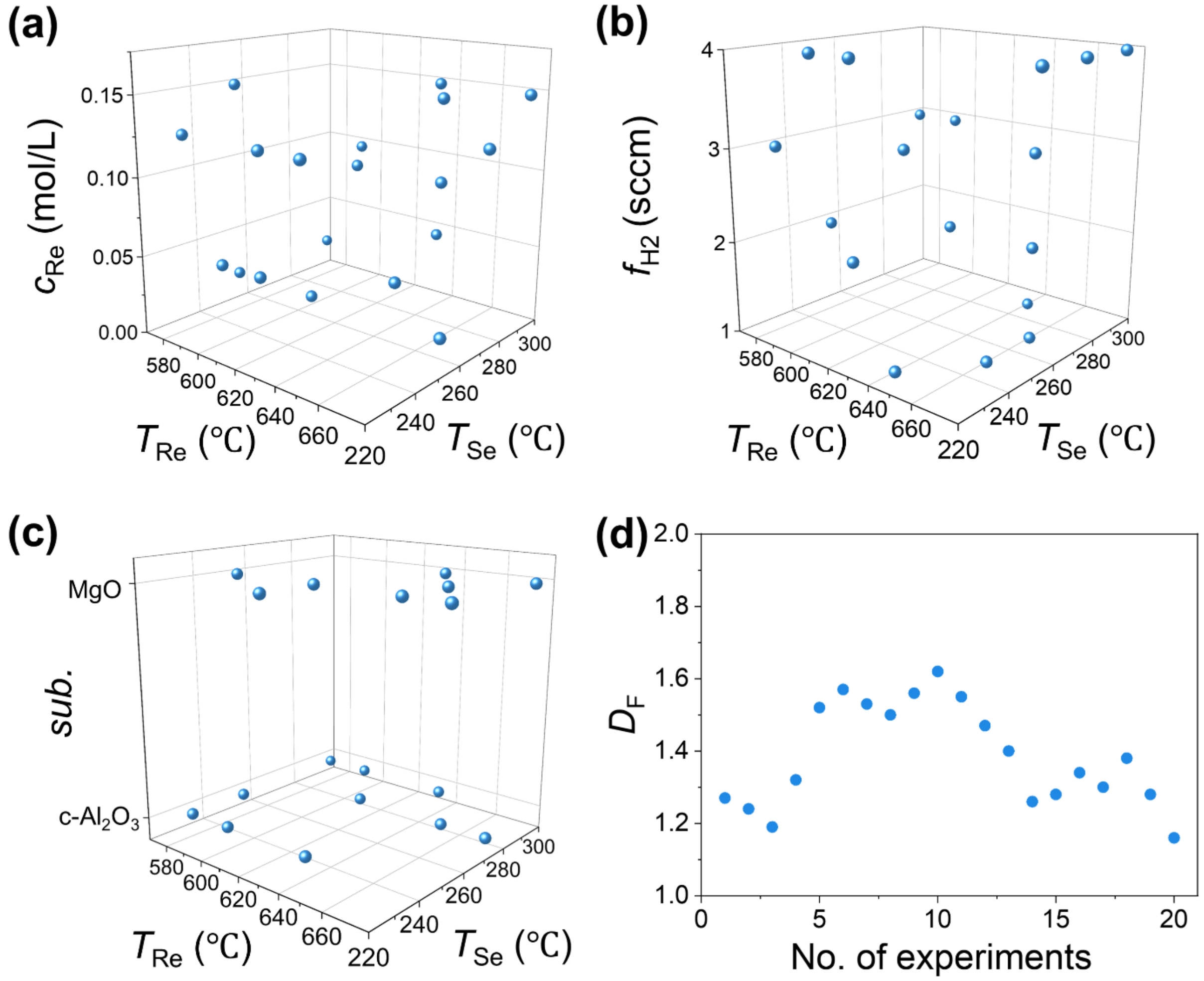

**Fig. S2.** Distribution of the initial dataset obtained through the Latin hypercube sampling (LHS). (a-c) Distribution of the initial dataset in three-dimensional parameter space: $T_{Re}$, $T_{Se}$, and rhenium concentration source ($c_{Re}$) (a); $T_{Re}$, $T_{Se}$, and $f_{H2}$ (b); $T_{Re}$, $T_{Se}$, and substrate type (*sub.*) (c). (d) Distribution in target $D_F$ space. Notebley, an initial dataset of 20 samples was constructed using LHS, enabling uniform coverage of the multidimensional CVD parameter space within the ranges defined in Supplementary Tables S1–S3.

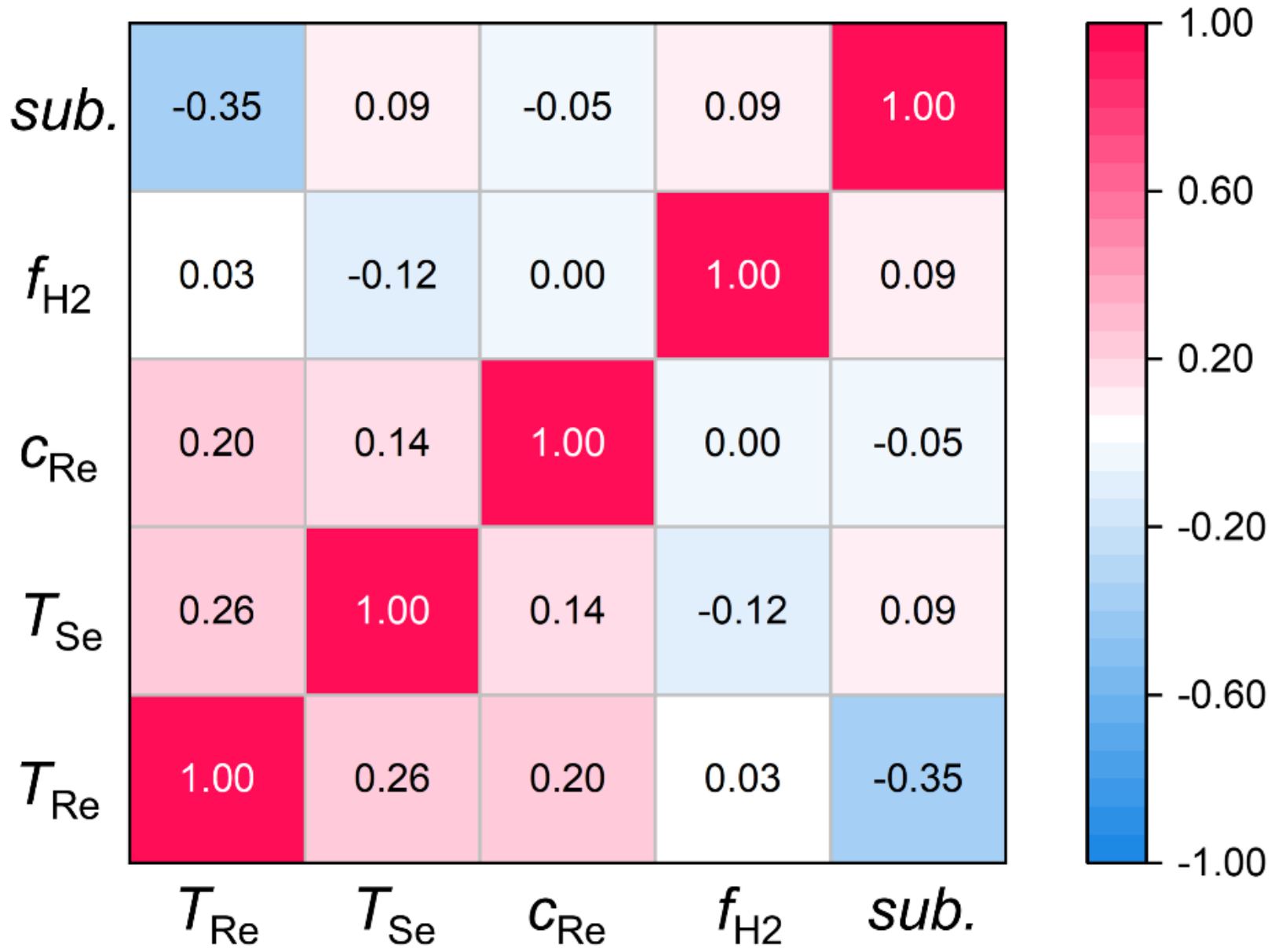

**Fig. S3.** Graphical representation of Pearson's correlation matrix for the five growth parameters. Deeper-colored squares indicate higher correlations. The results indicate no obvious linear relationship between the five input variables, demonstrating high independence of these features.

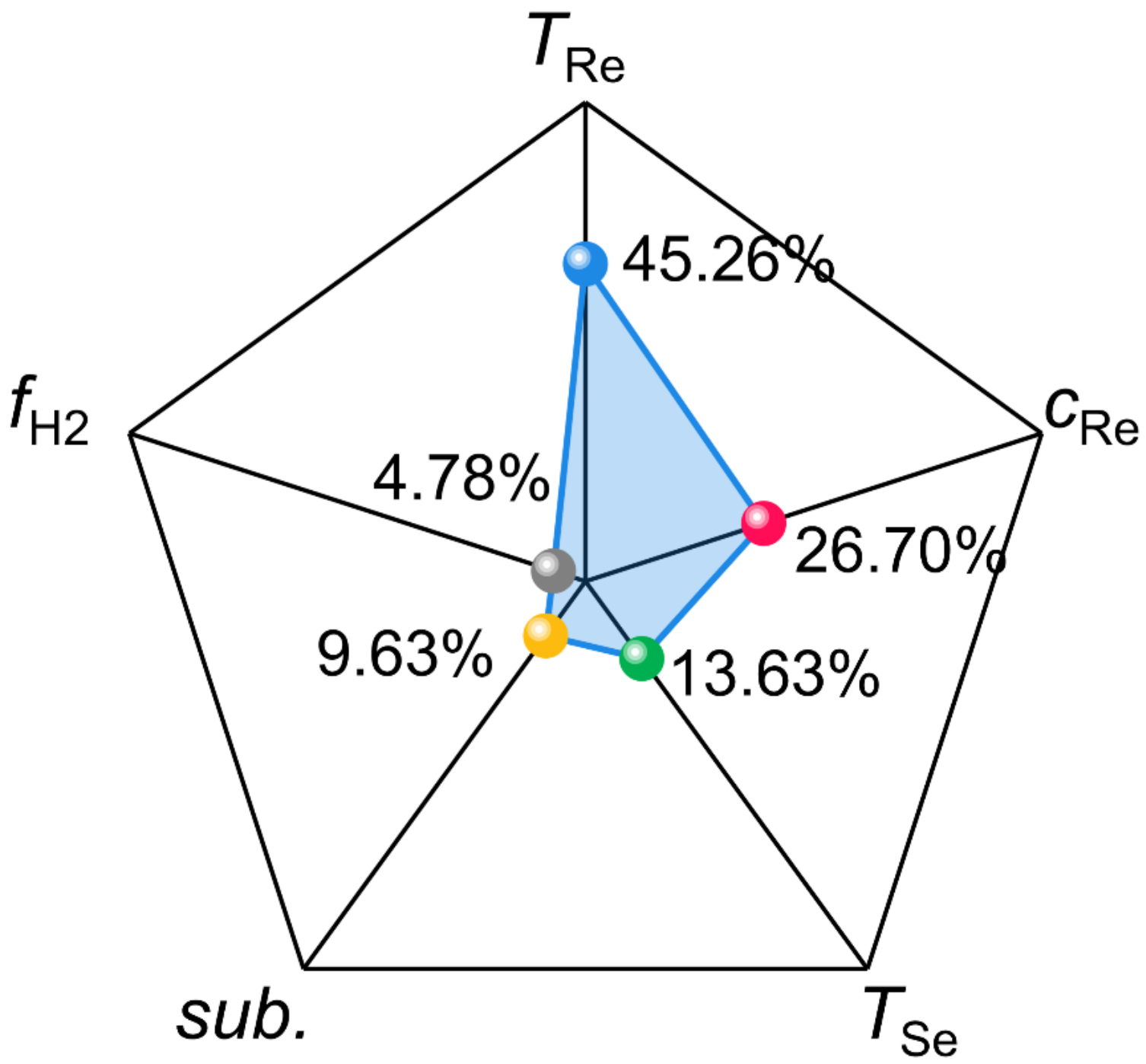

**Fig. S4.** Feature importance of five CVD process parameters on $D_F$ determined by normalized SHAP values based on the GPR model.

To interpret the BO process and identify the most informative variables for visualization, we quantified the relative contributions of the five CVD process parameters to $D_F$ using normalized SHAP values based on the GPR model. The results show that $T_{Re}$, $c_{Re}$, and $T_{Se}$ are the three most influential parameters in determining the predicted $D_F$ . Therefore, these three variables were selected as the focus of the BO landscape visualization. They are presented as the coordinate axes in the representative three-dimensional visualization shown in **Fig. 1**c of the main text, and are further resolved into pairwise two-dimensional (2D) contour maps in Figs. S6-S9 to provide a clearer view of the predicted $D_F$ distributions and the associated uncertainties during different optimization steps.

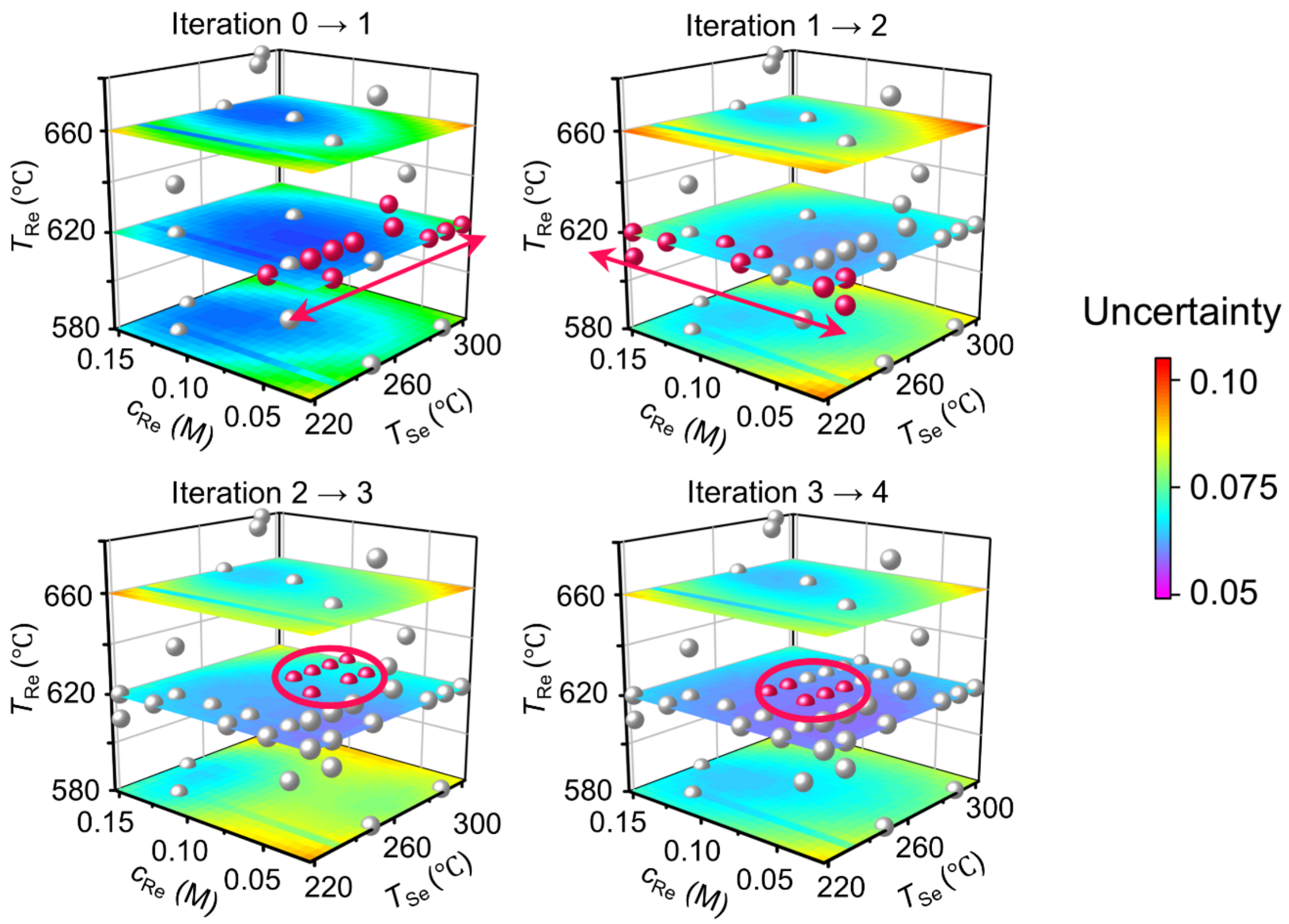

**Fig. S5.** Visualization of the BO process by exhibiting the three-dimensional CVD parameter-$D_F$ plot and the sampling strategies in every iteration $s$→($s$+1) ($s$=0,1,2,3). CVD parameters: $T_{Re}$, $c_{Re}$, and $T_{Se}$. Color bar: uncertainty of the predicted $D_F$. The contour maps and gray dots represent the CVD parameter-$D_F$ relation and the sampling points in iteration $s$, while the red dots represent the sampling points in iteration ($s$+1). This figure should be viewed together with Figure 1c in order to better understand the round-by-round sampling strategy of the active learning algorithm.

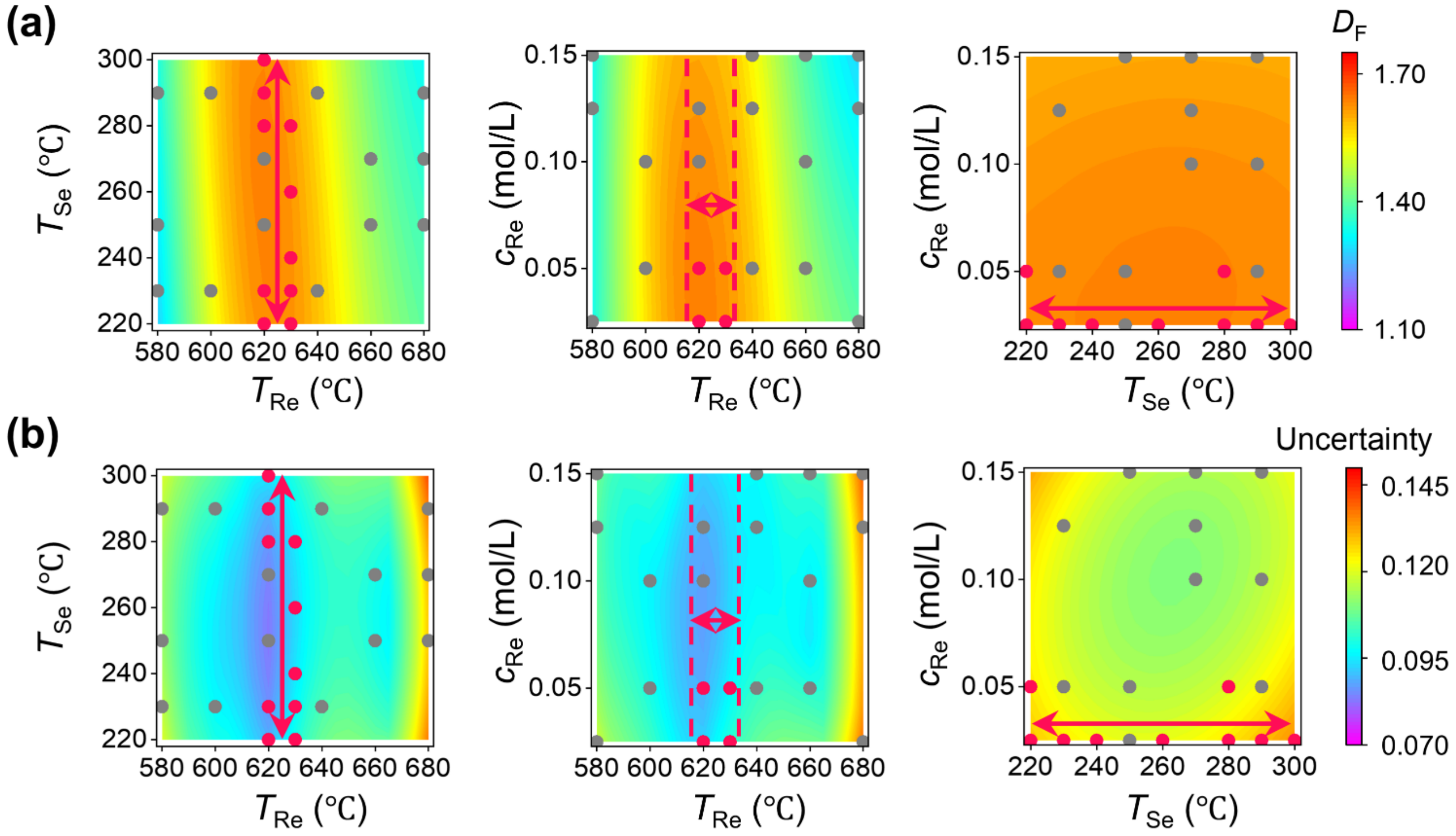

**Fig. S6.** Visualization of the predicted $D_F$ value (a) and the corresponding uncertainty (b) in iteration 0→1 by 2D contour maps. CVD parameter pairs: $T_{Re}$-$T_{Se}$, $T_{Re}$-$c_{Re}$, and $T_{Se}$-$c_{Re}$. Color bar: predicted $D_F$ in panel (a) and uncertainty of the predicted $D_F$ in panel (b). The contour maps and gray dots represent the CVD parameter-$D_F$ relation and the sampling points in iteration 0, while the red dots represent the sampling points in iteration 1.

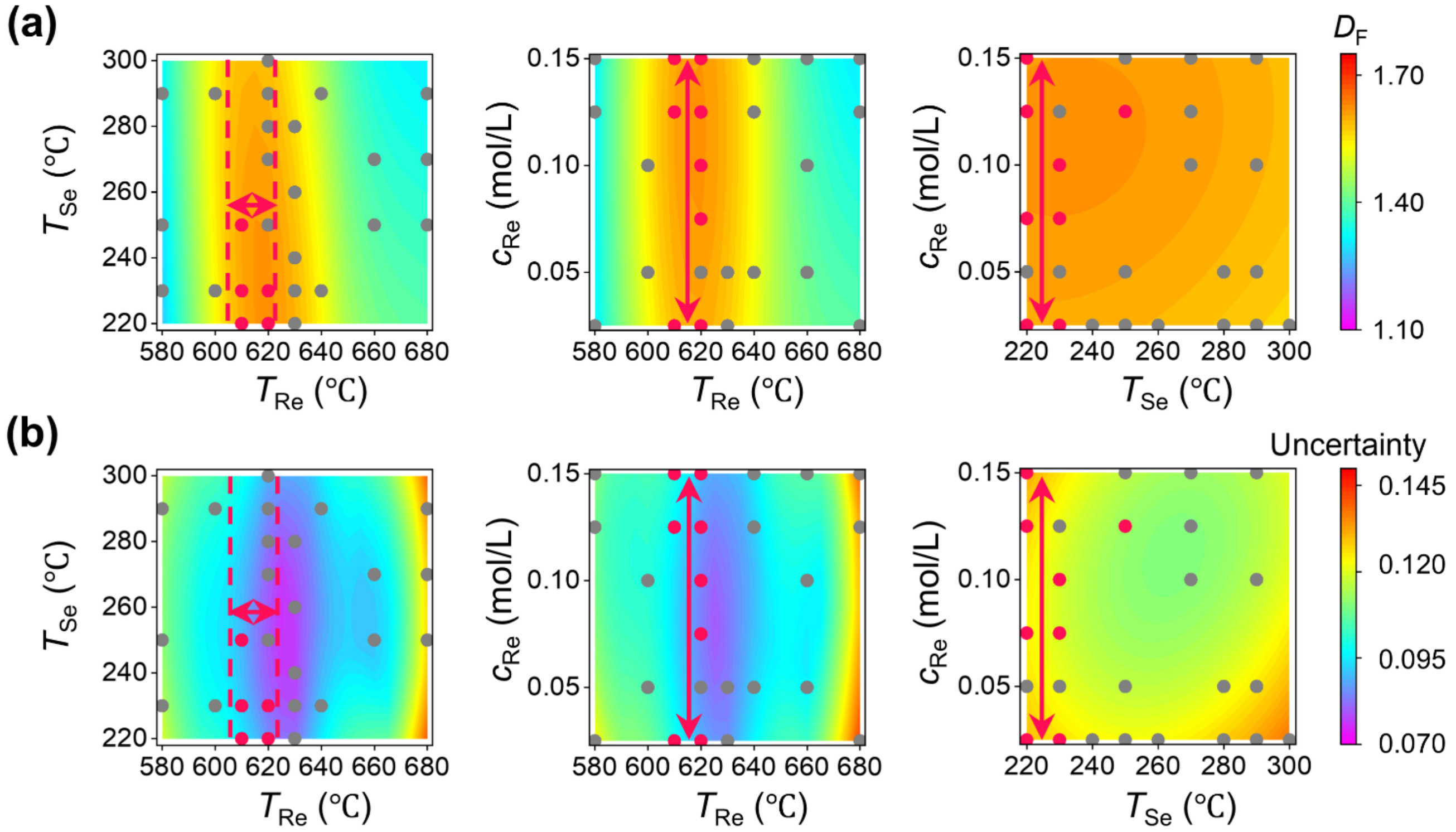

**Fig. S7.** Visualization of the predicted $D_F$ value (a) and the corresponding uncertainty (b) in iteration 1→2 by 2D contour maps. CVD parameter pairs: $T_{Re}$-$T_{Se}$, $T_{Re}$-$c_{Re}$, and $T_{Se}$-$c_{Re}$. Color bar: predicted $D_F$ in panel (a) and uncertainty of the predicted $D_F$ in panel (b). The contour maps and gray dots represent the CVD parameter-$D_F$ relation and the sampling points in iteration 1, while the red dots represent the sampling points in iteration 2.

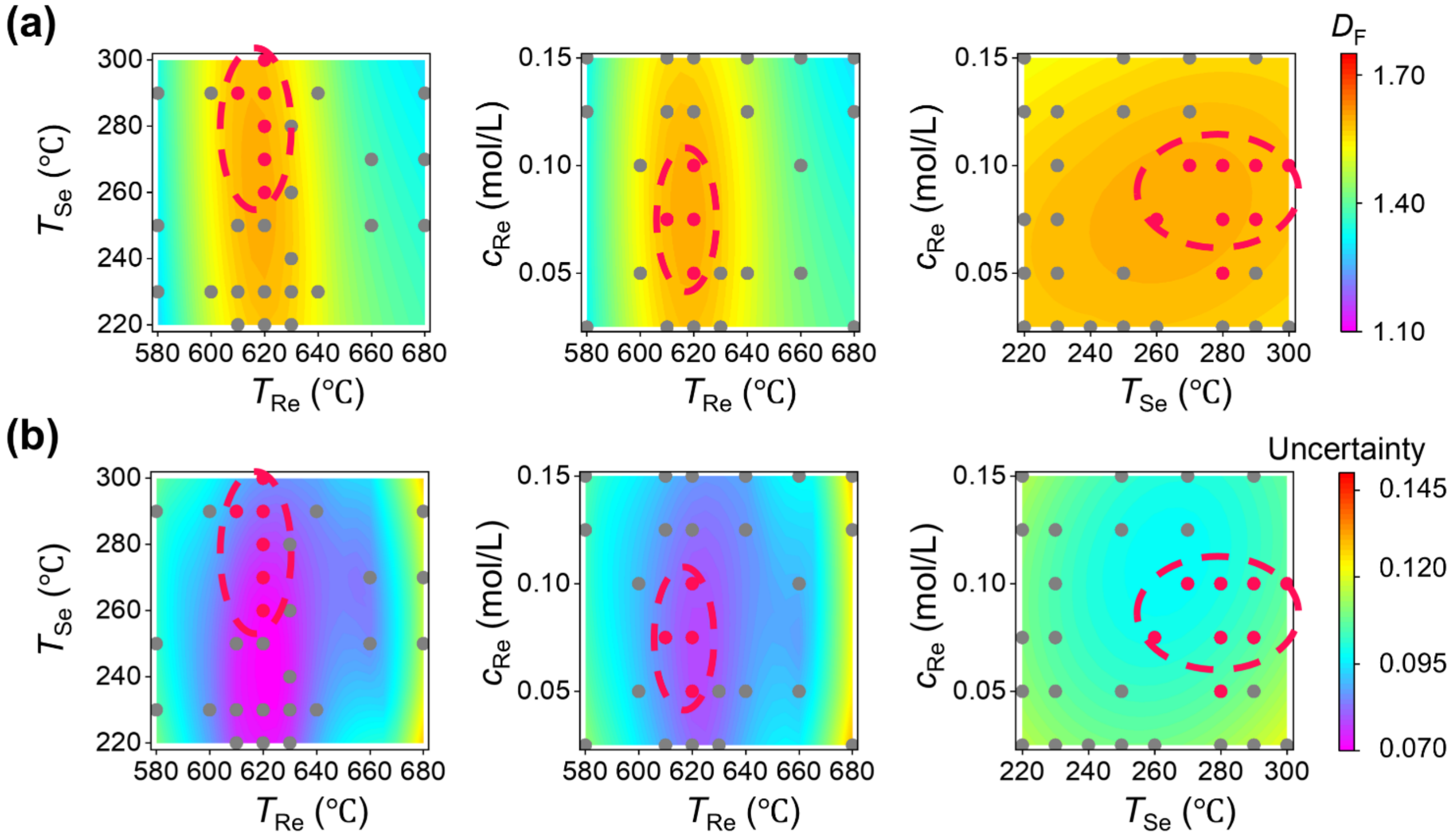

**Fig. S8.** Visualization of the (a) predicted $D_F$ value and (b) the corresponding uncertainty in iteration 2→3 by 2D contour maps. CVD parameter pairs: $T_{Re}$-$T_{Se}$, $T_{Re}$-$c_{Re}$, and $T_{Se}$-$c_{Re}$. Color bar: predicted $D_F$ in panel (a) and uncertainty of the predicted $D_F$ in panel (b). The contour maps and gray dots represent the CVD parameter-$D_F$ relation and the sampling points in iteration 2, while the red dots represent the sampling points in iteration 3.

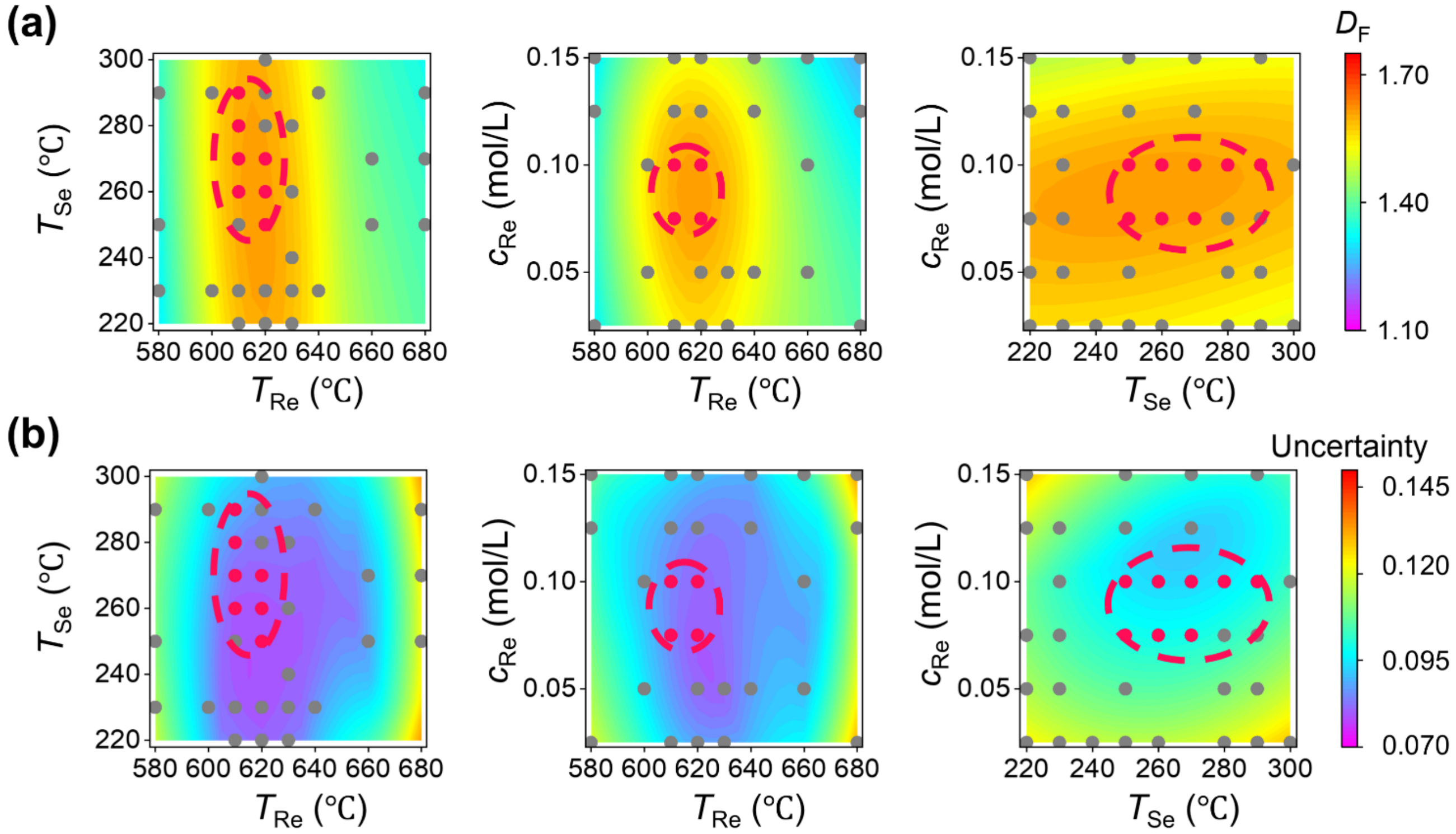

**Fig. S9.** Visualization of the predicted $D_F$ value (a) and the corresponding uncertainty (b) in iteration 3→4 by 2D contour maps. CVD parameter pairs: $T_{Re}$-$T_{Se}$, $T_{Re}$-$c_{Re}$, and $T_{Se}$-$c_{Re}$. Color bar: predicted $D_F$ in panel (a) and uncertainty of the predicted $D_F$ in panel (b). The contour maps and gray dots represent the CVD parameter-$D_F$ relation and the sampling points in iteration 3, while the red dots represent the sampling points in iteration 4.

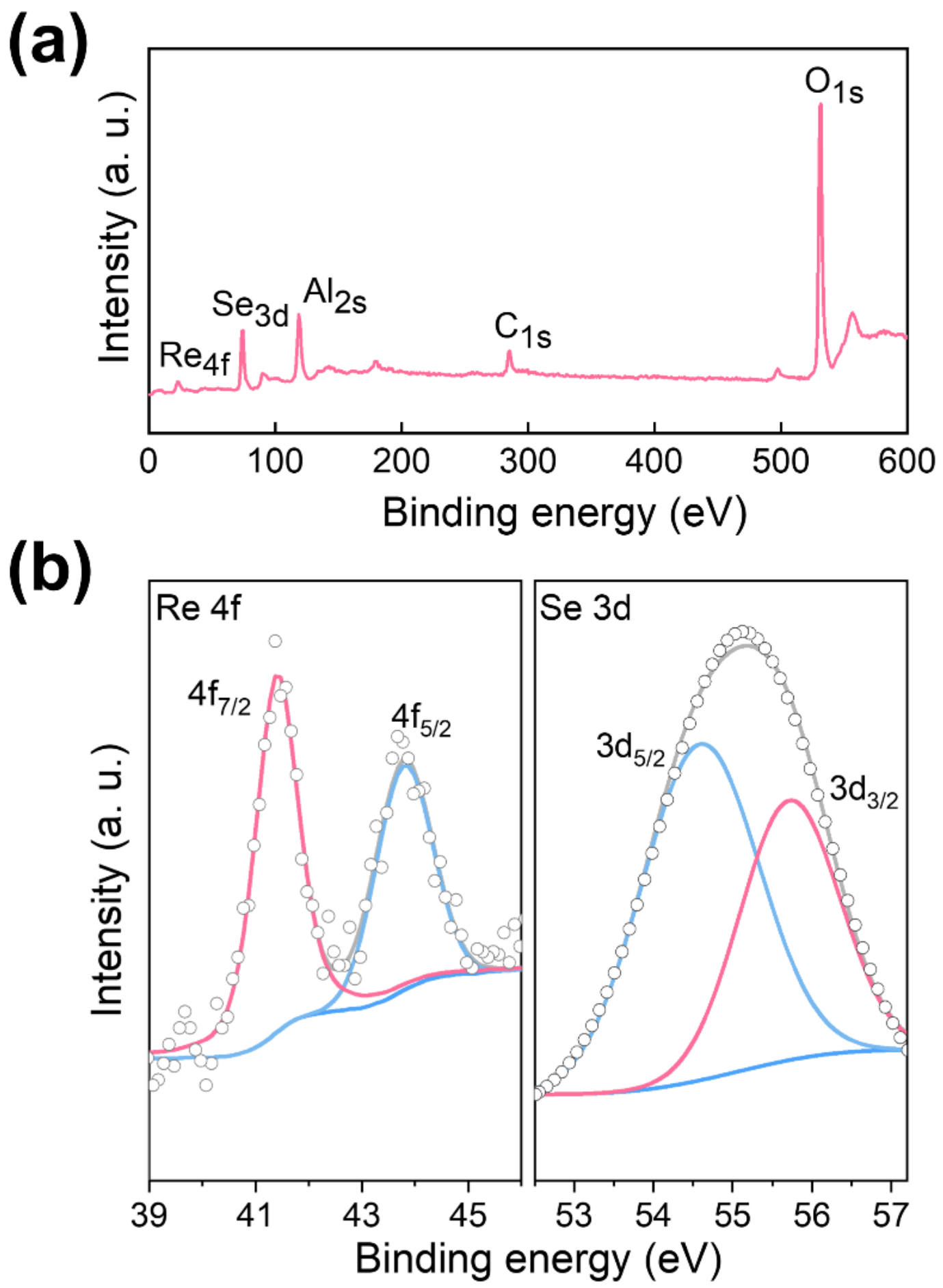

**Fig. S10.** XPS spectra of 2D $ReSe_2$ dendrites on *c*-$Al_2O_3$ substrate. (a) Full XPS spectra. (b) High-resolution XPS spectra of Se 3d and Re 4f peaks. The peaks of Se $3d_{3/2}$, Se $3d_{5/2}$, Mo $3d_{3/2}$, Re $4f_{5/2}$, and Re $4f_{7/2}$ are marked by dashed lines. The binding energies of $4f_{7/2}$ and $4f_{5/2}$ orbits for Re were located at ~41.8 and 44.2 eV, respectively, while the binding energies of $3d_{5/2}$ and $3d_{3/2}$ orbits for Se were located at ~55.0 and 56.1 eV, respectively, agreeing well with the XPS spectra of mechanically exfoliated $ReSe_2$ samples [12].

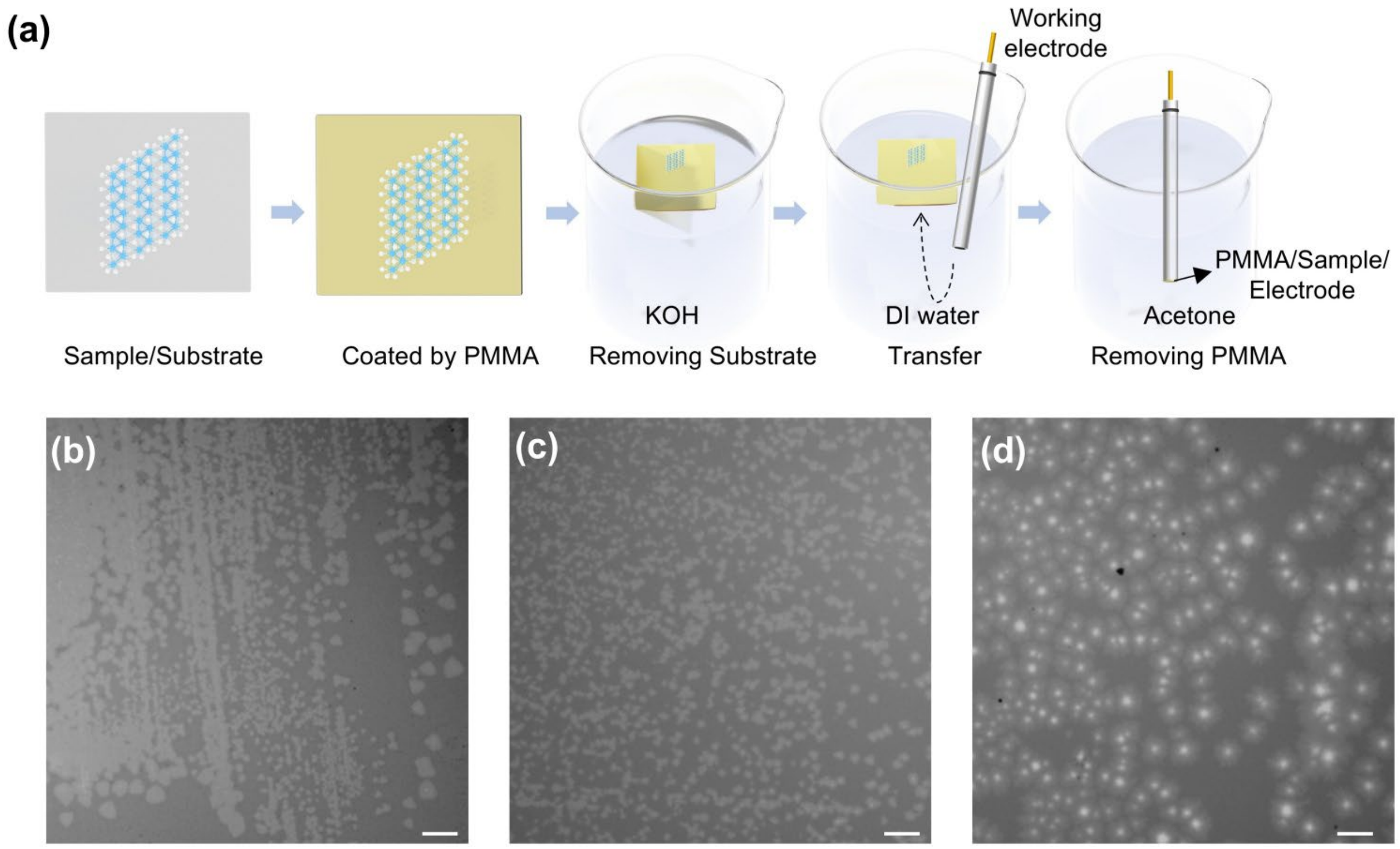

**Fig. S11.** Schematic showing the transfer of 2D $ReSe_2$ on the electrode and typical images of samples after transfer. (a) Schematic showing the transfer procedure of the CVD-grown samples via a chemical wet etching method. scale bars: 20 μm. (b-d) Typical optical images of 2D $ReSe_2$ crystals transferred on the electrode.

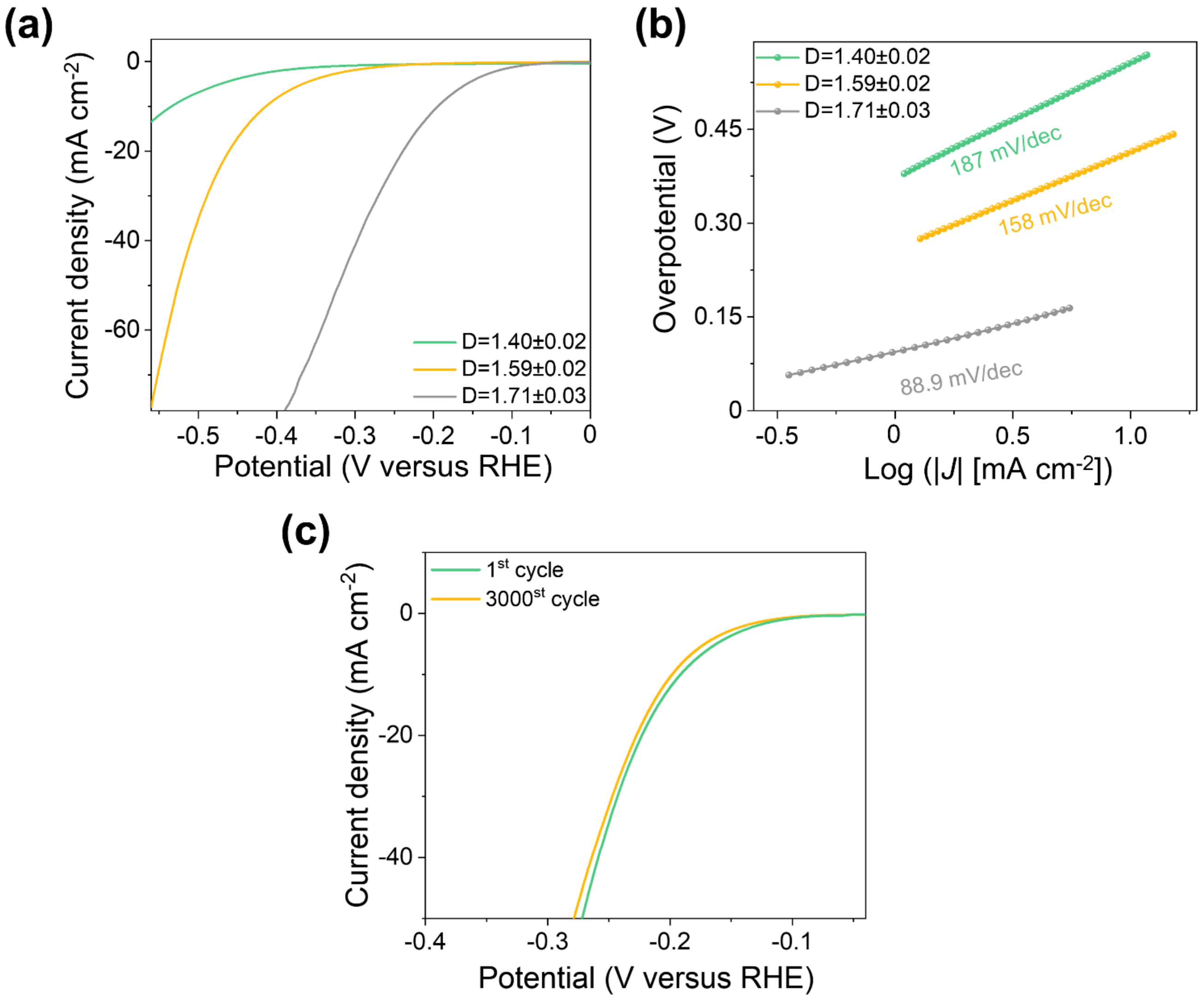

**Fig. S12.** HER performance of 2D $ReSe_2$ with different $D_F$. (a) Cathodic LSV curves of $ReSe_2$ with $D_F$ of 1.40±0.02, 1.59±0.02, and 1.71±0.03. (b) The corresponding Tafel slopes. (c) LSV curve of 2D $ReSe_2$ with a $D_F$ of 1.71 ± 0.03 before and after 3000 cycle tests.

Cathodic LSV curves show that $ReSe_2$ with a $D_F$ of 1.71 ± 0.03 exhibits an overpotential of only 195 mV at a current density of 10 mA cm$^{-2}$, representing a reduction of approximately 534 mV compared to $ReSe_2$ with a $D_F$ of 1.40 ± 0.02 (Fig. S12a). Its Tafel slope is 88.9 mV dec$^{-1}$, indicating a Volmer-Heyrovsky mechanism with the Heyrovsky reaction as the rate-determining step (Fig. S12b). The increase in $D_F$ promotes the rate-determining step to transform from the Volmer reaction ($H_3O^+ + e^- \rightarrow H_2O + H^*$) to the Heyrovsky reaction ($H_3O^+ + e^- \rightarrow H_2O + H_2$). In addition, the polarization curve of $ReSe_2$ with a $D_F$ of 1.71 ± 0.03 exhibits a negligible negative shift after 3000 cycles, demonstrating excellent long-term stability (Fig. S12c).

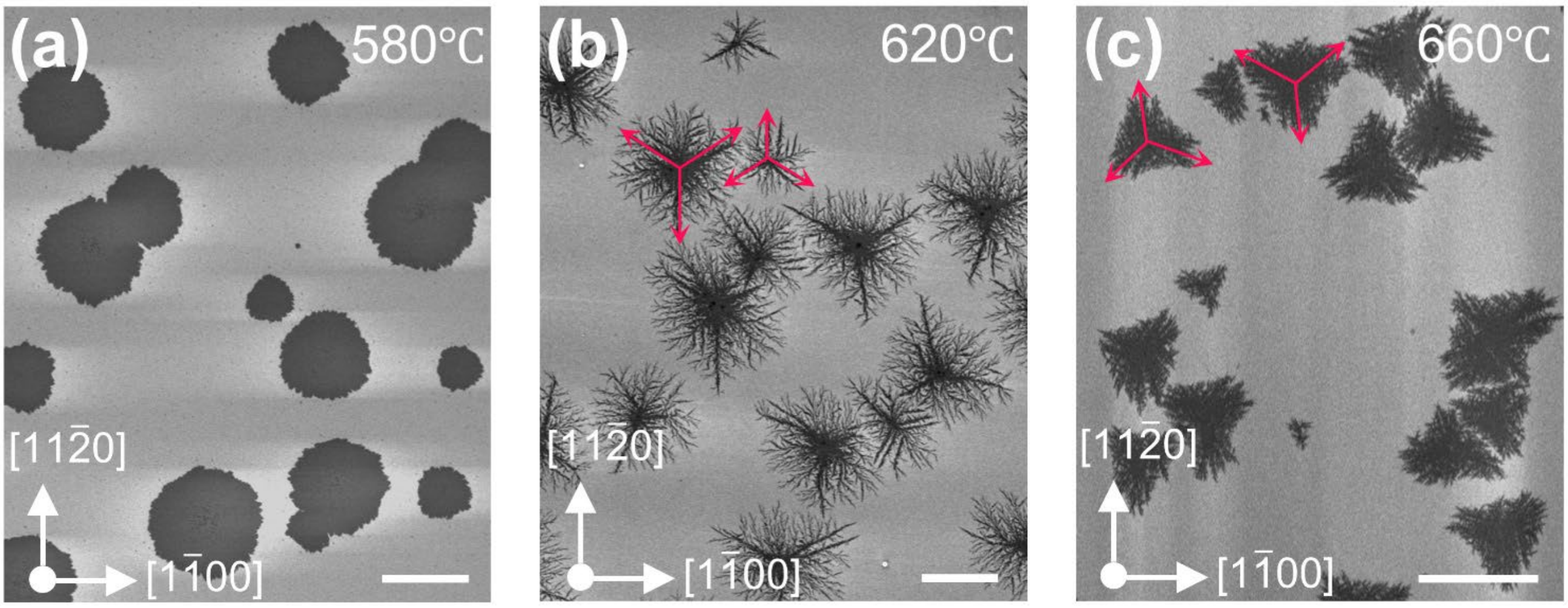

**Fig. S13.** Temperature-dependent morphology evolution and substrate symmetry-dictated orientation of 2D $ReSe_2$ dendrites on *c*- $Al_2O_3$. (a-c) SEM images of 2D $ReSe_2$ grown at different $T_{Re}$: 580 °C (a); 620 °C (b); and 660 °C (c). The other CVD process parameters are fixed at $c_{Re}$ = 0.10 mol/L, *sub.*= *c*-$Al_2O_3$, $T_{Se}$ = 250°C, $f_{H2}$ = 2 sccm. Scale bars (from left to right): 10 µm, 5 µm, 5 µm.

Fig. S13 together with Fig. 1g further supports two key experimental observations. First, for $ReSe_2$ grown on *c*-$Al_2O_3$ with three-fold rotational symmetry ($C_{3v}$), two preferential orientations twisted by 60° are present (Fig. S13b and c), as revealed by the angle concentration at 0° and 60° when measuring the smallest misorientation angle ($\alpha_1$) between *c*-$Al_2O_3$[11$\bar{2}$0] and the primary branches of $ReSe_2$ (Fig. 1g). Moreover, the minimum angles between the primary branches $\alpha_2$ and between the primary and secondary branches $\alpha_3$ are concentrated at 120° and 60°, respectively, showing an orientation preference that corresponds to the $C_{3v}$ symmetry of *c*-$Al_2O_3$. The susbstrate symmetry-dictated dendrite morphology is also present on MgO with four-fold rotational symmetry ($C_{4v}$), which is displayed in Fig. S14.

Second, the morphology shows a strong dependence on $T_{Re}$. As $T_{Re}$ increases from 580 to 660 °C, $ReSe_2$ domains evolve from nearly isotropic compact crystals to anisotropic dendrites. However, the highest branching degree occurs at $T_{Re}$ = 620 °C, with $D_F$ ≈1.7, whereas further increasing $T_{Re}$ to 660 °C lowers $D_F$ to ≈1.5, experimentally confirming the non-monotonic influence of $T_{Re}$ inferred from SHAP analysis, which will be discussed in detail in Fig. 2d.

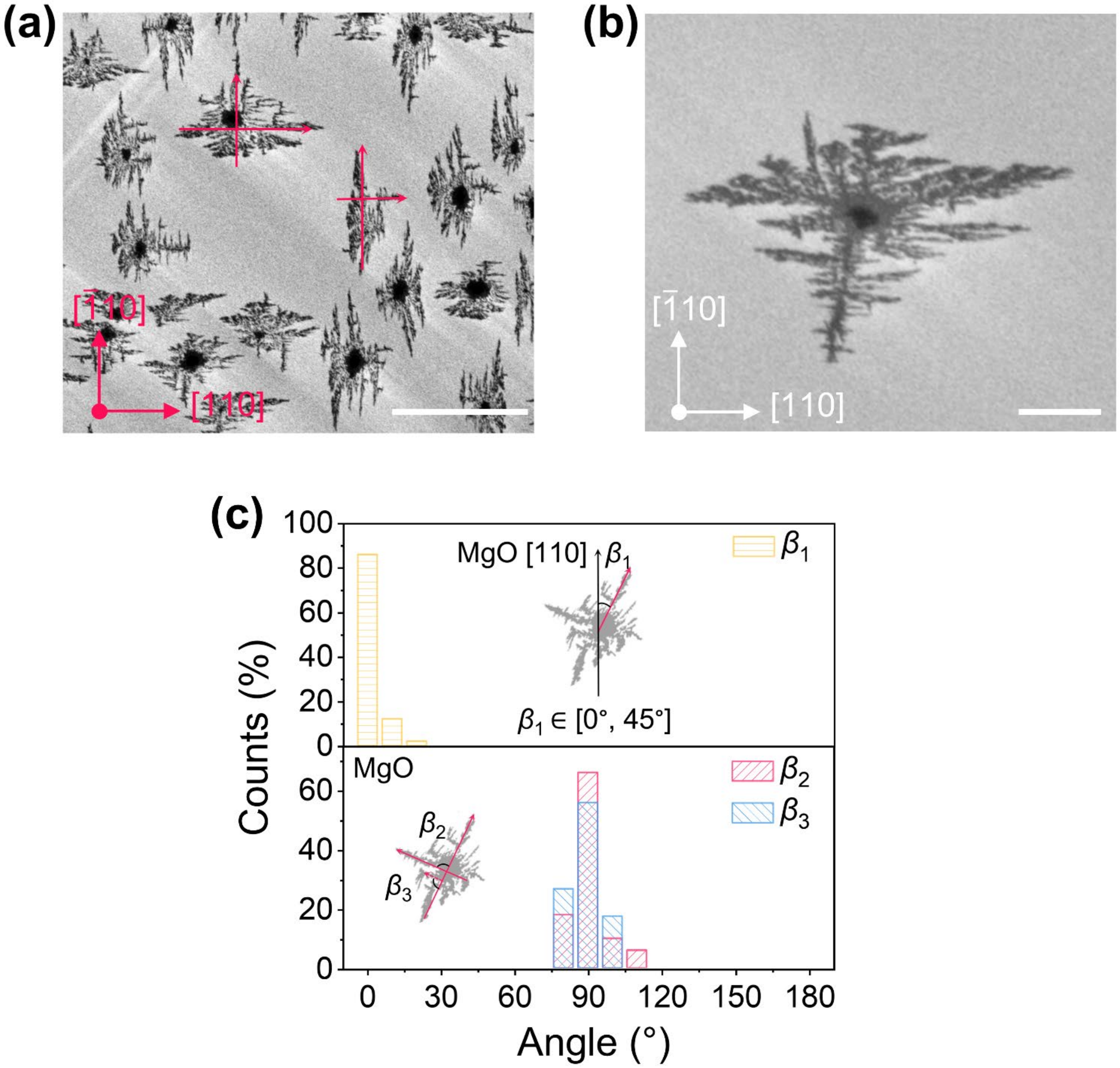

**Fig. S14.** Substrate symmetry-dictated orientation of 2D $ReSe_2$ dendrites grown on MgO. (a) The backbones of 2D $ReSe_2$ dendrites are highlighted with magenta arrows. Scale bar: 5 μm. (b) Zoomed-in SEM image of 2D $ReSe_2$ dendrites. Scale bar: 1 μm. (c) Histogram showing the angle distribution of $\beta_1$, $\beta_2$, and $\beta_3$ of 2D $ReSe_2$ grown on MgO(001). $\beta_1$: the minimum angle between MgO[110] and the primary branch of $ReSe_2$. $\beta_2$: the minimum angle between the primary branches of $ReSe_2$. $\beta_3$: the minimum angle between the primary and secondary branches of $ReSe_2$. Insets schematically illustrate $\beta_1$, $\beta_2$, and $\beta_3$.

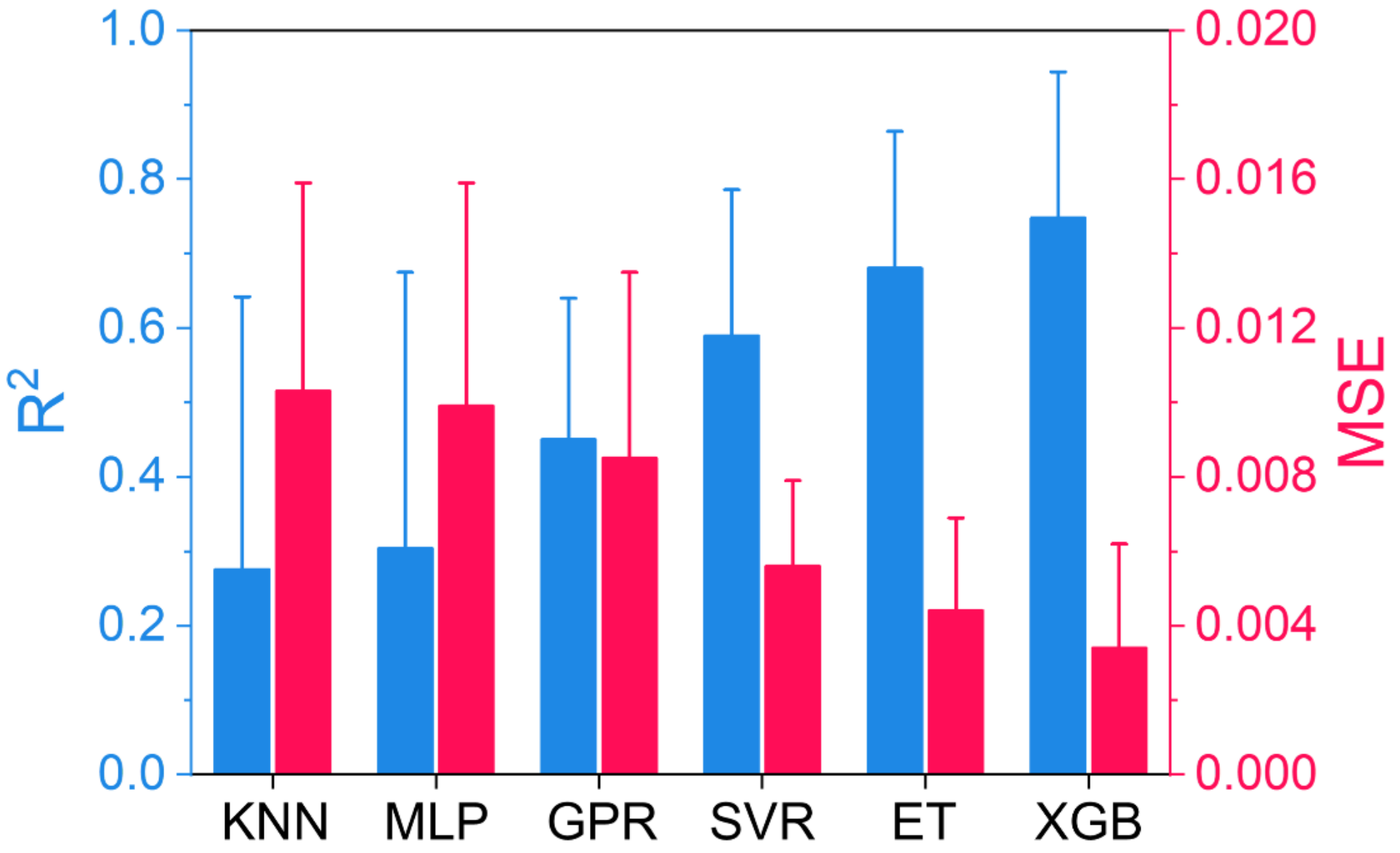

**Fig. S15.** Performances of six machine learning (ML) models, evaluated by $R^2$ and MSE using the 60 data points collected in Module 1 for training.

Six representative regression models covering different algorithmic architectures were selected to identify a suitable model for constructing the CVD parameter-$D_F$ mapping. These models include an instance-based k-nearest neighbors (KNN) regressor, a network-based multilayer perceptron (MLP), kernel-based GPR and support vector regressor (SVR) models, and tree-based extra trees (ET) regressor and XGBoost regressors. Among all the models, XGBoost achieved the highest $R^2$ with the lowest MSE, displaying the best fitting performance.

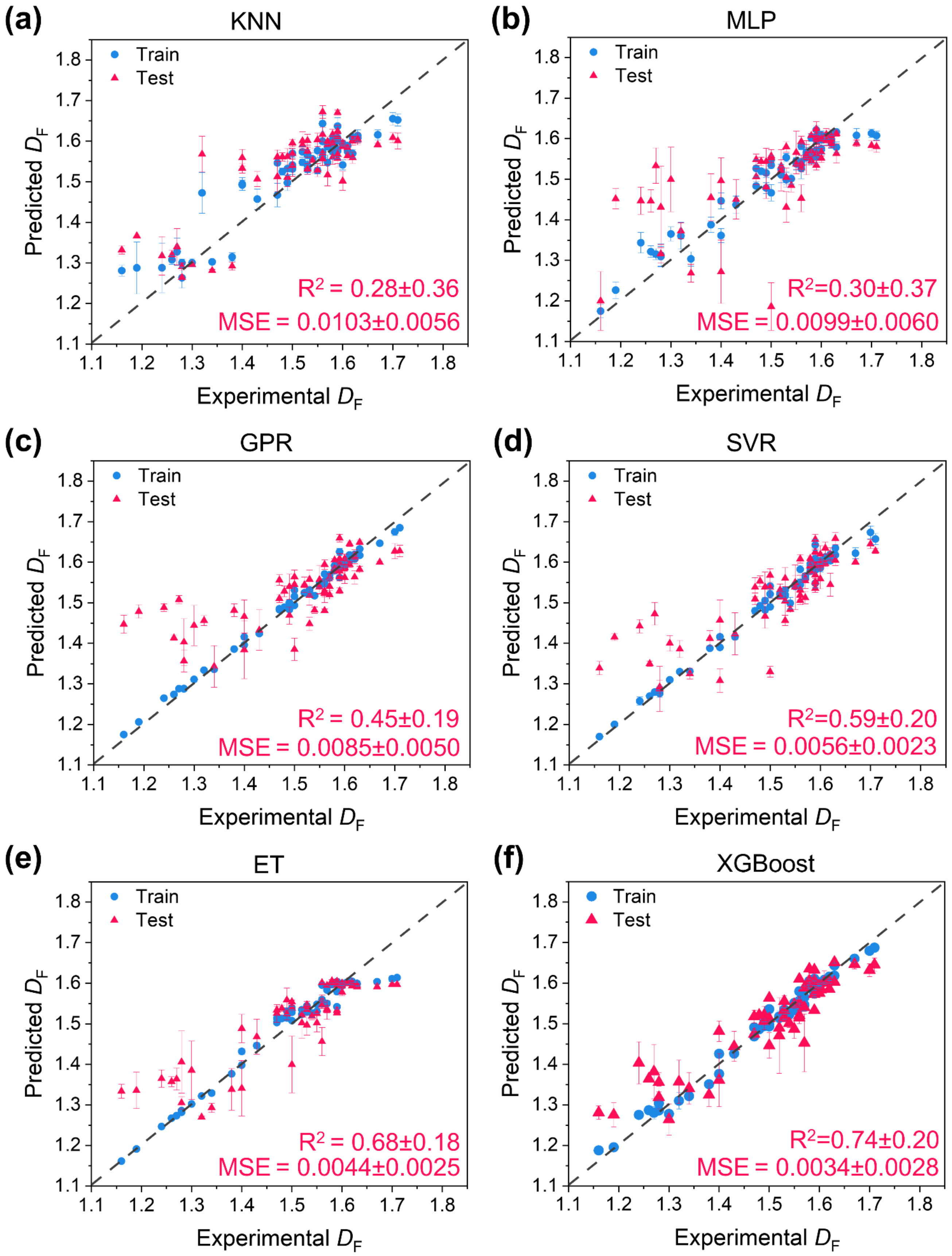

**Fig. S16.** Detailed regression performance of the six models using the dataset collected during active learning . Performance of KNN (a), MLP (b), GPR (c), SVR (d), ET (e), and XGBoost (f).

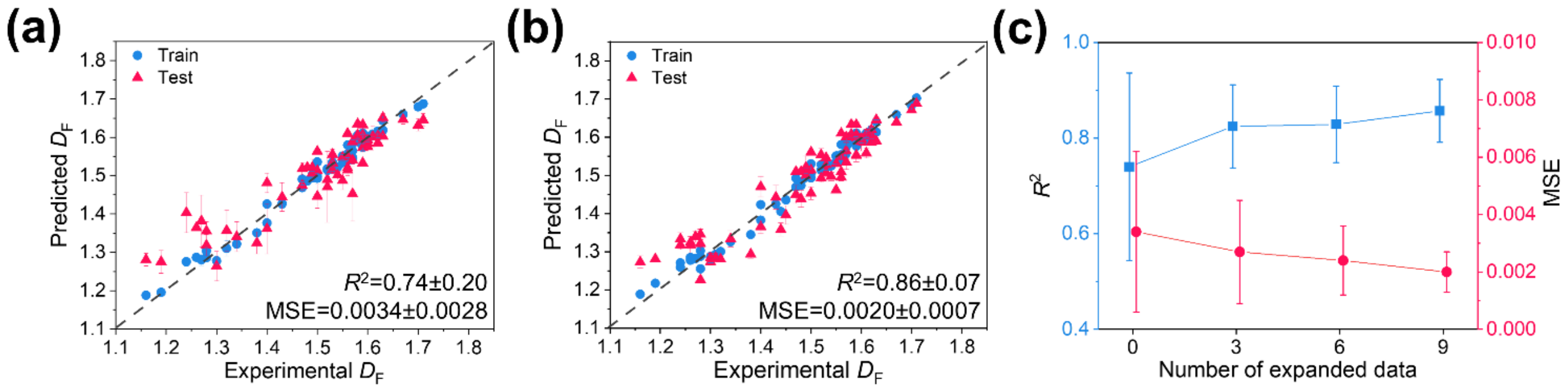

**Fig. S17.** Iterative improvement of XGBoost prediction by prediction accuracy (*PA*)-guided data augmentation. (a) Prediction performance of the XGBoost model before data augmentation. (b) Prediction performance of the XGBoost model after 3 rounds of data augmentation. (c) Evolution of $R^2$ and MSE values of the XGBoost model during the 3 rounds of *PA*-guided data augmentation process. Three supplementary experiments were added in each round, leading to a total of nine additional data points and a stepwise improvement in model performance.

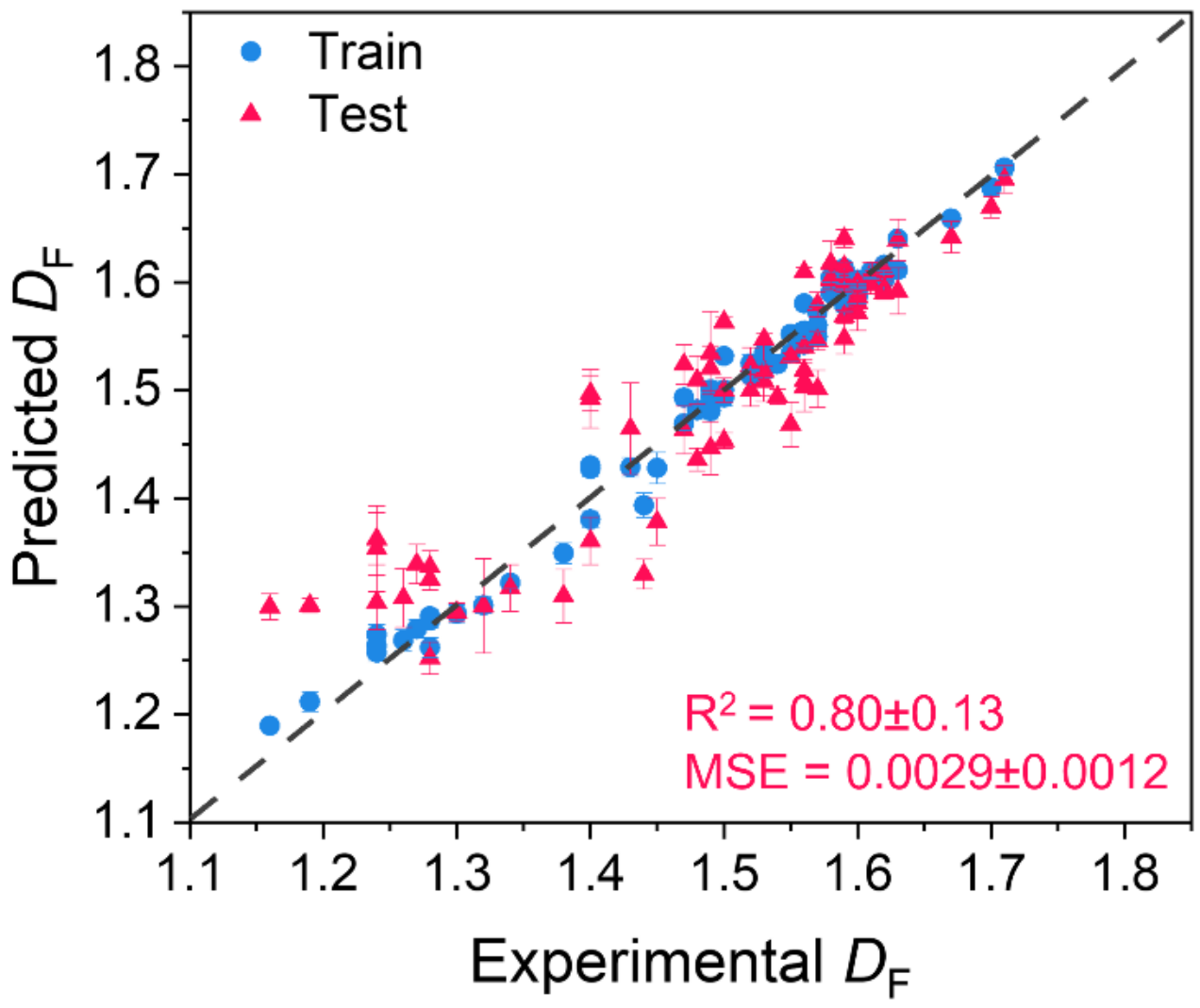

**Fig. S18.** Limited improvement of XGBoost prediction by one-shot data augmentation.

The XGBoost model was retrained after adding nine supplementary data points in a single round instead of 3 iterative *PA*-guided rounds (Table S10). This one-shot augmentation only reduces the MSE by 14.7% and increases $R^2$ by 8.1%, indicating that iterative model updating and stepwise compensation of high-error regions are more effective for improving global prediction performance.

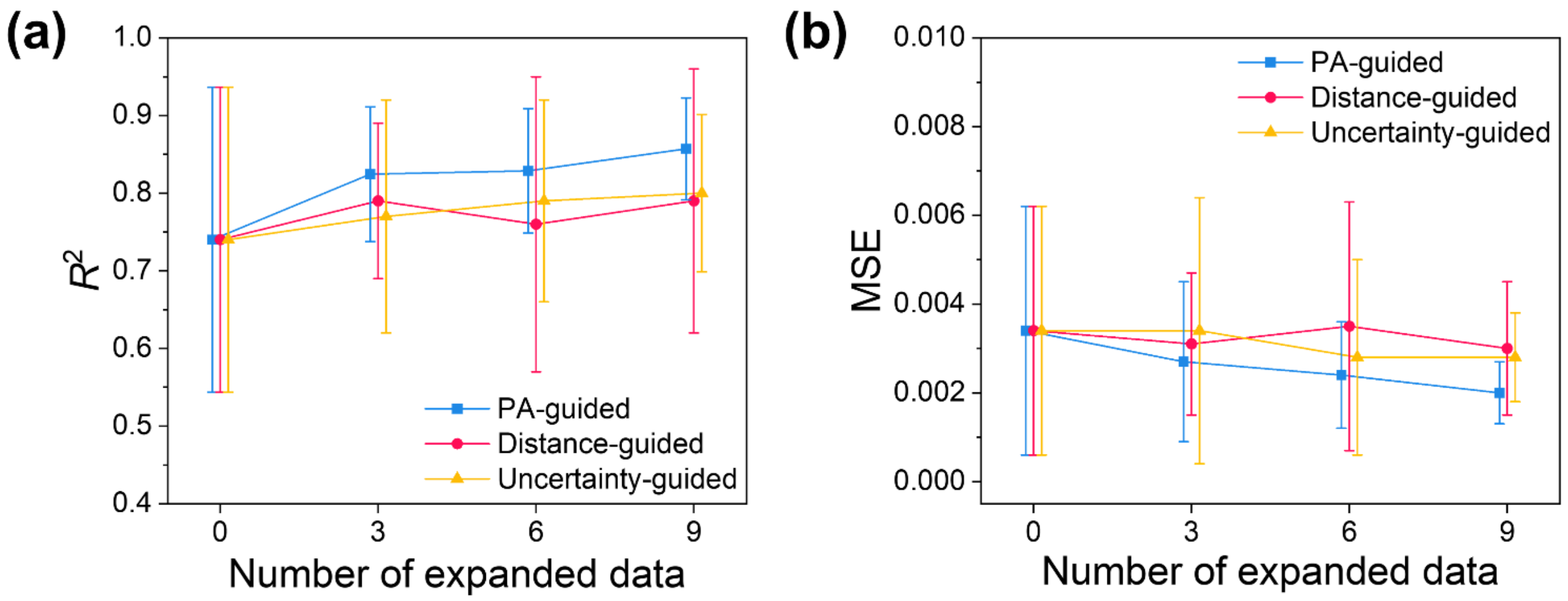

**Fig. S19.** Comparison of the data-augmentation efficiency of *PA*-guided, maximin distance-based, and uncertainty-based sampling strategies. (a) Evolution of $R^2$ values during the 3 rounds data augmentation process. (b) Corresponding evolution of the MSE. Each curve represents one sampling strategy. Three samples were added in each augmentation round, resulting in a total of nine added samples over three rounds.

To quantify the efficiency of the *PA*-guided data augmentation strategy with other sampling methods, we compared it with maximin distance-based sampling and uncertainty-based sampling strategy in Fig. S19. Maximin distance sampling method selects new data points by maximizing their minimum distance to the existing dataset, which can be expressed as:

$$x^* = arg \max_{x \in X_{cand}} \min_{x_i \epsilon D} d(x, x_i), \tag{S11}$$

where $\mathrm{X}_{cand}$ is the candidate space, $D$ is the existing dataset, and $d(x, x_i)$ is the distance between a candidate point and an existing data point. Uncertainty-based sampling method selects candidate points with the largest variation among predictions from the 25 models constructed through 5-repeated 5-fold cross-validation:

$$x^* = arg \max_{x \in X_{cand}} \sqrt{\frac{1}{25} \sum_{i=1}^{25} (y_{pred.}^{(i)} - \overline{y_{pred.}})^2}, \tag{S12}$$

where $y_{pred.}^{(i)}$ is the prediction of $i$-th model and $\overline{y_{pred.}}$ is the mean prediction of all models.

All three strategies started from the same dataset of 60 Bayesian-optimization experiments and selected three additional samples per augmentation round from the identical candidate space, resulting in a total of nine added samples over three rounds. After adding nine additional data points, $R^2$ of the PA-guided strategy, maximin distance sampling, and uncertainty-based sampling increased from 0.74 to 0.86, 0.79, and 0.80, respectively. Meanwhile, MSE decreased from $3.4 \times 10^{-3}$ to $2.0 \times 10^{-3}$, $3.0 \times 10^{-3}$, and $2.8 \times 10^{-3}$, respectively. Thus, the *PA*-guided strategy achieved an approximately 10% higher $R^2$ and an approximately 30% lower MSE than the two reference strategies under the same experimental budget.

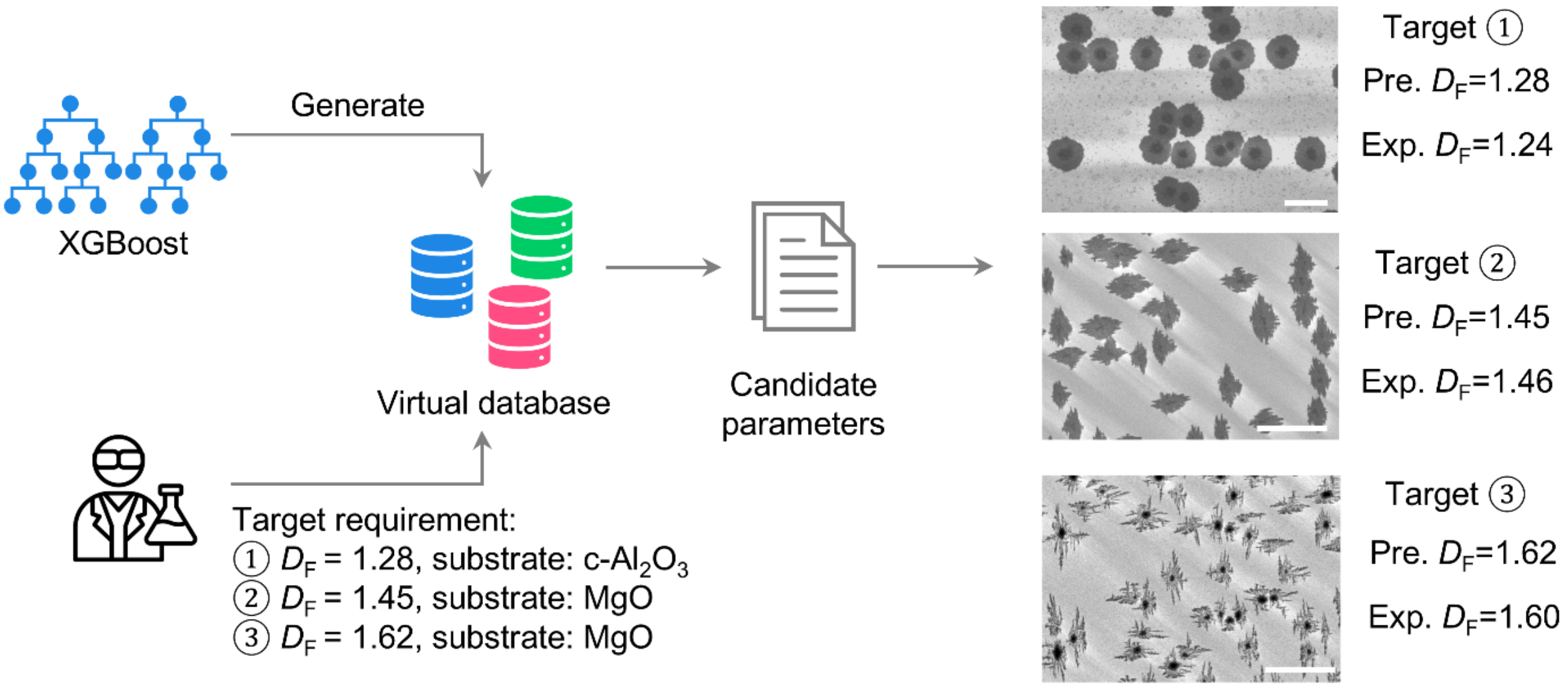

**Fig. S20.** Customized synthesis of 2D $ReSe_2$ dendrites through inverse search of the learned synthesis space. Scale bars: 5 μm.

Based on the XGBoost model enhanced by the $D_F$ prediction accuracy-guided data augmentation strategy, a virtual database can be generated, which contains paired data with growth parameters and $D_F$. Researchers can pre-set the target $D_F$ and substrate, then search for the corresponding growth parameter combinations in the virtual database for customized synthesis. For instance, we defined three $D_F$ targets, 1.28, 1.45, and 1.62, respectively and required the substrates to be *c*-$Al_2O_3$, MgO, and MgO. The combinations of growth parameters suggested by the XGBoost model are shown in Table S11. The experimental $D_F$ values of the 2D $ReSe_2$ crystals grown using the CVD parameters recommended by the ML model are very close to $D_F$ pre-defined by researchers with errors smaller than 0.05.

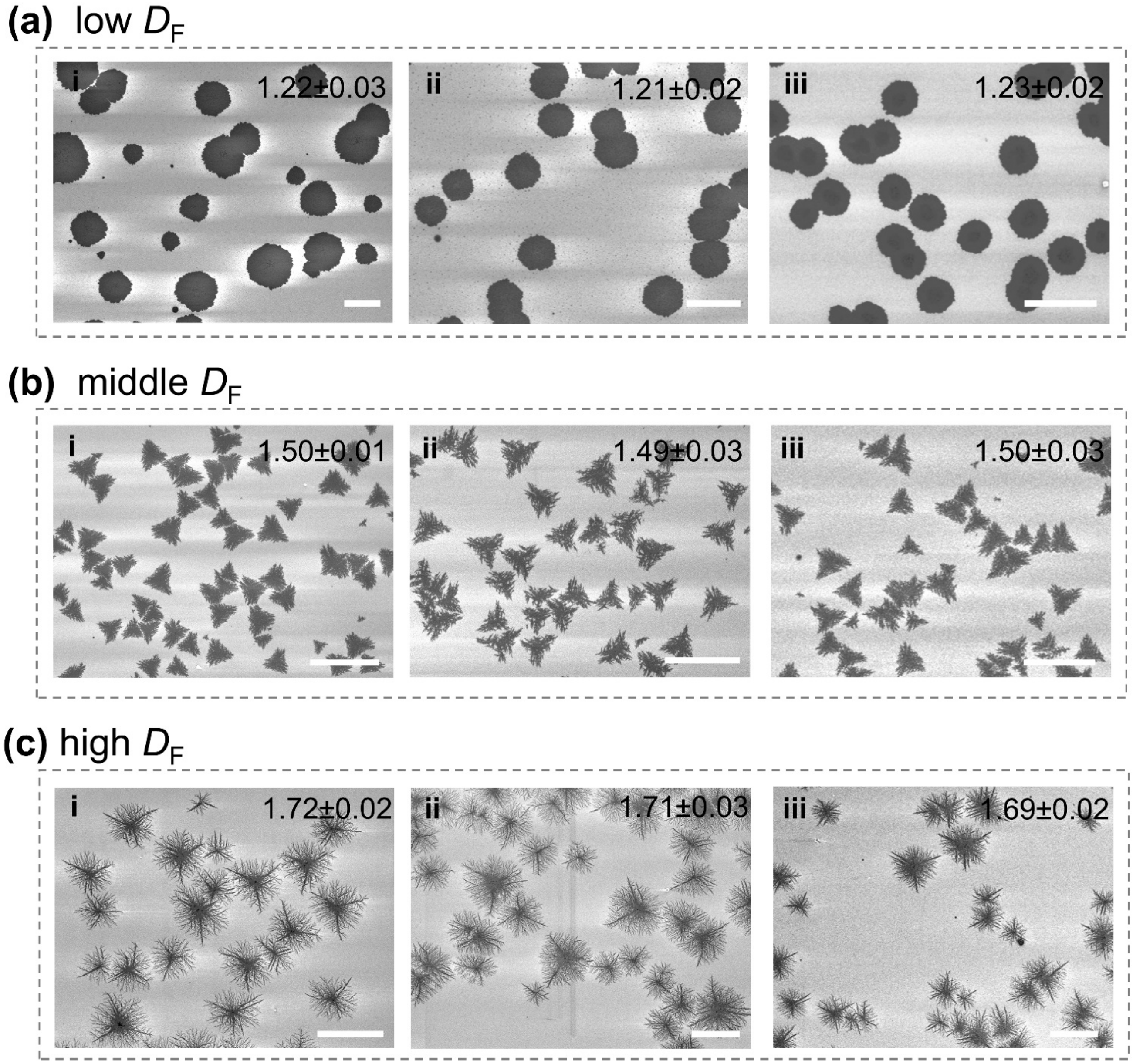

**Fig. S21.** Reproducibility of representative optimized $ReSe_2$ dendrite synthesis recipes. SEM images showing morphology characterizations of $ReSe_2$ dendrites synthesized from three optimized recipes corresponding to (a) low, (b) medium, and (c) high $D_F$. For each recipe, three independently repeated growth experiments were performed under the same synthesis conditions.The detailed recipe and the calculated $D_F$ for every experiment are displayed in Table S14. Scale bars in (a) : 10 µm. Scale bars in (b,c) : 5 µm.

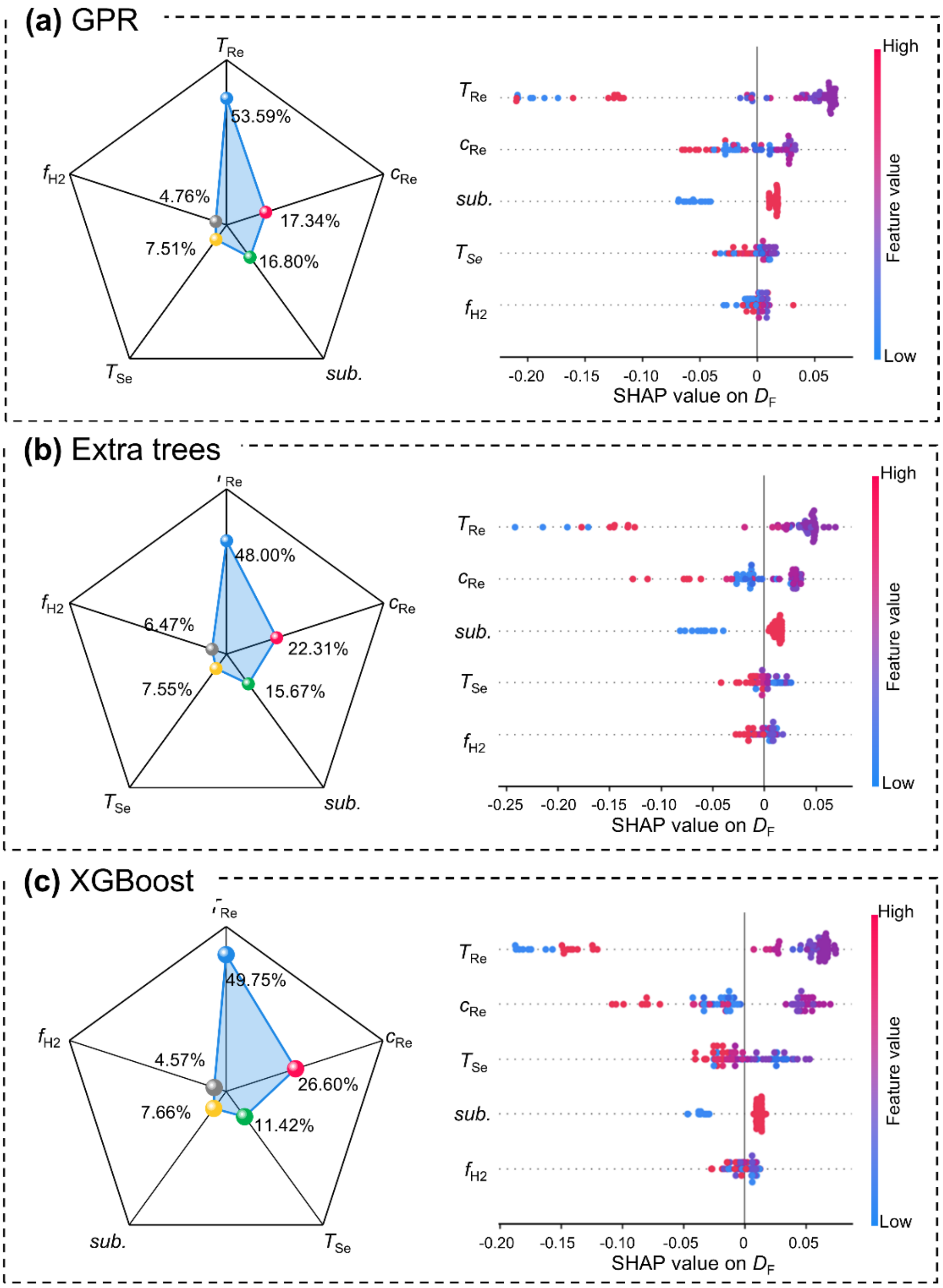

**Fig. S22.** Model dependence of SHAP-based feature importance for $ReSe_2$ dendrite growth. Feature importance and beeswarm plot of SHAP values obtained from (a) GPR, (b) Extra Trees, and (c) XGBoost models.

Since SHAP explains a specific trained model, the exact feature-importance ranking can vary with the regression model. To evaluate this model dependence, we compared the SHAP results obtained from GPR, Extra Trees, and XGBoost, as shown in Fig. S22.

The results show that the exact ranking of less important variables varies slightly among different models. However, $T_{Re}$ and $c_{Re}$ consistently remain among the most important features across all models. More importantly, all models reveal the same non-monotonic contribution pattern: both low and high $T_{Re}$ values make negative contributions to $D_F$, whereas only intermediate $T_{Re}$ values enhance $D_F$. A similar non-monotonic influence on the model output is also observed for $c_{Re}$. Therefore, the main conclusion that $T_{Re}$ and $c_{Re}$ dominate $D_F$ of $ReSe_2$ dendrites is robust and does not depend on a specific regression model.

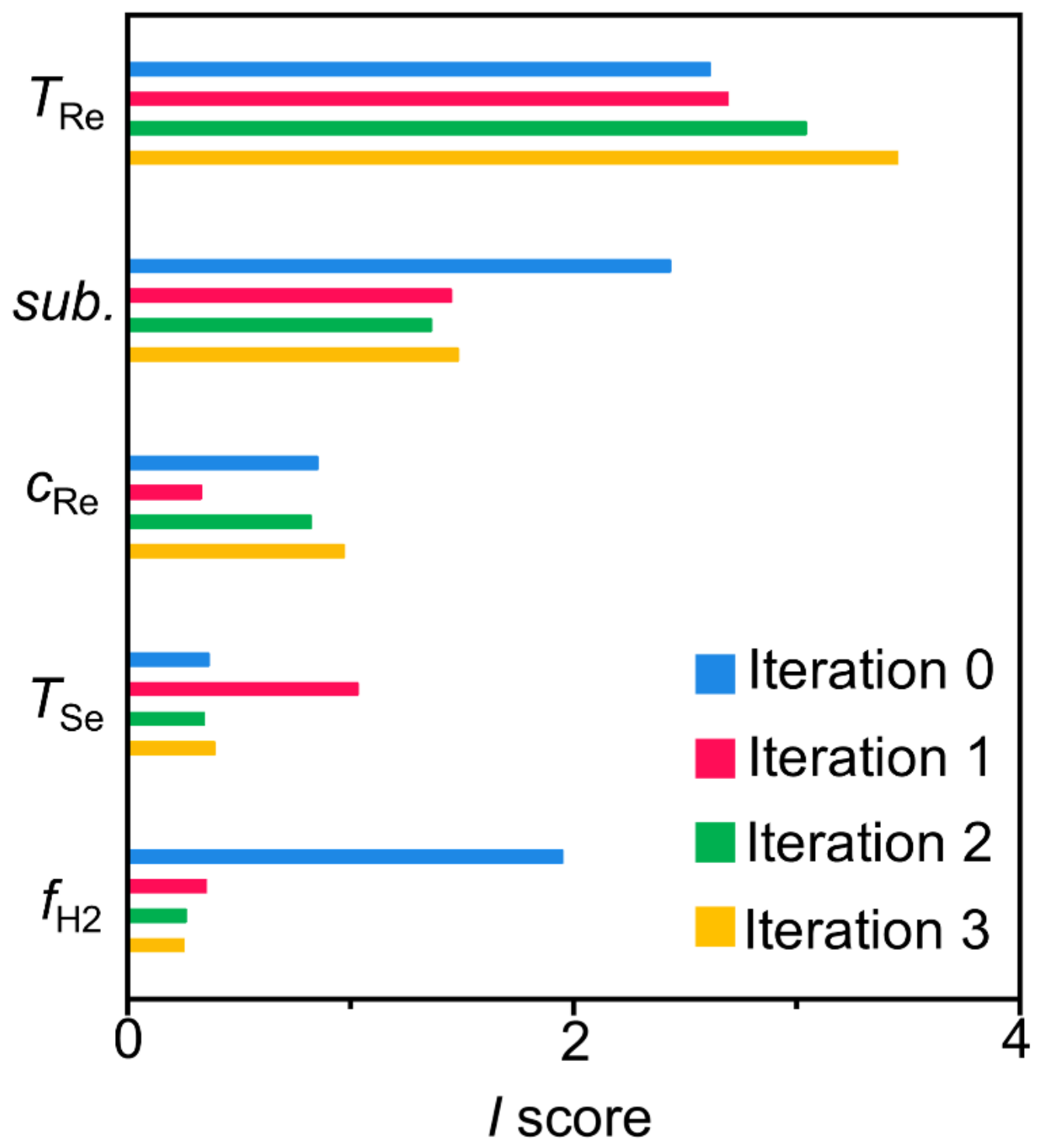

**Fig. S23.** Histogram of *I* scores for the five CVD process parameters in the active learning module.

Here, We first intend to elaborate on the definition of the *I* score. To quantify the contribution independence of each CVD parameter to $D_F$, we leveraged SHAP feature value analysis from collaborative game theory and proposed a measure, called *I* score, which is defined as the ratio between the SHAP main effect and the total SHAP interaction effects:

$$I\text{ score} = \frac{\left|\mathrm{SHAP}_{x_i\text{ main}}\right|}{\sum_{i\neq j}\left|\mathrm{SHAP}_{x_i\text{-}x_j\text{ interaction}}\right|}, \\ = \frac{\left|\mathrm{SHAP}_{x_i\text{-}x_i\text{ main}}\right|}{\sum_{i\neq j}\left|\mathrm{SHAP}_{x_i\text{-}x_j\text{ interaction}}\right|}, \tag{S13}$$

where $\mathrm{SHAP}_{x_i\text{ main}}$ represents the SHAP main effect value of feature $x_i$, indicating the contribution of $x_i$ to the model output when acting alone. It is equal to the SHAP interaction value of $x_i$ with itself, denoted as $\mathrm{SHAP}_{x_i\text{-}x_i\text{ main}}$. $\mathrm{SHAP}_{x_i\text{-}x_j\text{ interaction}}$ represents the SHAP interaction value of two different features $x_i$ and $x_j$, evaluating the synergistic contribution of $x_i$ and $x_j$ to the model output (i≠j) (See Supplementary methods, "Model interpretability- (ii) Quantification of the independence level of feature's impact on $D_F$"). The higher the *I* score is, the stronger the independence of this feature's impact on $D_F$ becomes.

The SHAP interaction matrix in **Fig. 2e** of the main text provides an example for this analysis. In this matrice, the diagonal elements represent the SHAP main effects of individual parameters, while the off-diagonal elements represent pairwise SHAP interaction effects. The bottom right panel of Fig. 2e shows the calculated *I* score for each CVD parameter. $T_{Re}$ and *sub.* have the highest *I* scores, indicating their independent contribution to $D_{F,}$ while the *I* scores of $T_{Se}$ and $f_{H2}$ are low, suggesting their strong entanglement with other parameters that jointly influence $D_F$.

We have also calculated I scores of the five CVD parameters for each iteration of the active learning process, as summarized in Fig. S20. The results are very similar with Fig. 2e and can further explain the convergence behavior of different parameters in the BO optimization process. $T_{Re}$ and *sub.* maintain the two highest *I* scores among the five parameters in all iterations, which indicates their strong independence level of impact on $D_F$ and is consistent with their fast convergence from iteration 1 in **Fig. 1b,c** in the main text. The *I* scores of $c_{Re}$ are situated in the middle in most iterations, while those of $T_{Se}$ and $f_{H2}$ are at the bottom, in line with the moderate convergence rate of $c_{Re}$ and non-convergence behaviors of $T_{Se}$ and $f_{H2}$ in active learning.

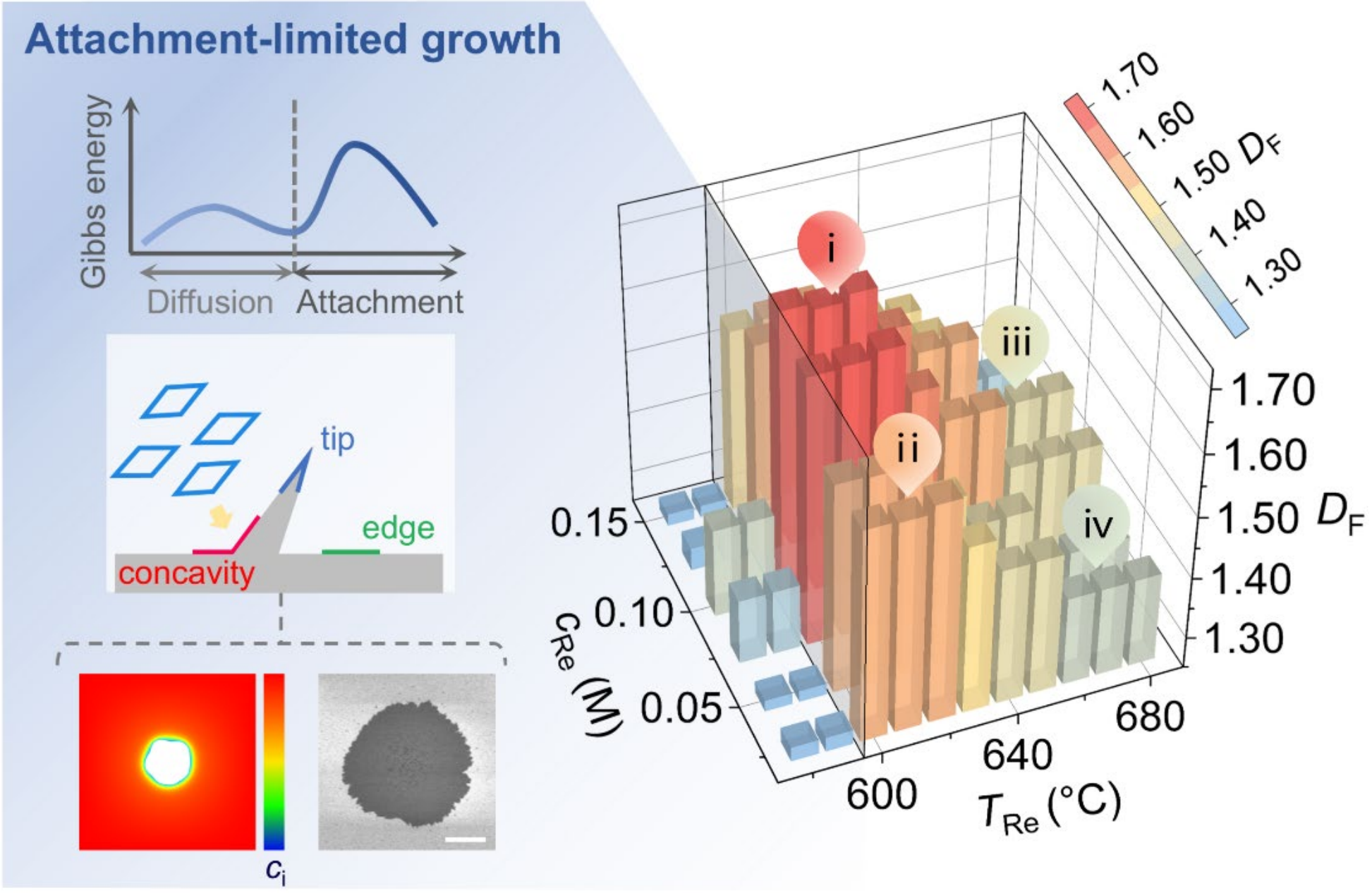

**Fig. S24.** Attachment-limited growth of 2D $ReSe_2$ when $T_{Re}$ is below 600°C.

**When $T_{Re}$ is below 600°C, 2D $ReSe_2$ is thermodynamically controlled by an attachment-limited growth**, meaning that the adsorption and bonding of the precursor species to the domain edges is the rate-dominant step. In this case, the surface energy of the system determines the preferential adsorption sites of the active species. Compared with crystal tips and edges, the attachment of precursor molecules to concavities reduces the surface energy the most. Therefore, any protrusions at domain edges will be quickly smoothed out under this mechanism, prohibiting the formation of dendritic structures thermodynamically. In addition, previous literature has reported that Re4 chains are prone to deflect integer multiples of ~60° on *c*-$Al_2O_3$ during crystal growth due to the capability of forming highly coherent grain boundaries with low energy[13,14]. This not only generates polycrystalline structures of $ReSe_2$ but also leads to a circular crystal shape due to isotropic growth rates in all lateral directions independent of either the substrate surface symmetry or the intrinsic crystal structure of the triclinic $ReSe_2$. This mechanistic deduction is further validated by experimental observations (the SEM image displaying the typical product morphology obtained under these conditions) and the phase-field simulation, as presented in the bottom left panel of Fig. S24.

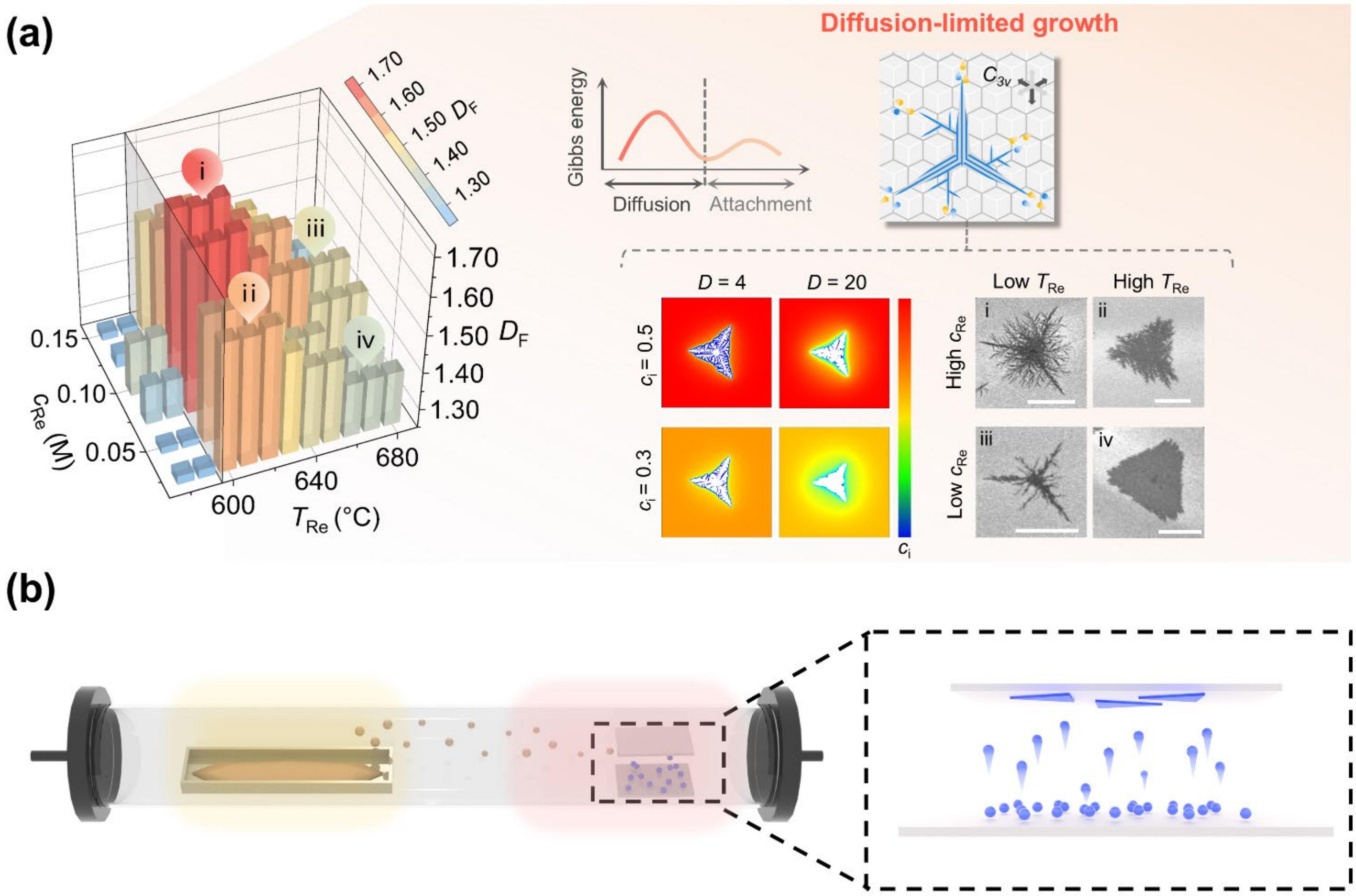

**Fig. S25.** Diffusion-limited growth of 2D $ReSe_2$ when $T_{Re}$ exceeds 600°C (a) and the corresponding space-confined CVD setup for $ReSe_2$ synthesis (b).

**When $T_{Re}$ exceeds 600°C, $ReSe_2$ swifts to kinetically controlled diffusion-limited growth**, in which the surface migration of precursor species becomes the rate-dominant step. In this regime, dendrite formation is governed by adatom diffusion and the local concentration gradient near the crystal edge. Our previous work has applied density functional theory calculations to unveil that the diffusion energy barrier of Re atoms on *c*-$Al_2O_3$ displays prominent anisotropy dependent on the lattice orientation of *c*-$Al_2O_3$[14]. Therefore, the macroscopic shape of polycrystalline dendrites and the preferential growth orientations of branches exhibit obvious association with the $C_{3v}$ symmetry of *c*-$Al_2O_3$, which agrees with the experimental observations in Fig. 1g.

It is noteworthy that, under diffusion-limited synthesis, the growth rate of 2D $ReSe_2$ at a certain temperature is affected by the concentration of precursor species absorbed on the substrate surface (commonly refers to Re concentration, because Se is in excess). In our **space-confined CVD setup** constructed by two substrates placed face-to-face, Re precursor ($NH_4ReO_4$ aqueous solution) is spin-coated onto the bottom blank substrate, which can be vaporized to adsorb onto the upper *c*-$Al_2O_3$ surface via heating for $ReSe_2$ dendrite growth (Fig. S25b).

**Under this setup, the Re concentration adsorbed on the upper substrate surface is jointly influenced by $T_{Re}$ and $c_{Re}$,** because $T_{Re}$ dictates both the heating temperature of the bottom substrate coated with Re source and the temperature of the upper substrate at which adsorption equilibrium occurs, while $c_{Re}$ impacts the supply amount of the Re source from the bottom substrate. **Therefore, different combinations of $T_{Re}$ and $c_{Re}$ can vary $D_F$ of dendrites** (as revealed by regions of (i), (ii), (iii), and (iv) marked on the left panel of Fig. S25a**.**

Take cases (i) and (ii) corresponding to low $T_{Re}$ with high $c_{Re}$ and low $T_{Re}$ with low $c_{Re}$ as an example to compare their products' $D_F$ (Fig. S25a, SEM images in the bottom right panel). Due to their same $T_{Re}$, the equilibrium Re adsorption concentration at the upper substrate surface (defined as $c_{equ.}$) is solely determined by $c_{Re}$, thereby resulting in a higher $c_{equ.}$ of case (i) than that of case (ii). Furthermore, the heating temperature of the upper substrate, which equals $T_{Re}$, affects the migration ability of Re atoms on the substrate surface, thus determining the thickness of the depletion zone at the $ReSe_2$ crystal edge. The lower the $T_{Re}$ temperature is, the narrower the depletion zone thickness becomes. Here, the depletion zone is the region with a negative concentration gradient of precursor species, extending from the edge of the $ReSe_2$ crystal to the location where the Re adsorption concentration reaches $c_{equ.}$. When a branch is accidentally generated at the edge of a $ReSe_2$ domain due to energy perturbation and inserted into the depletion layer, the Re concentration at the tip of the branch is higher than at its base due to the presence of a negative concentration gradient within the depletion layer, which makes the branch not to be smoothed out and disappear, but rather to grow longer and longer. This explains why dendrites can be produced in diffusion-limited growth. Since case (i) has a higher $c_{equ.}$ than case (ii) and both cases adopt a low $T_{Re}$, a sharper negative Re concentration gradient is produced in the depletion layer under case (i) than under case (ii). Therefore, the branch tip under case (i) is situated at a higher Re concentration than under case (ii), which gives $ReSe_2$ dendrites in case (i) a larger $D_F$. The $D_F$ variations by other combinations of $T_{Re}$ and $c_{Re}$ can be demonstrated in a similar way.

**We also applied the phase-field simulation to demonstrate the growth mechanism more quantitatively** (Fig. S25a, bottom-middle panels)**.** In this case, $T_{Re}$ mainly controls the adatom diffusion coefficient $D$ through an Arrhenius relationship, $D \propto exp(-\Delta G_D/k_B T_{Re})$, where $\Delta G_D$ is the diffusion barrier. $c_{Re}$ regulates the amount of Re precursor supplied from the bottom substrate and thereby determines the initial adatom concentration $c_i$ on the upper growth substrate. Different combinations of $T_{Re}$ and $c_{Re}$ lead to different simulated dendrites morphologies, which agree with the experimental observations well (SEM images on the right).

The above growth mechanism, which elaborates the **dependence of $D_F$ on multiple CVD parameters**, is established via the synergistic **integration of interpretable machine learning, multiscale characterizations, phase-field simulations and empirical domain knowledge**. It achieves excellent consistency with experimental observations and features both **qualitative interpretability and quantitative precision**, showcasing a novel paradigm of human–machine collaboration for mechanism unraveling.

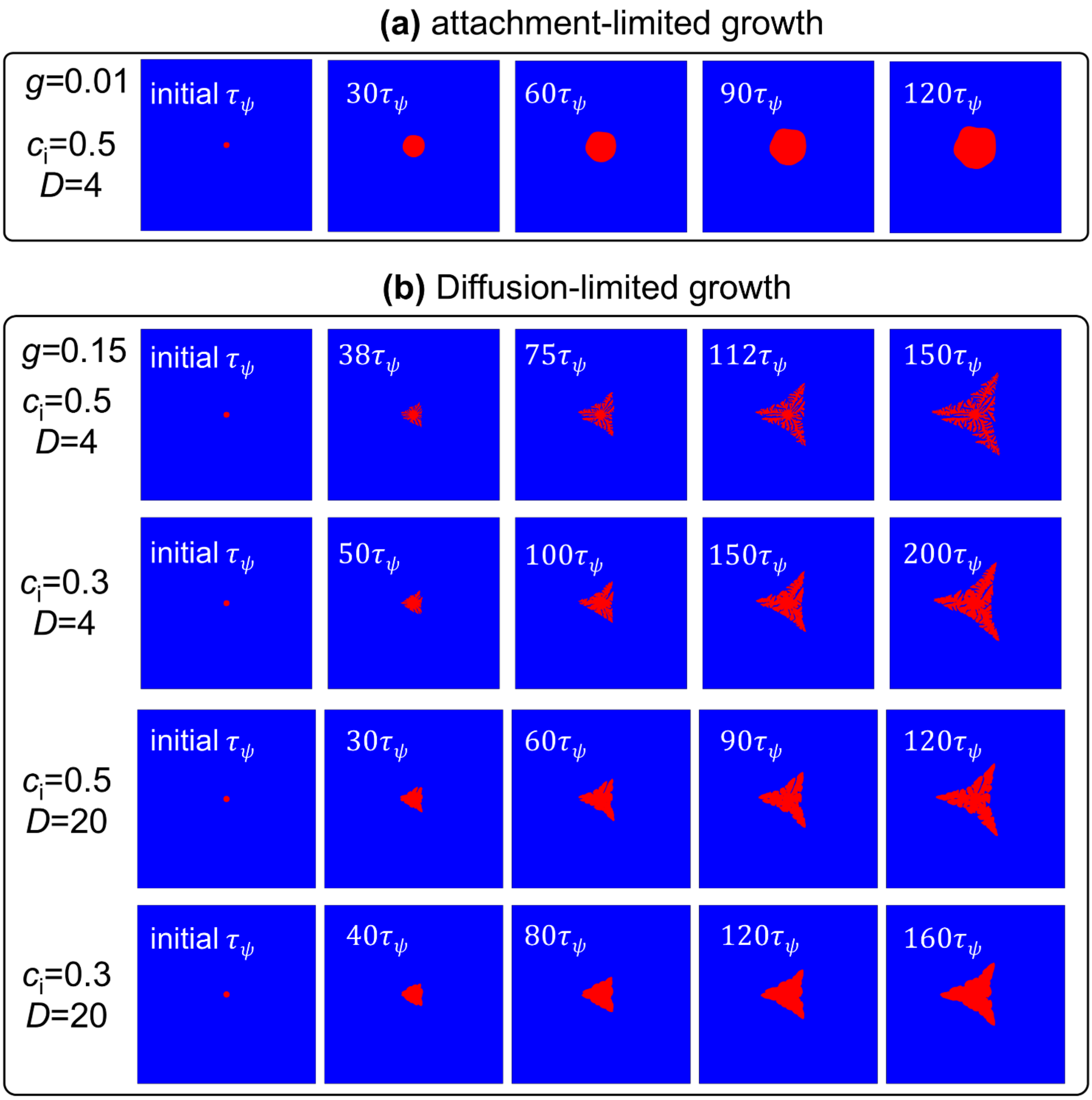

**Fig. S26.** Phase-field simulations of $ReSe_2$ morphological evolution. $\tau_\psi$ is the characteristic attachment time once the adatom arrives at the edge. The upper panel shows the temporal evolution of islands under attachment-limited growth, while the lower panel displays the temporal evolution of islands under diffusiont-limited growth. The initial simulation condition involves setting a circular nucleus with a radius of 0.1 $w_0$ at the center. The side length of the simulation domain is 600 $w_0$.

To simulate growth under thermodynamic control and kinetic control, the coefficient $g$ that quantifies the amplitude of the angular dependence of interfacial energy is modified in this work. Specifically, $g$=0.01 is adopted for adsorption-limited simulations, whereas $g$=0.15 is set for diffusion-limited simulations. In the attachment-limited growth regime, crystals maintain a stable circular morphology dominated by the energy term $\nabla\cdot\{[w^2(\theta)I_0+w(\theta)w'(\theta)I_1]\nabla\psi\}$ (details can found in Supplementary methods). In contrast, for diffusion-limited growth, the circular morphology can only be preserved in the early stage and eventually evolves into a dendritic structure. As the evolution time increases, $D_F$ of dendrites rises gradually (Fig. S26).

Based on the variable correspondence established in Fig. S25, $D$ and $c_i$ were varied in the diffusion-limited phase-field simulations to examine the effects of $T_{Re}$ and $c_{Re}$, respectively. When $c_i$ is fixed, a lower $D$ ($D$=4), corresponding to reduced adatom mobility at lower $T_{Re}$, restricts adatom migration toward the growth front and maintains a narrower depletion zone near the crystal edge, as shown by the concentration-field evolution in Fig. S27. Once a branch is accidentally generated at the edge of a $ReSe_2$ domain due to energy perturbation and inserted into the depletion layer, the narrow depletion layer at lower $D$ drives the branch to grow increasingly longer and leads to dendrites with higher $D_F$. This diffusion coefficient dependent trend is consistent with the experimental comparisons in **Fig. 2f** of the main text, where pairs i–ii and iii–iv show that varying $T_{Re}$ under comparable $c_{Re}$ conditions changes the dendritic branching degree.

The role of precursor supply was then examined by varying $c_i$ under a fixed $D$. A higher $c_i$, corresponding to increased Re precursor supply from larger $c_{Re}$, accelerates the advancement of the growth front. The rapid consumption of nearby adatoms generates a steeper concentration gradient ahead of the protruding branch tips. This positive feedback between adatom consumption, concentration-gradient formation and tip advancement promotes dendritic growth. This precursor supply dependent trend agrees with the experimental comparisons in **Fig. 2f** of the main text, where pairs i–iii and ii–iv show that varying $c_{Re}$ under comparable $T_{Re}$ conditions alters dendritic growth.

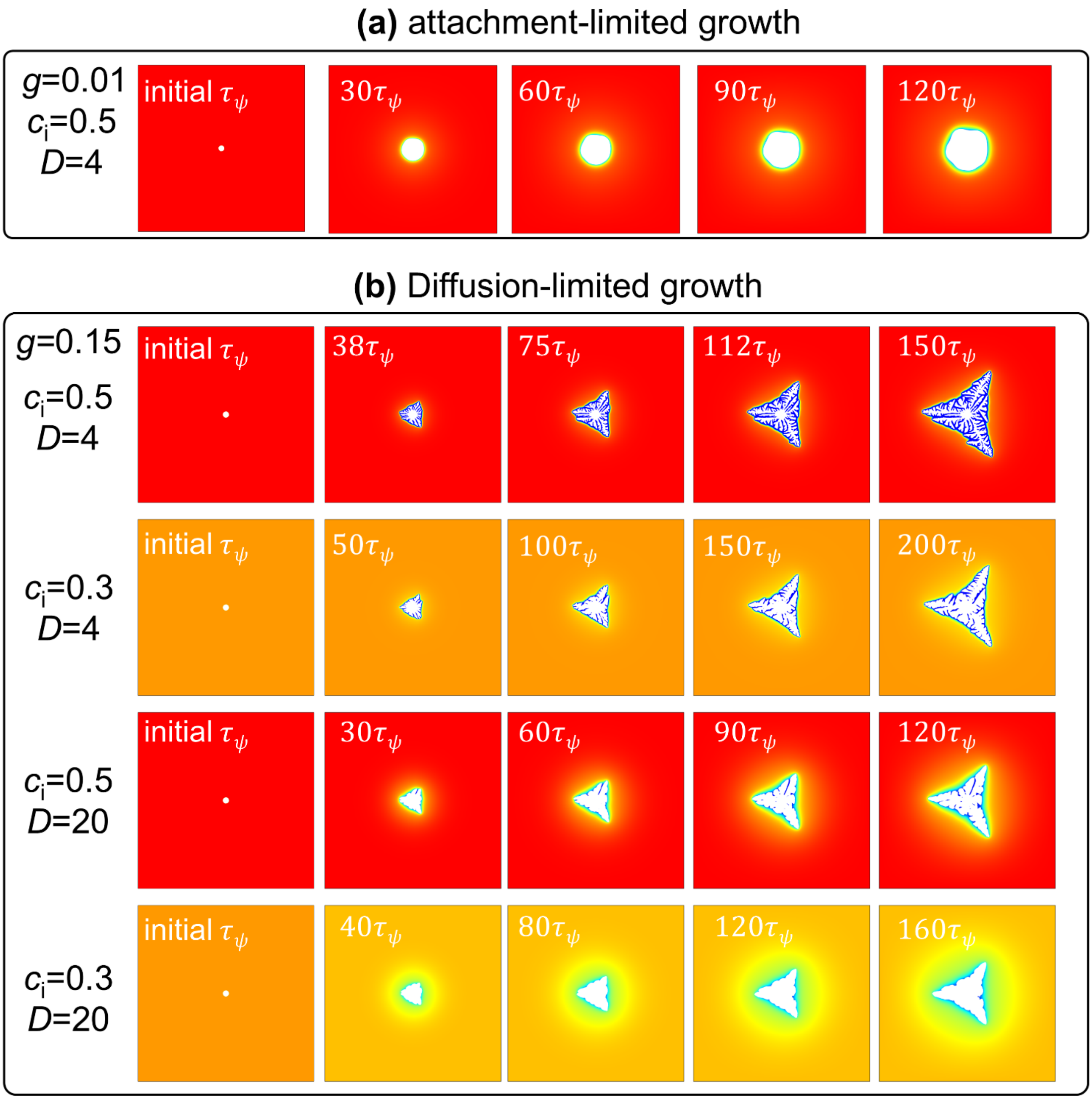

**Fig. S27.** Phase-field simulations of $ReSe_2$ concentration evolution. The side length of the simulation domain is 600 $w_0$.

## Supplementary tables

**Table S1.** Literature summary on AI applications in material synthesis.
AI applications are categorized into parameter optimization, customized synthesis, and mechanism deciphering.

| Ref. | Parameter optimization | Customized synthesis | Mechanism deciphering |
|---|---|---|---|
| Machine learning with knowledge constraints for process optimization of open-air perovskite solar cell manufacturing. Joule 2022;6:834–49. | √ | × | × |
| Machine intelligence-accelerated discovery of all-natural plastic substitutes. Nat Nanotechnol 2024;19:782–91. | × | √ | √ |
| Active learning guides discovery of a champion four-metal perovskite oxide for oxygen evolution electrocatalysis. Nat Mater 2024;23:108–15. | √ | × | × |
| Active learning streamlines development of high performance catalysts for higher alcohol synthesis. Nat Commun 2024;15:5844. | √ | × | × |
| Transforming the synthesis of carbon nanotubes with machine learning models and automation. Matter 2025;8:101913. | × | √ | × |
| Machine-learning design of ductile FeNiCoAlTa alloys with high strength. Nature 2025;643:119–24. | √ | × | × |
| Cross-material catalyst discovery via deep learning. Nat Mater 2026. | × | √ | √ |
| Autonomous closed-loop framework for reproducible perovskite solar cells. Nature 2026;653:707–14. | × | √ | √ |

Efficient process optimization, on-demand customized synthesis and multi-parameter mechanistic interpretation are three core goals of material synthesis with progressive logic. The first seeks optimal parameters for specific targets, the second captures global parameter-structure/property correlations, and the third uncovers the underlying physical nature behind these correlations.

However, most current studies only apply AI to one or two of these goals (Table S1). Such partial applications fail to fully unleash the potential of AI and cannot deliver a thorough paradigm shift for material synthesis. Against this background, our work constructs an all-chain AI-enabled unified framework covering all three objectives, to fully demonstrate how AI can comprehensively empower the entire workflow of material synthesis.

The three modules are closely connected and mutually supportive, which fits well with data-scarce scenarios. Active learning first accomplishes fast parameter optimization for specific targets. Based on the obtained dataset, merely a small number of additional samples are needed to fit global correlations and achieve on-demand synthesis, which greatly lowers experimental costs. Finally, data accumulated from process optimization and on-demand synthesis supports interpretable machine learning, which quantifies the importance and independence of each process parameter and help establish an accurate and reliable growth mechanism model combined with material characterization and physical-chemical principles. This constructed growth model also delivers domain knowledge to facilitate parameter selection and global correlation fitting for future similar synthesis tasks.

**Table S2**. The value range of $T_{Re}$ determined by the previous experiments. Scale bars: 5 μm.

| Exp. No. | $T_{Re}$ (°C) | $T_{Se}$ (°C) | $c_{Re}$ (mol/L) | $f_{H2}$ (sccm) | *sub.* | SEM | Raman | Grow or not |
|---|---|---|---|---|---|---|---|---|
| **1** | 560 | 250 | 0.025 | 1 | *c*-$Al_2O_3$ | 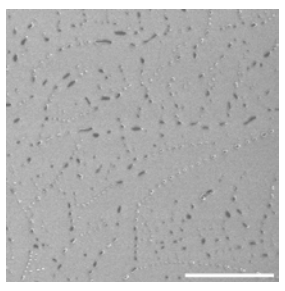 | 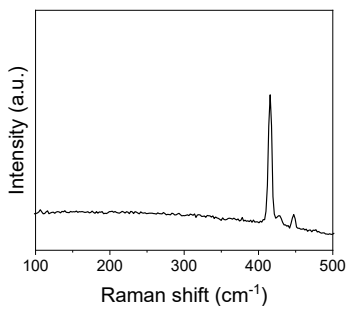 | No |
| **2** | 560 | 230 | 0.10 | 3 | MgO | 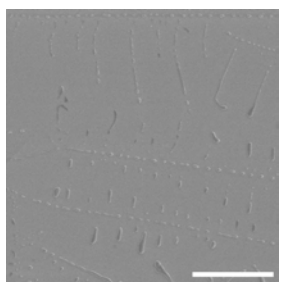 | 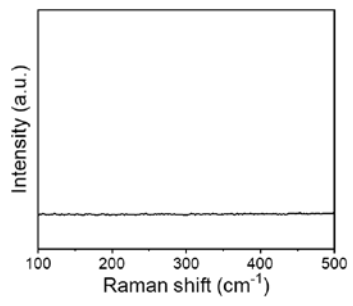 | No |
| **3** | 580 | 250 | 0.025 | 1 | c-$Al_2O_3$ | 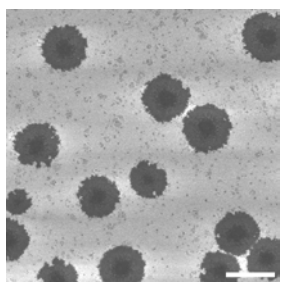 | 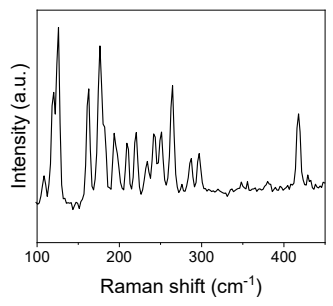 | Yes |
| **4** | 580 | 270 | 0.125 | 2 | MgO | 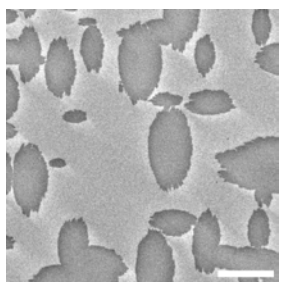 | 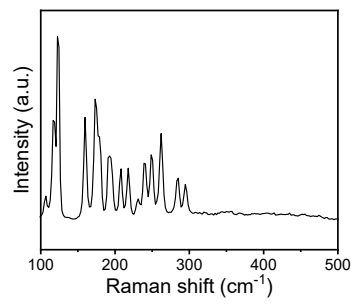 | Yes |
| **5** | 620 | 250 | 0.025 | 1 | *c*-$Al_2O_3$ | 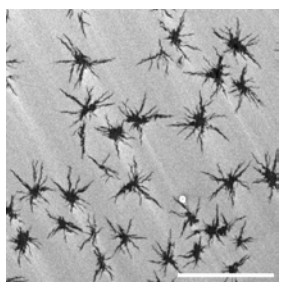 | 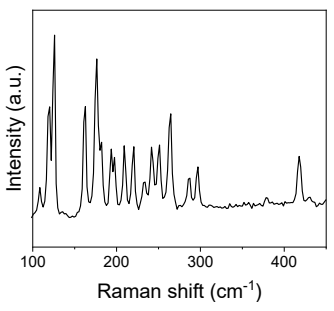 | Yes |
| **6** | 680 | 270 | 0.050 | 4 | MgO | 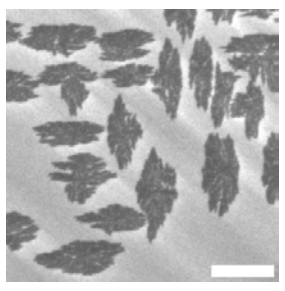 | 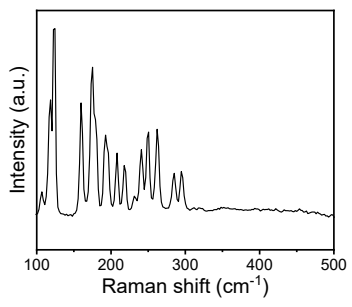 | Yes |
| **7** | 700 | 290 | 0.050 | 2 | *c*-$Al_2O_3$ | 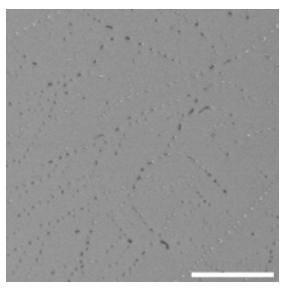 | 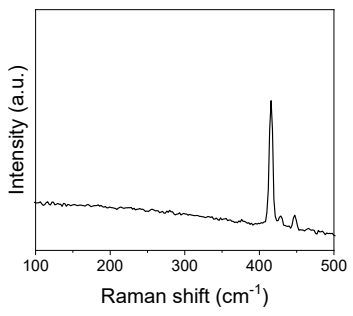 | No |

We conducted a series of binary classification experiments (can grow *vs.* cannot grow) to determine the range of $T_{Re}$. SEM and Raman spectroscopy were applied to identify whether 2D $ReSe_2$ was obtained under each condition. Table S2 shows some typical data. It displayed that, when $T_{Re}$ is between 580 and 680°C, $ReSe_2$ can be grown in many conditions. Therefore, the range of $T_{Re}$ is determined to be between 580 and 680°C.

**Table S3**. The value range of $f_{H2}$ determined by the previous experiments. Scale bars: 5 μm.

| Exp. No. | $T_{Re}$ (°C) | $T_{Se}$ (°C) | $c_{Re}$ (mol/L) | $f_{H2}$ (sccm) | *sub.* | SEM | Raman | Grow or not |
|---|---|---|---|---|---|---|---|---|
| **1** | 630 | 250 | 0.025 | 0 | MgO | 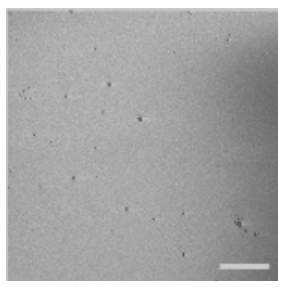 | 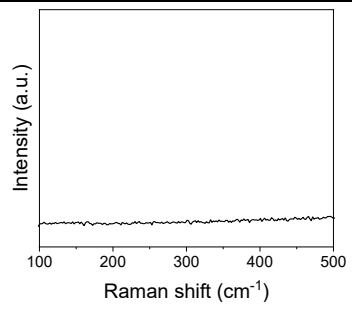  | No |
| **2** | 600 | 270 | 0.10 | 0 | $c$-$Al_2O_3$ | 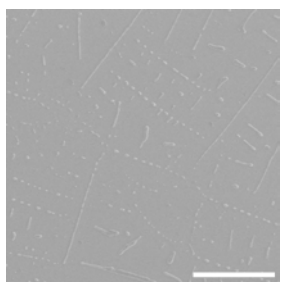 | 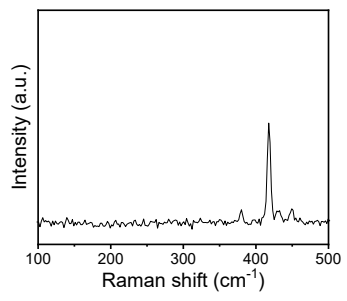  | No |
| **3** | 630 | 250 | 0.025 | 1 | MgO | 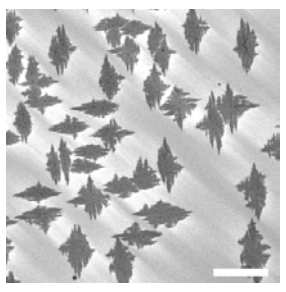 | 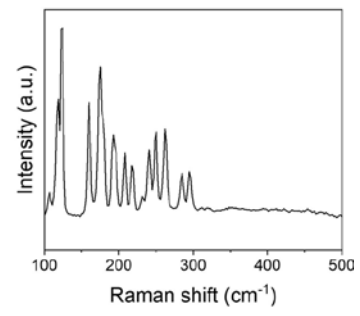  | Yes |
| **4** | 640 | 270 | 0.10 | 2 | $c$-$Al_2O_3$ | 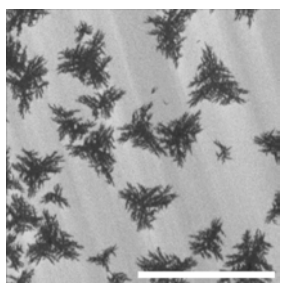 | 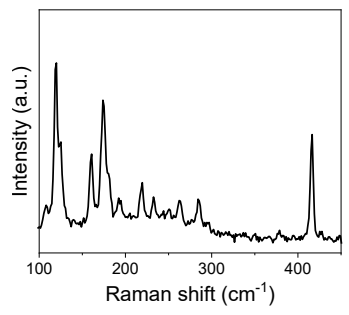  | Yes |
| **5** | 630 | 250 | 0.025 | 2 | MgO | 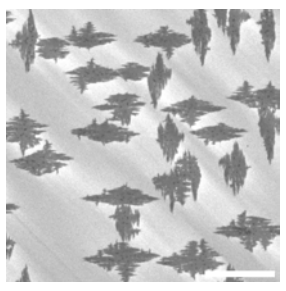 | 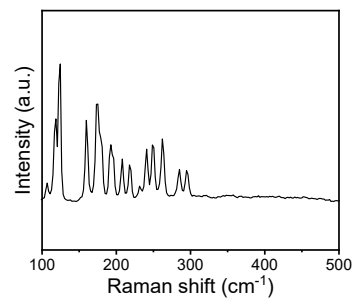  | Yes |
| **6** | 580 | 250 | 0.050 | 4 | $c$-$Al_2O_3$ | 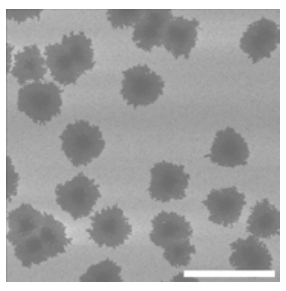 | 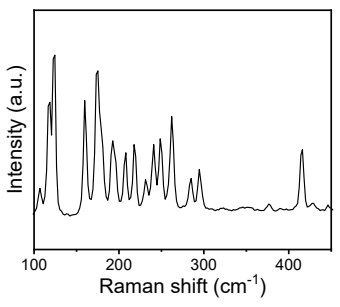  | Yes |
| **7** | 600 | 230 | 0.10 | 5 | MgO | 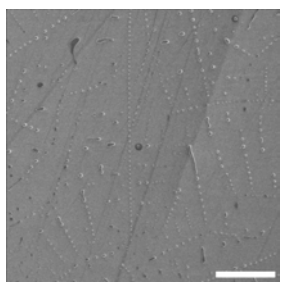 | 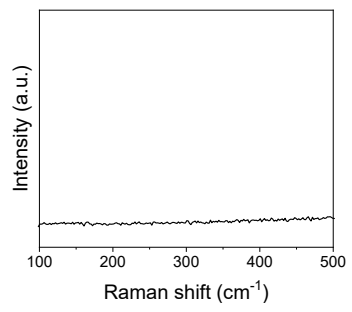  | No |

A similar method was applied to determine the range of $f_{H2}$. Table S3 shows that, when $f_{H2}$ is above 4 sccm or below 1 sccm, $ReSe_2$ cannot be achieved in many conditions. Therefore, the range of $f_{H2}$ is determined to be between 1 and 4 sccm. The ranges of the other three CVD parameters, $T_{Se}$, $c_{Re}$ and *sub.*, were determined by the same approach.

**Table S4**. The value range of CVD process parameters for active learning-guided CVD parameter optimization.

| $x_i$ | **Value range** | **Value interval** |
|---|---|---|
| $T_{Re}$ | 580-680°C | 10°C |
| $T_{Se}$ | 220-300°C | 10°C |
| $c_{Re}$ | 0.025-0.15 mol/L | 0.025 mol/L |
| $f_{H2}$ | 1-4 sccm | 1 sccm |
| *sub.* | c-$Al_2O_3$、MgO | \ |

Considering the experimental parameter range and appropriate step size for 2D $ReSe_2$ growth, the manual trial-and-error approach would require 4752 sets of experiments (11×9×4×6×2). With a daily throughput of 8-10 experiments, this would take approximately 1.5 years to complete. In stark contrast, the BO method enabled us to achieve process optimization with only 60 experiments (including the initial dataset), representing a ~98% reduction in experimental workload.

**Table S5.** Summary of CVD parameters from literature on 2D dendrite growth.

| Ref. | Materials | Morphology control objective | Key CVD parameters |
|---|---|---|---|
| Single-step chemical vapor deposition growth of platinum nanocrystal: monolayer $MoS_2$ dendrite hybrid materials for efficient electrocatalysis. Chem Mater 2020;32:8243–56. | $MoS_2$ | Dendrite | Precursor concentration |
| Synthesis of $MoS_2$ ribbons and their branched structures by chemical vapor deposition in sulfur-enriched environment. Appl Surf Sci 2017;409:396–402. | $MoS_2$ | Dendrite | Precursor concentration |
| Substrate effect on the growth of monolayer dendritic $MoS_2$ on $LaAlO_3$ (100) and its electrocatalytic applications. 2D Mater 2016;3:035001. | $MoS_2$ | Dendrite | Substrate |
| Dendritic, transferable, strictly monolayer $MoS_2$ flakes synthesized on $SrTiO_3$ single crystals for efficient electrocatalytic applications. ACS Nano 2014;8:8617–24. | $MoS_2$ | Dendrite | Growth temperature, substrate |
| Controllable synthesis of large-scale monolayer $MoS_2$ dendritic flakes with serrated edges and their multimodal microscopy and AFM characterizations. J Phys Chem C 2022;126:13449–57. | $MoS_2$ | Dendrite | Precursor concentration, growth temperature, Ar flow rate |
| Effect of substrate symmetry on the dendrite morphology of $MoS_2$ Film synthesized by CVD. Sci Rep 2017;7:15166. | $MoS_2$ | Dendrite | Substrate |
| Monolayer $MoS_2$ dendrites on a symmetry-disparate $SrTiO_3$ (001) substrate: formation mechanism and interface interaction. Adv Funct Mater 2016;26:3299–3305. | $MoS_2$ | Dendrite | Precursor concentration, Substrate |
| Fractal growth of 2D $NbSe_2$ for broadband nonlinear optical limiting. Adv Funct Mater 2024;34:2401490. | $NbSe_2$ | Dendrite | Precursor concentration, growth temperature, $H_2$ flow rate |
| Two-dimensional to three-dimensional growth of transition metal diselenides by chemical vapor deposition: interplay between fractal, dendritic, and compact morphologies. ACS Appl Mater Interfaces 2020;12:15885–92. | $MoSe_2$ | Dendrite | Precursor concentration, growth temperature |
| Transformation from dendritic to triangular growth of $WS_2$ via NaCl assisted low-pressure chemical vapor deposition. J Vac Sci Technol A 2024;42:042203. | $WS_2$ | Dendrite | Amount of NaCl |
| Controlled fractal growth of transition metal dichalcogenides. Nanoscale 2019;11:17065–72. | $WS_2$ | Dendrite | Cooling rate, Growth time |

It is worth noting that the "knowledge" defined in our data-knowledge dual driven ML work falls into three categories: (i) literature and preliminary experimental experience for CVD parameter selection, (ii) multiscale characterization results, and (iii) fundamental thermodynamic and kinetic growth principles.

First, prior knowledge from literature and pre-experimental material experience guides us to screen out five meaningful CVD parameters before constructing the active learning model, which eliminates countless irrelevant variables and drastically reduces experimental costs (Table S5).

Second, multiscale material characterizations and the fundamental thermodynamic and kinetic growth principles serve as important domain knowledge (which are discussed in detail in Fig. 1d-g and Fig. 2 in the main text). They tell us that polycrystalline crystals grow into isotropic round domains without substrate symmetry correlation below 600°C, while oriented dendrites linked to substrate symmetry emerge above 600°C. Such distinct morphological differences lay the foundation to distinguish attachment-limited and diffusion-limited growth regimes, which complements quantitative feature contribution results from SHAP analysis. This integrated knowledge constrains our growth model interpretation logic and further help determine the two key simulation parameters (adatom diffusion coefficient D and initial adatom concentration ci) adopted in phase-field modeling.

**Table S6.** Data points ranked by PA score in iteration 1 of data augmentation process.

| No. | $T_{Re}$ (°C) | $T_{Se}$ (°C) | $c_{Re}$ (mol/L) | $f_{H2}$ (sccm) | *sub.* | Exp. $D_F$ | Mean pred. $D_F$ | *PA* score |
|---|---|---|---|---|---|---|---|---|
| 1 | 580 | 250 | 0.025 | 2 | *c*-$Al_2O_3$ | 1.24 | 1.40 | 0.0295 |
| 2 | 600 | 290 | 0.1 | 3 | *c*-$Al_2O_3$ | 1.57 | 1.45 | 0.0188 |
| 3 | 580 | 230 | 0.125 | 3 | *c*-$Al_2O_3$ | 1.27 | 1.38 | 0.0170 |
| 4 | 680 | 290 | 0.15 | 4 | MgO | 1.16 | 1.28 | 0.0149 |
| 5 | 600 | 230 | 0.05 | 4 | *c*-$Al_2O_3$ | 1.52 | 1.47 | 0.0110 |
| 6 | 660 | 250 | 0.05 | 1 | MgO | 1.26 | 1.36 | 0.0110 |
| 7 | 580 | 250 | 0.15 | 2 | MgO | 1.19 | 1.28 | 0.0082 |
| 8 | 620 | 220 | 0.15 | 1 | *c*-$Al_2O_3$ | 1.4 | 1.48 | 0.0073 |
| 9 | 660 | 270 | 0.1 | 1 | MgO | 1.28 | 1.36 | 0.0063 |
| 10 | 640 | 290 | 0.15 | 1 | MgO | 1.4 | 1.36 | 0.0058 |
| 11 | 620 | 250 | 0.075 | 2 | *c*-$Al_2O_3$ | 1.7 | 1.63 | 0.0049 |
| 12 | 640 | 230 | 0.125 | 1 | *c*-$Al_2O_3$ | 1.55 | 1.49 | 0.0045 |
| 13 | 620 | 250 | 0.1 | 2 | *c*-$Al_2O_3$ | 1.71 | 1.65 | 0.0044 |
| 14 | 580 | 290 | 0.025 | 3 | *c*-$Al_2O_3$ | 1.32 | 1.36 | 0.0042 |
| 15 | 630 | 230 | 0.025 | 4 | *c*-$Al_2O_3$ | 1.5 | 1.56 | 0.0041 |
| 16 | 620 | 230 | 0.125 | 4 | MgO | 1.5 | 1.45 | 0.0038 |
| 17 | 680 | 250 | 0.025 | 4 | MgO | 1.38 | 1.32 | 0.0038 |
| 18 | 610 | 250 | 0.125 | 1 | *c*-$Al_2O_3$ | 1.59 | 1.53 | 0.0032 |
| 19 | 620 | 260 | 0.1 | 3 | *c*-$Al_2O_3$ | 1.58 | 1.63 | 0.0031 |
| 20 | 660 | 270 | 0.15 | 3 | *c*-$Al_2O_3$ | 1.3 | 1.26 | 0.0028 |
| 21 | 620 | 280 | 0.075 | 1 | *c*-$Al_2O_3$ | 1.56 | 1.61 | 0.0026 |
| 22 | 620 | 290 | 0.025 | 4 | *c*-$Al_2O_3$ | 1.47 | 1.52 | 0.0025 |
| 23 | 620 | 280 | 0.05 | 1 | *c*-$Al_2O_3$ | 1.56 | 1.51 | 0.0023 |
| 24 | 680 | 270 | 0.125 | 4 | *c*-$Al_2O_3$ | 1.28 | 1.32 | 0.0023 |
| 25 | 620 | 260 | 0.075 | 2 | *c*-$Al_2O_3$ | 1.59 | 1.63 | 0.0019 |
| 26 | 630 | 280 | 0.05 | 4 | *c*-$Al_2O_3$ | 1.48 | 1.52 | 0.0018 |
| 27 | 620 | 250 | 0.025 | 3 | MgO | 1.56 | 1.52 | 0.0017 |
| 28 | 620 | 280 | 0.025 | 4 | *c*-$Al_2O_3$ | 1.54 | 1.50 | 0.0016 |
| 29 | 610 | 220 | 0.15 | 1 | *c*-$Al_2O_3$ | 1.43 | 1.44 | 0.0016 |
| 30 | 660 | 270 | 0.1 | 2 | MgO | 1.34 | 1.34 | 0.0015 |
| 31 | 620 | 220 | 0.125 | 1 | *c*-$Al_2O_3$ | 1.49 | 1.52 | 0.0015 |
| 32 | 620 | 300 | 0.1 | 1 | *c*-$Al_2O_3$ | 1.62 | 1.59 | 0.0012 |
| 33 | 620 | 220 | 0.05 | 4 | *c*-$Al_2O_3$ | 1.57 | 1.54 | 0.0010 |
| 34 | 630 | 260 | 0.025 | 4 | *c*-$Al_2O_3$ | 1.52 | 1.49 | 0.0009 |
| 35 | 610 | 230 | 0.025 | 1 | *c*-$Al_2O_3$ | 1.53 | 1.56 | 0.0008 |
| 36 | 610 | 280 | 0.1 | 2 | *c*-$Al_2O_3$ | 1.58 | 1.61 | 0.0008 |
| 37 | 610 | 260 | 0.1 | 2 | *c*-$Al_2O_3$ | 1.61 | 1.60 | 0.0008 |
| 38 | 620 | 260 | 0.075 | 3 | *c*-$Al_2O_3$ | 1.63 | 1.60 | 0.0008 |

(continued on next page)

(continued from previous page)

| No. | $T_{Re}$ (°C) | $T_{Se}$ (°C) | $c_{Re}$ (mol/L) | $f_{H2}$ (sccm) | *sub.* | Exp. $D_F$ | Mean pred. $D_F$ | *PA* score |
|---|---|---|---|---|---|---|---|---|
| 39 | 620 | 230 | 0.05 | 2 | MgO | 1.53 | 1.53 | 0.0008 |
| 40 | 620 | 300 | 0.025 | 4 | *c*-$Al_2O_3$ | 1.49 | 1.51 | 0.0007 |
| 41 | 620 | 230 | 0.1 | 1 | *c*-$Al_2O_3$ | 1.67 | 1.65 | 0.0007 |
| 42 | 620 | 220 | 0.075 | 1 | *c*-$Al_2O_3$ | 1.6 | 1.58 | 0.0007 |
| 43 | 620 | 290 | 0.075 | 2 | *c*-$Al_2O_3$ | 1.58 | 1.60 | 0.0006 |
| 44 | 620 | 230 | 0.075 | 1 | *c*-$Al_2O_3$ | 1.63 | 1.65 | 0.0006 |
| 45 | 610 | 270 | 0.1 | 3 | *c*-$Al_2O_3$ | 1.57 | 1.59 | 0.0004 |
| 46 | 620 | 270 | 0.075 | 2 | *c*-$Al_2O_3$ | 1.62 | 1.60 | 0.0004 |
| 47 | 620 | 230 | 0.025 | 4 | *c*-$Al_2O_3$ | 1.59 | 1.57 | 0.0004 |
| 48 | 610 | 290 | 0.075 | 1 | *c*-$Al_2O_3$ | 1.6 | 1.58 | 0.0004 |
| 49 | 620 | 270 | 0.1 | 2 | *c*-$Al_2O_3$ | 1.62 | 1.60 | 0.0004 |
| 50 | 610 | 290 | 0.1 | 2 | *c*-$Al_2O_3$ | 1.6 | 1.59 | 0.0004 |
| 51 | 620 | 270 | 0.1 | 1 | *c*-$Al_2O_3$ | 1.59 | 1.61 | 0.0003 |
| 52 | 620 | 230 | 0.025 | 1 | *c*-$Al_2O_3$ | 1.56 | 1.55 | 0.0003 |
| 53 | 620 | 290 | 0.1 | 2 | *c*-$Al_2O_3$ | 1.59 | 1.61 | 0.0003 |
| 54 | 620 | 280 | 0.1 | 2 | *c*-$Al_2O_3$ | 1.61 | 1.60 | 0.0003 |
| 55 | 620 | 220 | 0.025 | 1 | *c*-$Al_2O_3$ | 1.5 | 1.51 | 0.0003 |
| 56 | 630 | 220 | 0.025 | 4 | *c*-$Al_2O_3$ | 1.53 | 1.52 | 0.0002 |
| 57 | 640 | 290 | 0.05 | 1 | *c*-$Al_2O_3$ | 1.47 | 1.47 | 0.0002 |
| 58 | 620 | 290 | 0.1 | 1 | *c*-$Al_2O_3$ | 1.59 | 1.59 | 0.0001 |
| 59 | 630 | 240 | 0.025 | 4 | *c*-$Al_2O_3$ | 1.55 | 1.55 | 0.0001 |
| 60 | 610 | 270 | 0.075 | 2 | *c*-$Al_2O_3$ | 1.6 | 1.59 | 0.0001 |

Three data points with the top three *PA* scores among the 60 pieces of data are highlighted in blue.

**Table S7.** Supplementary data points in iteration 1 of data augmentation process.

| No. | $T_{Re}$ (°C) | $T_{Se}$ (°C) | $c_{Re}$ (mol/L) | $f_{H2}$ (sccm) | *sub.* | Exp. $D_F$ |
|---|---|---|---|---|---|---|
| 1 | 580 | 250 | 0.05 | 3 | *c*-$Al_2O_3$ | 1.24 |
| 2 | 600 | 270 | 0.1 | 3 | *c*-$Al_2O_3$ | 1.61 |
| 3 | 580 | 250 | 0.1 | 3 | *c*-$Al_2O_3$ | 1.44 |

We added one data point in the vicinity of each of the top three data points in Table S6 based on their *PA* scores, as shown in Table S7. Here, we define "vicinity" as follows: 1) The change in $T_{Re}$, which is the feature that impacts $D_F$ the most (verified by SHAP analysis in Fig. S4), is less than 5%. 2) The change in all other input features is less than 2 units of the interval specified for this variable in Table S3. This standard has been maintained throughout the three rounds of data supplementation.

**Table S8.** Data points ranked by *PA* score in iteration 2 of data augmentation process.

| No. | $T_{Re}$ (°C) | $T_{Se}$ (°C) | $c_{Re}$ (mol/L) | $f_{H2}$ (sccm) | *sub.* | Exp. $D_F$ | Mean pred. $D_F$ | *PA* score |
|---|---|---|---|---|---|---|---|---|
| 1 | 580 | 250 | 0.15 | 2 | MgO | 1.19 | 1.31 | 0.0172 |
| 2 | 580 | 230 | 0.125 | 3 | *c*-$Al_2O_3$ | 1.27 | 1.37 | 0.0130 |
| 3 | 680 | 290 | 0.15 | 4 | MgO | 1.16 | 1.27 | 0.0129 |
| 4 | 580 | 250 | 0.05 | 3 | *c*-$Al_2O_3$ | 1.24 | 1.34 | 0.0110 |
| 5 | 580 | 250 | 0.1 | 3 | *c*-$Al_2O_3$ | 1.44 | 1.37 | 0.0093 |
| 6 | 580 | 250 | 0.025 | 2 | *c*-$Al_2O_3$ | 1.24 | 1.34 | 0.0092 |
| 7 | 640 | 230 | 0.125 | 1 | *c*-$Al_2O_3$ | 1.55 | 1.47 | 0.0075 |
| 8 | 680 | 250 | 0.025 | 4 | MgO | 1.38 | 1.30 | 0.0069 |
| 9 | 660 | 250 | 0.05 | 1 | MgO | 1.26 | 1.34 | 0.0068 |
| 10 | 610 | 250 | 0.125 | 1 | *c*-$Al_2O_3$ | 1.59 | 1.52 | 0.0060 |
| 11 | 660 | 270 | 0.1 | 1 | MgO | 1.28 | 1.35 | 0.0051 |
| 12 | 620 | 220 | 0.15 | 1 | *c*-$Al_2O_3$ | 1.4 | 1.47 | 0.0050 |
| 13 | 630 | 230 | 0.025 | 4 | *c*-$Al_2O_3$ | 1.5 | 1.56 | 0.0036 |
| 14 | 620 | 280 | 0.05 | 1 | *c*-$Al_2O_3$ | 1.56 | 1.50 | 0.0033 |
| 15 | 620 | 250 | 0.025 | 3 | MgO | 1.56 | 1.51 | 0.0032 |
| 16 | 620 | 260 | 0.1 | 3 | *c*-$Al_2O_3$ | 1.58 | 1.63 | 0.0032 |
| 17 | 680 | 270 | 0.125 | 4 | *c*-$Al_2O_3$ | 1.28 | 1.33 | 0.0027 |
| 18 | 620 | 260 | 0.075 | 2 | *c*-$Al_2O_3$ | 1.59 | 1.64 | 0.0022 |
| 19 | 620 | 220 | 0.05 | 4 | *c*-$Al_2O_3$ | 1.57 | 1.52 | 0.0022 |
| 20 | 620 | 290 | 0.025 | 4 | *c*-$Al_2O_3$ | 1.47 | 1.52 | 0.0021 |
| 21 | 620 | 280 | 0.075 | 1 | *c*-$Al_2O_3$ | 1.56 | 1.60 | 0.0021 |
| 22 | 580 | 290 | 0.025 | 3 | *c*-$Al_2O_3$ | 1.32 | 1.31 | 0.0021 |
| 23 | 620 | 250 | 0.075 | 2 | *c*-$Al_2O_3$ | 1.7 | 1.66 | 0.0021 |
| 24 | 620 | 250 | 0.1 | 2 | *c*-$Al_2O_3$ | 1.71 | 1.67 | 0.0020 |
| 25 | 600 | 270 | 0.1 | 3 | *c*-$Al_2O_3$ | 1.61 | 1.57 | 0.0020 |
| 26 | 620 | 230 | 0.125 | 4 | MgO | 1.5 | 1.46 | 0.0018 |
| 27 | 610 | 270 | 0.1 | 3 | *c*-$Al_2O_3$ | 1.57 | 1.61 | 0.0017 |
| 28 | 620 | 230 | 0.1 | 1 | *c*-$Al_2O_3$ | 1.67 | 1.63 | 0.0016 |
| 29 | 620 | 260 | 0.075 | 3 | *c*-$Al_2O_3$ | 1.63 | 1.59 | 0.0015 |
| 30 | 620 | 220 | 0.125 | 1 | *c*-$Al_2O_3$ | 1.49 | 1.53 | 0.0014 |
| 31 | 640 | 290 | 0.15 | 1 | MgO | 1.4 | 1.40 | 0.0011 |
| 32 | 660 | 270 | 0.1 | 2 | MgO | 1.34 | 1.31 | 0.0011 |
| 33 | 620 | 300 | 0.1 | 1 | *c*-$Al_2O_3$ | 1.62 | 1.59 | 0.0011 |
| 34 | 620 | 230 | 0.025 | 4 | *c*-$Al_2O_3$ | 1.59 | 1.56 | 0.0010 |
| 35 | 620 | 220 | 0.075 | 1 | *c*-$Al_2O_3$ | 1.6 | 1.57 | 0.0010 |
| 36 | 610 | 220 | 0.15 | 1 | *c*-$Al_2O_3$ | 1.43 | 1.44 | 0.0009 |
| 37 | 630 | 260 | 0.025 | 4 | *c*-$Al_2O_3$ | 1.52 | 1.49 | 0.0008 |
| 38 | 620 | 230 | 0.075 | 1 | *c*-$Al_2O_3$ | 1.63 | 1.64 | 0.0008 |

(continued on next page)

(continued from previous page)

| No. | $T_{Re}$ (°C) | $T_{Se}$ (°C) | $c_{Re}$ (mol/L) | $f_{H2}$ (sccm) | *sub.* | Exp. $D_F$ | Mean pred. $D_F$ | *PA* score |
|---|---|---|---|---|---|---|---|---|
| 39 | 620 | 280 | 0.025 | 4 | *c*-$Al_2O_3$ | 1.54 | 1.51 | 0.0008 |
| 40 | 600 | 230 | 0.05 | 4 | *c*-$Al_2O_3$ | 1.52 | 1.54 | 0.0007 |
| 41 | 630 | 280 | 0.05 | 4 | *c*-$Al_2O_3$ | 1.48 | 1.50 | 0.0006 |
| 42 | 620 | 270 | 0.1 | 1 | *c*-$Al_2O_3$ | 1.59 | 1.61 | 0.0006 |
| 43 | 600 | 290 | 0.1 | 3 | *c*-$Al_2O_3$ | 1.57 | 1.56 | 0.0006 |
| 44 | 610 | 280 | 0.1 | 2 | *c*-$Al_2O_3$ | 1.58 | 1.60 | 0.0006 |
| 45 | 620 | 300 | 0.025 | 4 | *c*-$Al_2O_3$ | 1.49 | 1.51 | 0.0005 |
| 46 | 620 | 290 | 0.075 | 2 | *c*-$Al_2O_3$ | 1.58 | 1.60 | 0.0005 |
| 47 | 660 | 270 | 0.15 | 3 | *c*-$Al_2O_3$ | 1.3 | 1.28 | 0.0005 |
| 48 | 640 | 290 | 0.05 | 1 | *c*-$Al_2O_3$ | 1.47 | 1.47 | 0.0004 |
| 49 | 620 | 270 | 0.075 | 2 | *c*-$Al_2O_3$ | 1.62 | 1.60 | 0.0004 |
| 50 | 620 | 290 | 0.1 | 2 | *c*-$Al_2O_3$ | 1.59 | 1.61 | 0.0004 |
| 51 | 620 | 220 | 0.025 | 1 | *c*-$Al_2O_3$ | 1.5 | 1.51 | 0.0003 |
| 52 | 620 | 230 | 0.05 | 2 | MgO | 1.53 | 1.52 | 0.0003 |
| 53 | 620 | 290 | 0.1 | 1 | *c*-$Al_2O_3$ | 1.59 | 1.61 | 0.0003 |
| 54 | 610 | 260 | 0.1 | 2 | *c*-$Al_2O_3$ | 1.61 | 1.62 | 0.0003 |
| 55 | 630 | 220 | 0.025 | 4 | *c*-$Al_2O_3$ | 1.53 | 1.52 | 0.0002 |
| 56 | 620 | 280 | 0.1 | 2 | *c*-$Al_2O_3$ | 1.61 | 1.60 | 0.0002 |
| 57 | 610 | 290 | 0.075 | 1 | *c*-$Al_2O_3$ | 1.6 | 1.59 | 0.0002 |
| 58 | 630 | 240 | 0.025 | 4 | *c*-$Al_2O_3$ | 1.55 | 1.54 | 0.0002 |
| 59 | 610 | 230 | 0.025 | 1 | *c*-$Al_2O_3$ | 1.53 | 1.54 | 0.0002 |
| 60 | 620 | 230 | 0.025 | 1 | *c*-$Al_2O_3$ | 1.56 | 1.56 | 0.0001 |
| 61 | 610 | 290 | 0.1 | 2 | *c*-$Al_2O_3$ | 1.6 | 1.6 | 0.0001 |
| 62 | 610 | 270 | 0.075 | 2 | *c*-$Al_2O_3$ | 1.6 | 1.6 | 0.0001 |
| 63 | 620 | 270 | 0.1 | 2 | *c*-$Al_2O_3$ | 1.62 | 1.62 | 0.000002 |

Three data points with the top three *PA* scores among the 63 pieces of data are highlighted in blue.

**Table S9.** Supplementary data points in iteration 2 of data augmentation process.

| No. | $T_{Re}$ (°C) | $T_{Se}$ (°C) | $c_{Re}$ (mol/L) | $f_{H2}$ (sccm) | *sub.* | Exp. $D_F$ |
|---|---|---|---|---|---|---|
| 1 | 600 | 250 | 0.15 | 2 | MgO | 1.45 |
| 2 | 580 | 250 | 0.125 | 2 | MgO | 1.3 |
| 3 | 660 | 250 | 0.1 | 2 | *c*-$Al_2O_3$ | 1.48 |

Here are the three data points added in the vicinity of the top three data points among the 63 pieces of data in Table S8.

**Table S10.** Data points ranked by *PA* score in iteration 3 of data augmentation process.

| No. | $T_{Re}$ (°C) | $T_{Se}$ (°C) | $c_{Re}$ (mol/L) | $f_{H2}$ (sccm) | *sub.* | Exp. $D_F$ | Mean pred. $D_F$ | *PA* score |
|---|---|---|---|---|---|---|---|---|
| 1 | 680 | 290 | 0.15 | 4 | MgO | 1.16 | 1.29 | 0.0171 |
| 2 | 580 | 250 | 0.05 | 3 | $c$-$Al_2O_3$ | 1.24 | 1.35 | 0.0123 |
| 3 | 580 | 250 | 0.025 | 2 | $c$-$Al_2O_3$ | 1.24 | 1.35 | 0.0122 |
| 4 | 660 | 250 | 0.05 | 1 | MgO | 1.26 | 1.36 | 0.0109 |
| 5 | 580 | 250 | 0.15 | 2 | MgO | 1.19 | 1.29 | 0.0097 |
| 6 | 660 | 270 | 0.1 | 1 | MgO | 1.28 | 1.37 | 0.0097 |
| 7 | 640 | 230 | 0.125 | 1 | $c$-$Al_2O_3$ | 1.55 | 1.48 | 0.0056 |
| 8 | 680 | 270 | 0.125 | 4 | $c$-$Al_2O_3$ | 1.28 | 1.34 | 0.0049 |
| 9 | 580 | 250 | 0.1 | 3 | $c$-$Al_2O_3$ | 1.44 | 1.38 | 0.0045 |
| 10 | 580 | 230 | 0.125 | 3 | $c$-$Al_2O_3$ | 1.27 | 1.33 | 0.0041 |
| 11 | 610 | 250 | 0.125 | 1 | $c$-$Al_2O_3$ | 1.59 | 1.53 | 0.0040 |
| 12 | 680 | 250 | 0.025 | 4 | MgO | 1.38 | 1.33 | 0.0039 |
| 13 | 630 | 230 | 0.025 | 4 | $c$-$Al_2O_3$ | 1.5 | 1.56 | 0.0036 |
| 14 | 620 | 220 | 0.15 | 1 | $c$-$Al_2O_3$ | 1.4 | 1.46 | 0.0034 |
| 15 | 620 | 220 | 0.05 | 4 | $c$-$Al_2O_3$ | 1.57 | 1.52 | 0.0032 |
| 16 | 620 | 250 | 0.025 | 3 | MgO | 1.56 | 1.51 | 0.0031 |
| 17 | 640 | 290 | 0.15 | 1 | MgO | 1.4 | 1.35 | 0.0028 |
| 18 | 660 | 270 | 0.1 | 2 | MgO | 1.34 | 1.35 | 0.0027 |
| 19 | 620 | 260 | 0.1 | 3 | $c$-$Al_2O_3$ | 1.58 | 1.63 | 0.0027 |
| 20 | 620 | 280 | 0.075 | 1 | $c$-$Al_2O_3$ | 1.56 | 1.61 | 0.0025 |
| 21 | 660 | 270 | 0.15 | 3 | $c$-$Al_2O_3$ | 1.3 | 1.30 | 0.0022 |
| 22 | 600 | 270 | 0.1 | 3 | $c$-$Al_2O_3$ | 1.61 | 1.57 | 0.0021 |
| 23 | 600 | 230 | 0.05 | 4 | $c$-$Al_2O_3$ | 1.52 | 1.56 | 0.0019 |
| 24 | 620 | 230 | 0.025 | 4 | $c$-$Al_2O_3$ | 1.59 | 1.55 | 0.0018 |
| 25 | 620 | 260 | 0.075 | 3 | $c$-$Al_2O_3$ | 1.63 | 1.59 | 0.0017 |
| 26 | 620 | 260 | 0.075 | 2 | $c$-$Al_2O_3$ | 1.59 | 1.63 | 0.0017 |
| 27 | 620 | 290 | 0.025 | 4 | $c$-$Al_2O_3$ | 1.47 | 1.51 | 0.0016 |
| 28 | 620 | 250 | 0.075 | 2 | $c$-$Al_2O_3$ | 1.7 | 1.66 | 0.0016 |
| 29 | 620 | 280 | 0.05 | 1 | $c$-$Al_2O_3$ | 1.56 | 1.52 | 0.0016 |
| 30 | 580 | 250 | 0.125 | 2 | MgO | 1.3 | 1.29 | 0.0016 |
| 31 | 610 | 270 | 0.1 | 3 | $c$-$Al_2O_3$ | 1.57 | 1.61 | 0.0015 |
| 32 | 620 | 300 | 0.1 | 1 | $c$-$Al_2O_3$ | 1.62 | 1.59 | 0.0012 |
| 33 | 600 | 250 | 0.15 | 2 | $c$-$Al_2O_3$ | 1.45 | 1.42 | 0.0011 |
| 34 | 620 | 220 | 0.125 | 1 | $c$-$Al_2O_3$ | 1.49 | 1.52 | 0.0010 |
| 35 | 620 | 230 | 0.1 | 1 | $c$-$Al_2O_3$ | 1.67 | 1.64 | 0.0009 |
| 36 | 620 | 250 | 0.1 | 2 | $c$-$Al_2O_3$ | 1.71 | 1.68 | 0.0009 |
| 37 | 610 | 280 | 0.1 | 2 | $c$-$Al_2O_3$ | 1.58 | 1.61 | 0.0009 |
| 38 | 630 | 280 | 0.05 | 4 | $c$-$Al_2O_3$ | 1.48 | 1.50 | 0.0009 |

(continued on next page)

(continued from previous page)

| No. | $T_{Re}$ (℃) | $T_{Se}$ (℃) | $c_{Re}$ (mol/L) | $f_{H2}$ (sccm) | *sub.* | Exp. $D_F$ | Mean pred. $D_F$ | *PA* score |
|---|---|---|---|---|---|---|---|---|
| 39 | 580 | 290 | 0.025 | 3 | $c$-$Al_2O_3$ | 1.32 | 1.30 | 0.0008 |
| 40 | 620 | 280 | 0.025 | 4 | $c$-$Al_2O_3$ | 1.54 | 1.51 | 0.0008 |
| 41 | 620 | 230 | 0.125 | 4 | MgO | 1.5 | 1.47 | 0.0008 |
| 42 | 630 | 260 | 0.025 | 4 | $c$-$Al_2O_3$ | 1.52 | 1.50 | 0.0007 |
| 43 | 630 | 220 | 0.025 | 4 | $c$-$Al_2O_3$ | 1.53 | 1.51 | 0.0007 |
| 44 | 620 | 230 | 0.075 | 1 | $c$-$Al_2O_3$ | 1.63 | 1.66 | 0.0007 |
| 45 | 640 | 290 | 0.05 | 1 | $c$-$Al_2O_3$ | 1.47 | 1.46 | 0.0006 |
| 46 | 610 | 230 | 0.025 | 1 | $c$-$Al_2O_3$ | 1.53 | 1.55 | 0.0006 |
| 47 | 620 | 230 | 0.05 | 2 | MgO | 1.53 | 1.55 | 0.0005 |
| 48 | 620 | 220 | 0.025 | 1 | $c$-$Al_2O_3$ | 1.5 | 1.52 | 0.0005 |
| 49 | 610 | 290 | 0.075 | 1 | $c$-$Al_2O_3$ | 1.6 | 1.58 | 0.0004 |
| 50 | 630 | 240 | 0.025 | 4 | $c$-$Al_2O_3$ | 1.55 | 1.53 | 0.0004 |
| 51 | 620 | 270 | 0.075 | 2 | $c$-$Al_2O_3$ | 1.62 | 1.60 | 0.0003 |
| 52 | 620 | 220 | 0.075 | 1 | $c$-$Al_2O_3$ | 1.6 | 1.58 | 0.0003 |
| 53 | 660 | 250 | 0.1 | 2 | $c$-$Al_2O_3$ | 1.48 | 1.46 | 0.0003 |
| 54 | 620 | 290 | 0.075 | 2 | $c$-$Al_2O_3$ | 1.58 | 1.60 | 0.0003 |
| 55 | 620 | 290 | 0.1 | 2 | $c$-$Al_2O_3$ | 1.59 | 1.61 | 0.0003 |
| 56 | 620 | 290 | 0.1 | 1 | $c$-$Al_2O_3$ | 1.59 | 1.61 | 0.0003 |
| 57 | 600 | 290 | 0.1 | 3 | $c$-$Al_2O_3$ | 1.57 | 1.58 | 0.0002 |
| 58 | 610 | 290 | 0.1 | 2 | $c$-$Al_2O_3$ | 1.6 | 1.59 | 0.0002 |
| 59 | 620 | 270 | 0.1 | 2 | $c$-$Al_2O_3$ | 1.62 | 1.61 | 0.0002 |
| 60 | 620 | 230 | 0.025 | 1 | $c$-$Al_2O_3$ | 1.56 | 1.55 | 0.0002 |
| 61 | 610 | 220 | 0.15 | 1 | $c$-$Al_2O_3$ | 1.43 | 1.43 | 0.0001 |
| 62 | 610 | 260 | 0.1 | 2 | $c$-$Al_2O_3$ | 1.61 | 1.62 | 0.0001 |
| 63 | 610 | 270 | 0.075 | 2 | $c$-$Al_2O_3$ | 1.6 | 1.60 | 0.0001 |
| 64 | 620 | 280 | 0.1 | 2 | $c$-$Al_2O_3$ | 1.61 | 1.60 | 0.0001 |
| 65 | 620 | 300 | 0.025 | 4 | $c$-$Al_2O_3$ | 1.49 | 1.50 | 0.00004 |
| 66 | 620 | 270 | 0.1 | 1 | $c$-$Al_2O_3$ | 1.59 | 1.59 | 0.00002 |

Three data points with the top three *PA* scores among the 66 pieces of data are highlighted in blue. The fourth-ranked data point is marked in orange and is also used for data supplementation due to the high similarity between the second- and third-ranked data points.

**Table S11.** Supplementary data points in iteration 3 of data augmentation process.

| No. | $T_{Re}$ (°C) | $T_{Se}$ (°C) | $c_{Re}$ (mol/L) | $f_{H2}$ (sccm) | *sub.* | Exp. $D_F$ |
|---|---|---|---|---|---|---|
| 1 | 660 | 270 | 0.15 | 2 | MgO | 1.28 |
| 2 | 580 | 250 | 0.05 | 1 | *c*-$Al_2O_3$ | 1.31 |
| 3 | 660 | 250 | 0.025 | 2 | MgO | 1.26 |

It is worth noting that, since the CVD parameters and $D_F$ values of the data points ranked second and third in Table S10 are very similar, we only supplemented one data point near these two data points in iteration 3. Additionally, we supplemented one data point in the vicinity of the fourth-ranked data point (marked in orange in Table S10), as displayed in the last row in Table S11.

**Table S12.** Nine supplementary data points added at once based on the *PA* score ranking of 60 data points in the active learning process.

| No. | $T_{Re}$ (°C) | $T_{Se}$ (°C) | $c_{Re}$ (mol/L) | $f_{H2}$ (sccm) | *sub.* | Exp. $D_F$ |
|---|---|---|---|---|---|---|
| 1 | 580 | 250 | 0.05 | 3 | *c*-$Al_2O_3$ | 1.24 |
| 2 | 600 | 270 | 0.1 | 3 | MgO | 1.4 |
| 3 | 580 | 250 | 0.1 | 3 | *c*-$Al_2O_3$ | 1.44 |
| 4 | 660 | 270 | 0.15 | 2 | MgO | 1.28 |
| 5 | 600 | 250 | 0.05 | 3 | MgO | 1.49 |
| 6 | 660 | 250 | 0.025 | 1 | MgO | 1.24 |
| 7 | 600 | 250 | 0.15 | 2 | MgO | 1.45 |
| 8 | 620 | 250 | 0.15 | 2 | *c*-$Al_2O_3$ | 1.56 |
| 9 | 660 | 250 | 0.1 | 2 | *c*-$Al_2O_3$ | 1.48 |

Here show the 9 data points supplemented in the vicinity of the top 9 data points based on their *PA* scores in Table S6. The fitting result is displayed in Fig. S18, verifying that one-time addition of nine data points yields poorer fitting performance relative to the progressive supplementary sampling scheme.

**Table S13**. CVD growth parameters suggested by the XGBoost model for three typical user-defined $D_F$ and substrates.

| No. | User-defined $D_F$ | User-defined *sub.* | Suggested $T_{Re}$ (°C) | Suggested $T_{Se}$ (°C) | Suggested $c_{Re}$ (mol/L) | Suggested $f_{H2}$ (sccm) | Experimental $D_F$ |
|---|---|---|---|---|---|---|---|
| 1 | 1.28 | *c*-$Al_2O_3$ | 580 | 250 | 0.05 | 0.01 | 1.28 |
| 2 | 1.45 | MgO | 640 | 230 | 0.025 | 0.02 | 1.46 |
| 3 | 1.62 | MgO | 620 | 250 | 0.1 | 0.02 | 1.60 |

**Table S14.** Repeated growth experiments and statistical $D_F$ values for representative optimized recipes.

| $D_F$ | $T_{Re}$ (°C) | $c_{Re}$ (M) | $T_{Se}$ (°C) | $f_{H2}$ (sccm) | *sub.* | Exp. i | Exp. ii | Exp. iii |
|---|---|---|---|---|---|---|---|---|
| Low $D_F$ | 580 | 0.05 | 250 | 1 | *c*-$Al_2O_3$ | 1.22±0.03 | 1.21±0.02 | 1.23±0.02 |
| Middle $D_F$ | 680 | 0.10 | 250 | 2 | *c*-$Al_2O_3$ | 1.50±0.01 | 1.49±0.03 | 1.50±0.03 |
| High $D_F$ | 620 | 0.10 | 250 | 2 | *c*-$Al_2O_3$ | 1.72±0.03 | 1.71±0.03 | 1.69±0.02 |

## Supplementary references

[1] Jiang J, Wang X, Wang D, Chai Y, Yang Y, Ding L, et al. Chirality-transferred epitaxy of circular polarization-sensitive $ReS_2$ monolayer single crystals. Nat Commun 2025;16:7119.

[2] Choi Y, Kim K, Lim SY, Kim J, Park JM, Kim JH, et al. Complete determination of the crystallographic orientation of $ReX_2$ (X = S, Se) by polarized Raman spectroscopy. Nanoscale Horizons 2020;5:308–315.

[3] Zhao H, Wu J, Zhong H, Guo Q, Wang X, Xia F, et al. Interlayer interactions in anisotropic atomically thin rhenium diselenide. Nano Res 2015;8:3651–3661.

[4] Liu Z, Rolston N, Flick AC, Colburn TW, Ren Z, Dauskardt RH, et al. Machine learning with knowledge constraints for process optimization of open-air perovskite solar cell manufacturing. Joule 2022;6:834–849.

[5] Paleyes A, Mahsereci M, Lawrence N. Emukit: A Python toolkit for decision making under uncertainty. In: Proceedings of the 22nd Python in Science Conference 2023; 68–75

[6] Seeger M. Gaussian processes for machine learning. Int J Neural Syst 2004;14:69–106.

[7] Wang Z, Jegelka S. Max-value entropy search for efficient Bayesian optimization. In: Proceedings of the 34th International Conference on Machine Learning. 2017, 5530–5543.

[8] Chen T, Guestrin C. XGBoost: A scalable tree boosting system. In: Proceedings of the 22nd ACM SIGKDD International Conference on Knowledge Discovery and Data Mining. 2016, 785–794.

[9] Lundberg SM, Erion G, Chen H, DeGrave A, Prutkin JM, Nair B, et al. From local explanations to global understanding with explainable AI for trees. Nat Mach Intell 2020;2:56–67.

[10] Muschalik M, Baniecki H, Fumagalli F, Kolpaczki P, Hammer B, Hüllermeier E. shapiq: Shapley Interactions for Machine Learning. In: Proceedings of the 38th International Conference on Neural Information Processing Systems 2024; 4141:130324

[11] Kenneth Falconer. In: Fractal geometry: mathematical foundations and applications. 2nd ed. Wiley, 2003, 59–75.

[12] Corbet CM, Sonde SS, Tutuc E, Banerjee SK. Improved contact resistance in $ReSe_2$ thin film field-effect transistors. Appl Phys Lett 2016;108:162104.

[13] Chen X, Lei B, Zhu Y, Zhou J, Gao M, Liu Z, et al. Diverse spin-polarized in-gap states at grain boundaries of rhenium dichalcogenides induced by unsaturated Re-Re bonding. ACS Mater Lett 2021;3:1513–1520.

[14] Hu P, Zhang H, Li A, Sheng L, Jiang J, Yu Y, et al. Lateral and vertical morphology engineering of low-symmetry, weakly-coupled 2D $ReS_2$. Adv Funct Mater 2023;33:2210502.